\documentclass[12pt,twoside]{mitthesis}

\usepackage{setspace}
\usepackage{lgrind}
\usepackage[table,xcdraw]{xcolor}
\usepackage{cmap}
\usepackage[T1]{fontenc}
\usepackage{indentfirst}
\usepackage{breqn}

\usepackage{tablefootnote}
\usepackage[perpage]{footmisc}

\usepackage{amsmath}
\usepackage{amssymb}
\usepackage{array}
\usepackage{caption}
  \usepackage{subcaption}
\usepackage{bm}
\usepackage{enumitem}
\usepackage{graphicx}
  \usepackage{epstopdf}
\graphicspath{ {./figs/} }
\usepackage{svg}

\usepackage{tabularx}
\usepackage{lipsum}

\usepackage{float}

\usepackage{cite}

\usepackage[colorlinks=true,bookmarks=true]{hyperref}
  \usepackage{bookmark}
  \usepackage[capitalise]{cleveref}
  \hypersetup{breaklinks=true}
\usepackage[numbers,sort&compress]{natbib}
\usepackage{physics}
\usepackage{pict2e}

\usepackage{slashed}
\usepackage{siunitx}
\usepackage{supertabular}
\usepackage{textcomp}
\usepackage{url}
\usepackage{mathbbol}
\usepackage{changes}

\begin{document}
\hypersetup{pageanchor=false}
\pagenumbering{roman}
%
%
%
%
%
%
%
%
%
%
%


\title{The Main Electrode System of the Nab Experiment and the Analysis of the Performance in the Measurement of the Fierz Term b}

\makeatletter
\author{Huangxing Li} \let\Author\@author

\degree{Doctor of Philosophy}

\degreemonth{Dec}
\degreeyear{2021} \let\Year\@degreeyear
\thesisdate{Dec 1, 2021}
\makeatother


\supervisor{Stefan Bae$\beta$ler}{Professor}

\chairman{N/A}{N/A}

\makeatletter
\def\maketitle{\begin{titlepage}
\doublespacing
\vspace{0.8in}
\large
{\LARGE\bf \@title \par}
\@author \\
\@prevdegrees
\par
A Dissertation Presented to the Graduate Faculty \\
of the University of Virginia in Candidacy for the Degree of \\
\@degree
\par
University of Virginia \\
\@degreemonth, \@degreeyear
\vspace{0.8in}
\end{titlepage}}
\makeatother
\maketitle





\pagestyle{empty}
\newpage
\vspace*{\fill}
\noindent \textcopyright Copyright by \Author {} \Year \\
All Rights Reserved

\cleardoublepage

\vspace*{30px}
\pdfbookmark[0]{\LARGE{Abstract}}{Abstract}
\vspace*{3em}
\section*{Abstract}
%
%


\noindent


The Nab collaboration will study free neutron beta decay at the Spallation Neutron Source at Oak Ridge National Lab. A neutron decays into a proton, an electron and an anti-neutrino in this process, where the energy of the outgoing protons and electrons are collected to determine (1) the electron-antineutrino correlation co-efficient $a$ to the precision of $\abs{\Delta a/a} \leq 10^{-3}$ and (2) the Fierz interference term $b$ to the precision of $\abs{\Delta b} \leq 3 \cross 10^{-3}$. 

From the measurement of $a$, we can calculate the axial-vector to vector coupling ratio $\lambda$. Together with the neutron lifetime measurement, we could also calculate the upper right element of the Cabbibo-Kobayashi-Maskawa Matrix, and test the unitarity of that matrix. The measurement of $b$ could shed light on the physics beyond the Standard Model since $b$ is predicted to be 0 by the $V-A$ structure of weak interaction in the Standard Model.

This thesis presents the design of Nab electrode system, and the solution to a major systematic effect in the measurement of $a$: the requirement of having a low electrical field environment in the neutron decay region. The electrode system has been built and installed successfully, and as our characterization and its analysis shows, the electrode system meets the required specifications.

This thesis also gives a systematic uncertainty study for $b$ measurement, and provides a table of the requirements for the $b$ measurement. The analysis of the systematic uncertainties of the Nab-$b$ measurement shows that the main challenge for the measurement goal of $b$ is the calibration and characterization of the detector, which is achievable. The total systematic uncertainty, as discussed in this thesis, is $\Delta b_{\text{sys}} \sim 2.2\cross10^{-3}$. The statistical uncertainty, as discussed in this thesis, is at the level of $\Delta b_{\text{stat}} \sim 10^{-3}$ given $10^8$ decay events. This conforms to the overall goal of $\Delta b < 3\cross 10^{-3}$. In the Nab experiment, we expect a decay rate of 1600\,Hz, and the run time for $10^8$ decay events is less than two days. 
\cleardoublepage

\pagenumbering{roman}
\pagestyle{plain}
\vspace*{30px}
\pdfbookmark[0]{\LARGE{Acknowledgments}}{Acknowledgments}
\vspace*{3em}
\section*{Acknowledgments}

I would like to thank my advisor Professor Stefan Bae\ss{}ler and Professor Dinko Po\u{c}ani\'{c}. The Ph.D. work would have been much more difficult for me without their detailed guidance and kind instructions. They enlightened me greatly in many aspects, from academic research to everyday life.  

I would like to thank my parents for their consideration and support to my life during this challenging and rewarding Ph.D. study time.

I would also like to thank all my colleagues and friends.

\cleardoublepage


\hypersetup{pageanchor=true}

\hypersetup{linkcolor=black}
\pagestyle{plain}


\cleardoublepage
\pdfbookmark[0]{\contentsname}{Contents}
\tableofcontents
\cleardoublepage
\pdfbookmark[1]{List of Figures}{List of Figures}
\listoffigures
\cleardoublepage
\pdfbookmark[1]{List of Tables}{List of Tables}
\listoftables
\cleardoublepage


\hypersetup{linkcolor=red}

\makeatletter
\def\thickhline{%
  \noalign{\ifnum0=`}\fi\hrule \@height \thickarrayrulewidth \futurelet
   \reserved@a\@xthickhline}
\def\@xthickhline{\ifx\reserved@a\thickhline
               \vskip\doublerulesep
               \vskip-\thickarrayrulewidth
             \fi
      \ifnum0=`{\fi}}
\makeatother
\newlength{\thickarrayrulewidth}
\setlength{\thickarrayrulewidth}{2\arrayrulewidth}

\pagenumbering{arabic}


\chapter{Introduction}
\label{C1}

\section{Introduction to Neutron Beta Decay}
\label{C1S1}
\noindent
In 1920, Earnest Rutherford proposed the nucleus model, which consists of positively charged protons and another type of neutrally charged particles\cite{10.2307/93888}. Rutherford suggested this neutrally charged particle to be a bound state of a proton and an electron. In 1930, Walther Bothe and Herbert Becker found a special type of radiation was made by hitting beryllium, boron, or lithium with alpha rays\cite{1930ZPhy...66..289B}. The radiation was electrically neutral, and was identified to be gamma radiation by Bothe and Becker. After carrying out more research on this radiation, James Chadwick found that the idea of it to be gamma radiation was not correct, and that the neutral radiation actually consists of particles with about the same mass as the proton. In 1932, James Chadwick proposed the idea that these particles were neutrons\cite{Chadwick:1932ma}. In 1934 and 1935, Chadwick and Goldhaber performed future experiments and reported the first accurate mass of neutron\cite{1935RSPSA.151..479C}\cite{Chadwick1934}.

The first observation of radioactivity was made by Henri Becquerel in 1896. Based on the penetration depth, Ernest Rutherford classified the radioactive decays by their decay particles, and distinguished three types: alpha, beta, and gamma decay. In 1900, Henri Becquerel measured the mass to charge ratio of beta rays and identified them to be made from electrons. The energy spectrum of the electrons in beta decay remained confusing and intriguing for researchers at that time, since beta decay was first thought to be a two body process: a neutron was thought to decay into an electron and a proton. According to this interpretation, the outgoing electrons from beta decay should have the total energy of $E_e=(m_n-m_p)c^2$ where $m_n$ and $m_p$ are the rest masses of the neutron and proton, respectively. In 1914, James Chadwick showed that the beta spectrum was continuous\cite{Chadwick:262756}, which was in contradiction to this argument. Another puzzle in beta decay was the angular momentum balance: molecular band spectra showed  that the spin is integral for nuclei of even mass number and half-integral for nuclei of odd mass number. In beta decay, the mass number of nucleus remains unchanged, which means that the change in the nuclear spin must be an integer. Since the outgoing electrons have a spin of 1/2, this contradicts the requirement of angular momentum conservation.

In 1930, Wolfgang Pauli suggested that an extremely light neutral particle was emitted in neutron beta decay in addition to the electron: the neutrino (as it was called later by Enrico Fermi), to account for the requirement of energy and angular momentum conservation. In 1934, Fermi proposed his model of beta decay\cite{1934ZPhy...88..161F}. In his model, Fermi suggested an interaction that couples an electron, a proton, a neutron, and a neutrino.

\section{Fermi's Model of Neutron Beta Decay}
\label{C1S2}
\noindent
In 1934, Fermi proposed the relativistic vector type interaction model for neutron beta decay, which is also called ``four-fermion interaction''. In modern notation, this interaction could be written as\cite{PhysRev.109.193}:
\begin{equation}
 \mathcal{L}_{int} = C(\bar{\psi}_{p}\gamma_{\mu} \psi_{n})(\bar{e}\gamma^{\mu} \nu_{e}) + h.c.
 \label{firsteq}
\end{equation}
Furthermore, Fermi mentioned a possible generalization of the this interaction, he realized that weak decays could be described by:
\begin{equation}
\label{eqLint}
\mathcal{L}_{int} = \sum_{i} C_{i}(\bar{\psi}_{p}O_{i} \psi_{n})(\bar{e}O_{i} \nu_{e}) + h.c.,
\end{equation}
with $O_{i}$ representing different types of interactions:
\begin{align}
 \label{Oeq1}
 &O_{S} = \mathbb{1}\,,&   &\text{Scalar\,,}\\
 &O_{V} = \gamma_{\mu}\,,&   &\text{Vector\,,}\\
 &O_{T} = \sigma_{\mu\nu} = \frac{i}{2} [\gamma_{\mu}, \gamma_{\nu}]\,,&   &\text{Tensor\,,}\\
 &O_{A} = \gamma_{5}\gamma_{\mu}\,,&   &\text{Axial-vector\,,}\\
 &O_{P} = \gamma_{5}\,,&    &\text{Pseudo-scalar\,.}
 \label{Oeq5}
\end{align}

In 1956, Tsung-Dao Lee and Chen-Ning Yang proposed that parity could be violated in the weak interaction\cite{1956PhRv..104..254L}, and this was verified by Chien-shiung Wu's team in 1957\cite{1957PhRv..105.1413W}. This means that Eq.\,\eqref{firsteq} cannot be correct. According to their suggestion, the Langrangian is a specific linear combination of terms given in Eq.\,\eqref{eqLint} - \eqref{Oeq5}. This linear combination can be rewritten as:
\begin{equation}
\begin{split}
\mathcal{L}_{int} &= \sum_{i} C_{i}(\bar{\psi}_{p}O_{i} \psi_{n})(\bar{e}O_{i}\nu_{e}) +\sum_{i} C^{'}_{i}(\bar{\psi}_{p}O_{i}\psi_{n})(\bar{e}O_{i}\gamma_{5}\nu_{e}) + h.c.\\
&= \sum_{i} (\bar{\psi}_{p}O_{i}\psi_{n})(C_{i}\bar{e}O_{i}\nu_{e}+C_{i}^{'}\bar{e}O_{i}\gamma_{5}\nu_{e}) + h.c. \label{eq:2}
\end{split}
\end{equation}
This theory shows the completed Lagrangian under Fermi's initial four-fermion structure. In the non-relativistic limit, scalar and vector interactions result in the Fermi transitions, and tensor and axial-vector interactions result in the Gamow-Teller transitions. The pseudo-scalar interaction is suppressed in the non-relativistic limit, and can be ignored in most of the decay rate calculations at that time, including in Jackson's paper\cite{PhysRev.106.517}.

In 1957, Maurice Goldhaber and his colleague investigated beta decay of $^{152m}$Eu and found that all outgoing neutrinos are left-handed \cite{PhysRev.109.1015}. Later in 1958, Richard Feynman and Murray Gell-Mann proposed the $V-A$ form of the weak interaction\footnote{The $V-A$ theory was actually studied and suggested earilier by Ennackal Chandy George Sudarshan and Robert Marshak, as Feynman 
stated in 1963: “The $V-A$ theory that was discovered by Sudarshan and Marshak, publicized by Feynman and Gell-Mann —”}\cite{PhysRev.109.193}:
\begin{equation}
\begin{split}
\mathcal{L}_{int} &= C_{V}(\bar{\psi}_{p}\gamma_{\mu}\psi_{n})[\bar{e}(\gamma^{\mu}(1-\gamma_{5})\nu_{e}] + C_{A}(\bar{\psi}_{p}\gamma_{5}\gamma_{\mu}\psi_{n})[\bar{e}(\gamma^{\mu}(1-\gamma_{5})\nu_{e}] + h.c.
\end{split}
\end{equation}
Defining $G_{F} =  {\sqrt{2}}C_{V}$ and $\lambda = \frac{C_{A}}{C_{V}}$, the above Lagrangian could be rewritten as:
\begin{equation}
\begin{split}
\mathcal{L}_{int} &= \frac{G_{F}}{\sqrt{2}}[\bar{\psi}_{p}\gamma_{\mu}(1-\lambda\gamma_{5})\psi_{n}][\bar{e}\gamma^{\mu}(1-\gamma_{5})\nu_{e}] + h.c.
\label{4fequation}
\end{split}
\end{equation}

In 1937, Markus Fierz discussed the beta decay rate\cite{1937ZPhy..104..553F}. In his paper, Fierz proposed the modification term $(1+a\frac{\bm{p_{e}} \cdot \bm{p_{\nu}}}{E_{e}E_{\nu}}+b\frac{m_{e}}{E_{e}})$ of the beta decay rate, and discussed that if only the electron spectrum is observed, the term $a\frac{\bm{p_{e}} \cdot \bm{p_{\nu}}}{E_{e}E_{\nu}}$ is unobservable, only the $b\frac{m_{e}}{E_{e}}$ term will survive. In 1957, John David Jackson, Sam Bard Treiman, and Henry William Wyld, Jr. investigated the time reversal invariance in beta decay based of the full interaction Lagrangian in Eq.\,\eqref{eq:2} and wrote the neutron beta decay rate alphabet soup as\cite{PhysRev.106.517,JACKSON1957206}:
\begin{align}
\nonumber
\frac{dw}{dE_{e}d\Omega_{e}d\Omega_{\nu}} =&\frac{F(Z,E_{e})}{{(2\pi)}^{5}}  p_{e}E_{e}(E_{0}-E_{e})^{2}\xi\\
&\cross[1+a\frac{\bm{p_{e}} \cdot \bm{p_{\nu}}}{E_{e}E_{\nu}}+b\frac{m_{e}}{E_{e}}+ \langle \bm{\sigma}_{n} \rangle\cdot(A\frac{\bm{p_{e}}}{E_{e}}+B\frac{\bm{p_{\nu}}}{E_{\nu}}+D\frac{\bm{p_{e}}\cross\bm{p_{\nu}}}{E_{e}E_{\nu}})]\,,\label{eq:3}
\end{align}
with $\xi, a, b, A, B$ as following\footnote{
\begin{align}
\gamma =(1-\alpha^2 Z^2)^{\frac{1}{2}}\;, \quad
  \lambda_{J'J} =
    \begin{cases}
      1 & J'=J-1\\
      \frac{1}{J+1} & J'=J\\
      -\frac{1}{J+1} & J'=J+1
    \end{cases}   
\;, \quad
  \Lambda_{J'J} =
    \begin{cases}
      1 & J'=J-1\\
      -\frac{1(2J-1)}{J+1} & J'=J\\
      -\frac{J(2J-1)}{(J+1)(2J+3)} & J'=J+1
    \end{cases}  
    \;.
\end{align}Here $\alpha$ is the fine-structure constant, $Z$ is the atomic number of the decaying particle, $M_{F}$ and $M_{GT}$ are the Fermi nuclear matrix element and Gamow-Teller matrix element, $F(Z,E)$ is the Fermi function which comes from the correction of the electron energy, $J$ and $J'$ are the angular momentum of the nuclei before and after decay, respectively}:
\begin{align}
\nonumber
\xi = &\abs{M_{F}}^{2}(\abs{C_{S}^{\vphantom{\prime}}}^2+\abs{C_{V}}^2+\abs{C'_{S}}^2+\abs{C'_{V}}^2)\\
+&\abs{M_{GT}}^{2}(\abs{C_{T}}^2+\abs{C_{A}}^2+\abs{C'_{T}}^2+\abs{C'_{A}}^2)\,,\\
\nonumber
a\xi = &\abs{M_{F}}^{2}\bigg\{[-\abs{C_{S}}^2+\abs{C_{V}}^2-\abs{C'_{S}}^2+\abs{C'_{V}}^2]\mp \frac{\alpha Zm}{p_{e}} 2 \text{Im} (C_{S}C^{*}_{V}+C'_{S}{C'}^{*}_{V})\bigg\}\\
+&\frac{\abs{M_{GT}}^{2}}{3}\bigg\{\abs{C_{T}}^2-\abs{C_{A}}^2+\abs{C'_{T}}^2-\abs{C'_{A}}^2 \pm \frac{\alpha Zm}{p_{e}} 2 \text{Im} (C_{T}C^{*}_{A}+C'_{T}{C'}^{*}_{A} )\bigg\}\,,\\
\label{absoupparas}
b\xi=&\pm 2 \gamma \text{Re}[{\abs{M_{F}}}^{2}(C_{S} C^{*}_{V}+C'_{S} {C'}^{*}_{V})+{\abs{M_{GT}}}^{2}(C_{T} C^{*}_{A}+C'_{T} {C'}^{*}_{A})]\,,\\
\nonumber
A\xi = &\abs{M_{GT}}^{2}\lambda_{J'J} [\pm 2\text{Re}(C_{T}{C'}^{*}_{T}-C_{A}{C'}^{*}_{A}) + \frac{\alpha Zm}{p_{e}} 2 \text{Im} (C_{T} {C'}^{*}_{A}+{C'}_{T} {C}^{*}_{A})\\
+&\delta_{J'J}M_{F}M_{GT}{\sqrt{\frac{J}{J+1}}}[2\text{Re} (C_{S}{C'}^{*}_{T}+C'_{S}{C'}^{*}_{T}-C_{V}{C'}^{*}_{A}-C'_{V}C_{A}^{*})\\
\nonumber
\pm &\frac{\alpha Zm}{p_{e}}2\text{Im} (C_{S}{C'}^{*}_{A}+C'_{S}{C'}^{*}_{A}-C_{V}{C'}^{*}_{T}-C'_{V}C_{T}^{*})\,,\\
\nonumber
B\xi = &2\text{Re}\bigg\{\abs{M_{GT}}^{2}\lambda_{J'J}[\frac{\gamma m}{E_{e}}(C_{T}{C'}^{*}_{A}+C'_{T}C_{A}^{*})\pm(C_{T}{C'}^{*}_{T}+C_{A}{C'}^{*}_{A})]\\
-&\delta_{J'J}M_{F}M_{GT}\sqrt{\frac{J}{J+1}}[(C_{S}{C'}^{*}_{T}+{C'}_{S}C_{T}^{*}+C_{V}{C'}^{*}_{A}+{C'}_{V}C_{A}^{*})\\
\nonumber
\pm& \frac{\gamma m}{E_{e}}(C_{S}{C'}^{*}_{A}+{C'}_{S}C_{A}^{*}+C_{V}{C'}^{*}_{T}+{C'}_{V}C_{T}^{*})]\bigg\}\,,\\
D\xi = &2\text{Im}\bigg\{\delta_{J'J}\abs{M_{F}}\abs{M_{GT}}\sqrt{\frac{J}{J+1}}(C_{S}{C}^{*}_{T}-C_{V}{C}^{*}_{A}+C'_{S}{C'}_{T}^{*}-C'_{V}C_{A}^{*})\bigg\}
\end{align}
Note that if the weak interaction takes $V-A$ form \eqref{4fequation}, then $a= \frac{1-\abs{\lambda}^2}{1+3\abs{\lambda}^2}$, and Fierz interference term $b$ vanishes.

\section{Neutron Beta Decay in the Standard Model}
\label{C1S3}
\begin{figure}[h!]
\centering
\includegraphics[scale=0.24]{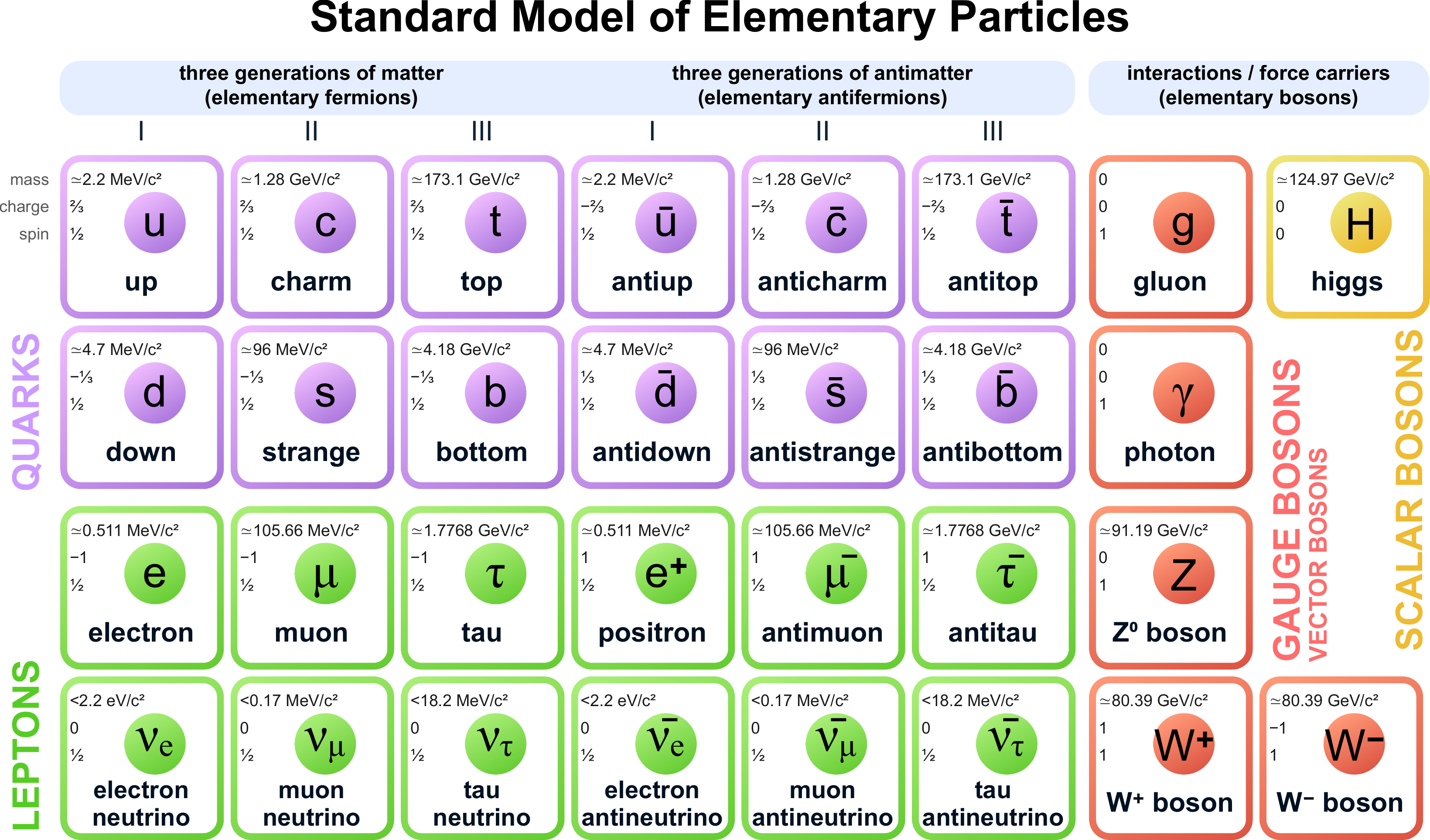}
 \caption{The Particle Table of the SM.}
 \label{Standard_Model_of_Elementary_Particles_w_anti}
\end{figure}

\noindent
The Standard Model of elementary particle physics (SM) is the most fundamental physics model established in 20th century, which had been experimentally tested to a very high precision by many measurements. In the SM, the fundamental particles are classified as fermions with half integer spin and bosons with integer spin. Fermions are the building blocks of matter and bosons are the mediators of interactions among fermions. As shown in Fig.\,\ref{Standard_Model_of_Elementary_Particles_w_anti}, the SM includes twelve types of fermions and five types of bosons. 
Fermions are further divided into quarks and leptons, with six types of quarks and six types of leptons. In the SM, neutron has the quark content of one up quark and two down quarks, and proton has the quark content of two up quarks and one down quark. Under the SM, the neutron beta decay is described as the transition of a down quark into an up quark by emitting a W$^{-}$ boson, which then decays into an electron-antineutrino pair. 

the SM preserves the $V-A$ form of the weak interaction\cite{pich2012standard}:
\begin{equation}
\begin{split}
\mathcal{L}_{cc} &= -\frac{g}{2\sqrt{2}}\{W^{\dagger}_{\mu}[\sum_{ij}\bar{u}_{i}\gamma^{\mu}(1-\gamma_{5})V_{ij}d_{j}+\sum_{l}\bar{\nu}_{l}\gamma^{\mu}(1-\gamma_{5})l]+h.c.\}\,.
\end{split}
\end{equation}
The Cabibbo-Kobayashi-Maskawa (CKM) matrix element $V_{ij}$ arises because the weak eigenstates are not the same as mass eigenstates for quarks.
\begin{figure}[ht]
\centering
\begin{minipage}[b]{0.45\linewidth}
\includegraphics[scale=0.6]{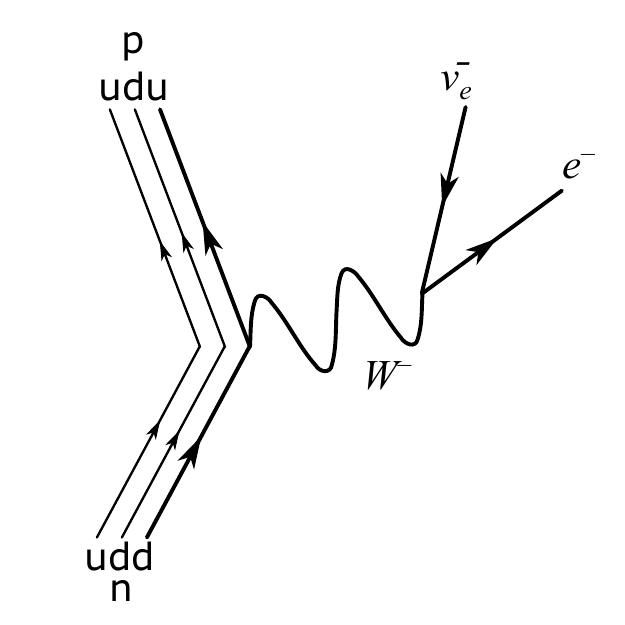}
\caption{Beta decay in the SM.}
\label{fig:minipage1}
\end{minipage}
\quad
\begin{minipage}[b]{0.45\linewidth}
\includegraphics[scale=0.6]{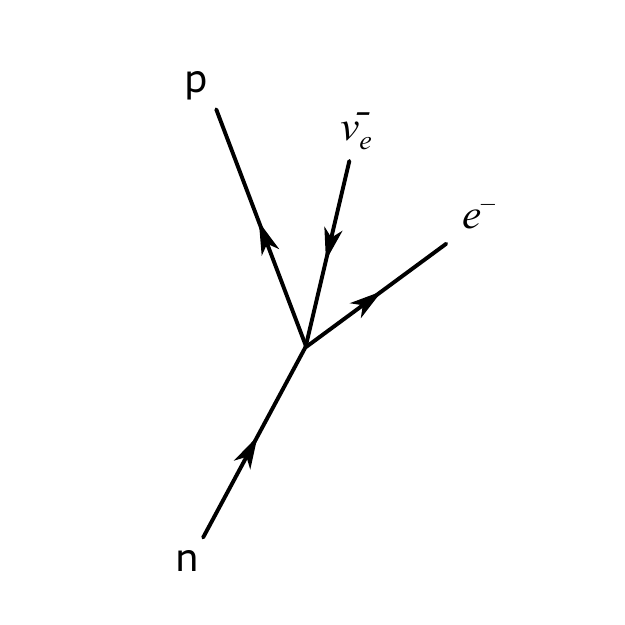}
\caption{Beta decay in four-fermion Theory.}
\label{fig:minipage2}
\end{minipage}
\end{figure}

From the Feynman diagram in Fig.\,\ref{fig:minipage1}, the transition matrix for neutron beta decay is:
\begin{equation}
T_{fi}=\frac{g^{2}V_{ud}}{8}\bar{e}\gamma^{\mu}(1-\gamma^{5})\nu_{e}D_{\mu\nu}\bra{p}\bar{u}\gamma^{\mu}(1-\gamma^{5})d+h.c.\ket{n}\,.
\end{equation}
The propagator of W boson $D_{\mu\nu}$ is:
\begin{equation}
D_{\mu\nu}=\frac{g_{\mu\nu}-\frac{q_{\mu}q_{\nu}}{m^{2}_{W}}}{q^{2}-m^{2}_{W}}\,,
\end{equation}
here $q_{\mu}$ is the 4-momentum of the propagator. When the energy scale is much less than the mass of the W boson ($\abs{q^{2}} \ll m^{2}_{W}$), the W boson propagator approximates to:
\begin{equation}
D_{\mu\nu}\approx -\frac{g_{\mu\nu}}{m^{2}_{W}}\,.
\end{equation}
Due to the complication of the nucleon structure, it is not possible to compute the term $\bra{p}\bar{u}\gamma^{\mu}(1-\gamma^{5})d\ket{n}$ with sufficient precision. So instead of calculating the term, we define the form factors $C_A$ and $C_V$ as following:
\begin{equation}
\begin{split}
\bra{p}\bar{u}\gamma^{\mu}(1-\gamma^{5})d+h.c.\ket{n} = \bar{\psi_{p}}(C_{V}\gamma^{\mu} - C_{A}\gamma^{\mu}\gamma^{5})\psi_{n}\,.
\end{split}
\end{equation}
Defining:
\begin{equation}
\begin{split}
\frac{C_{V}}{C_{A}} &= \lambda\,,
\end{split}
\end{equation}
the transition amplitude is then:
\begin{equation}
T_{fi}=\frac{g^{2}C_{V}V_{ud}}{8 M^{2}_{W}}{\sqrt{2}}[\bar{e}\gamma^{\mu}(1-\gamma^{5})\nu_{e}][\bar{\psi_{p}}\gamma_{\mu}(1 - \lambda\gamma^{5})\psi_{n}]\,.
\label{sdmodel}
\end{equation}
This transition amplitude returns to the form of four-fermion theory at low energy.

\section{Testing the Standard Model with Neutron Beta Decay}
\label{C1S4}
\noindent
Regardless of its great success in many experimental predictions, the SM is still being considered as incomplete. The first defect comes from the absence of gravitational interaction. As the first fundamental interaction to be described mathematically, gravity has been studied by physicists for many centuries. Starting from Newton's law of universal gravitation to Einstein's theory of general relativity, the macroscopic theory of gravity has been well-established and tested to very high precision. Yet the SM doesn't include this interaction, and so far, all attempts to extend it to include gravity are speculative, at best. The second defect comes from the neutrino mass. According to the SM, neutrinos are considered to be massless. However, the result from neutrino oscillation experiment\cite{PhysRevLett.81.1562, DAVIS199413, PhysRevD.83.073006} indicates that neutrinos are massive. Another important defect is that the SM does not include dark matter and dark energy, which are believed to make of 95\% the energy content of the universe.

As an interesting topic for very long time, neutron beta decay could provide a lot of detailed information about the SM, and shed light to physics beyond the SM. The $V-A$ structure of the weak interaction is established step by step through seminal and influential experiments, yet there is not an natural explanation for all the neutrinos being left handed and parity not being conserved. The study of neutron beta decay could hopefully also provide some information on these topics. 

\subsection{The Ratio of Vector to Axial-Vector Coupling Strength}
\label{C1S4S1}
\noindent
For unpolarized neutron ($\langle \bm{\sigma_{n}} \rangle = 0$), the decay rate Eq.\,\eqref{eq:3} could be written as:
\begin{equation}
\begin{split}
\frac{dw}{dE_{e}d\Omega_{e}d\Omega_{\nu}} &=G^{2}_{F} \abs{V_{ud}}^2(1+3\abs{\lambda}^2)\frac{F(Z,E_{e})}{{(2\pi)}^{5}}  p_{e}E_{e}(E_{0}-E_{e})^{2}(1+a\frac{\bm{p_{e}} \cdot \bm{p_{\nu}}}{E_{e}E_{\nu}}+b\frac{m_e}{E_e})\\
&\propto F(Z,E_{e})  p_{e}E_{e}(E_{0}-E_{e})^{2}(1+a\frac{\bm{p_{e}} \cdot \bm{p_{\nu}}}{E_{e}E_{\nu}}+b\frac{m_e}{E_e})\,.\label{eq:5}
\end{split}
\end{equation}
From the low energy approximation in Eq.\,\eqref{sdmodel}, to the leading order, for free neutron beta decay we have:
\begin{equation}
\begin{split}
\xi &= G^{2}_{F} \abs{V_{ud}}^2(1+3\abs{\lambda}^2)\,,\;\;\;\;\;\;  a = \frac{1-\abs{\lambda}^2}{1+3\abs{\lambda}^2}\,, \;\;\;\;\;\;  b = 0\,.\label{eq:6}
\end{split}
\end{equation}
The electron-antineutrino correlation coefficient $a$ can be extracted from Eq.\,\eqref{eq:5}, and used to evaluate the the ratio of vector to axial-vector coupling strength $\abs{\lambda}$.

\subsection{The Unitarity of Cabibbo-Kobayashi-Maskawa (CKM) matrix}
\label{C1S4S2}
\noindent
In the SM, the strong-interaction mass eigenstates of the quarks are not the same as eigenstates of weak interactions, and these sets of eigenstates are transformed into each other by the CKM matrix (denoted as $V$):
\begin{equation}
\begin{pmatrix}
d'\\
s'\\
b'
\end{pmatrix}
=
\begin{pmatrix}
V_{ud} & V_{us} & V_{ub}\\
V_{cd} & V_{cs} & V_{cb}\\
V_{td} & V_{ts} & V_{tb}\\
\end{pmatrix}
\begin{pmatrix}
d\\
s\\
b
\end{pmatrix}\,.
\end{equation}
An necessary property of the CKM matrix is its unitarity:
\begin{equation}
VV^{\dagger} =  \mathbb{1}\,.
\end{equation}
In particular, for the elements in the first row, we should have:
\begin{equation}
\abs{V_{ud}}^2+\abs{V_{us}}^2+\abs{V_{ub}}^2 = 1\,.\label{eq:1}
\end{equation}
Together with other experiments, the study of neutron beta decay can provide a method to measure $\abs{V_{ud}}^{2}$, and verify the unitarity of the CKM matrix. As of now, the best determination of the CKM matrix gives\cite{}:
\begin{equation}
\abs{V_{ud}}^2+\abs{V_{us}}^2+\abs{V_{ub}}^2 = 0.9985 \pm 0.0005\,,
\end{equation}
which fails the test of unitarity at the 3$\sigma$ level.

Integrating the decay rate Eq.\,\eqref{eq:5} assuming the validity of the SM ($b = 0$), we have:
\begin{dmath}
\frac{1}{\tau}= \int d\omega = \frac{G^{2}_{F} \abs{V_{ud}}^2(1+3\abs{\lambda}^2) }{{(2\pi)}^{5}}\int F(Z,E_{e})p_{e}E_{e}(E_{0}-E_{e})^{2}(1+a\frac{\bm{p_{e}} \cdot \bm{p_{\nu}}}{E_{e}E_{\nu}})dE_{e}d\Omega_{e}d\Omega_{\nu}
=\frac{G^{2}_{F} \abs{V_{ud}}^2(1+3\abs{\lambda}^2) m_{e}^{5} }{2\pi^{3}} f\,,
\end{dmath}
with
\begin{equation}
f= \frac{1}{m_{e}^{5}}\int F(Z,E_{e})p_{e}E_{e}(E_{0}-E_{e})^{2}d E_{e}\,,\\
\end{equation}
and $\tau$ is the mean lifetime of free neutron.

With $\lambda$ from the measurement of the electron-antineutrino correlation coefficient $a$ in Nab, $\tau$ from the measurement of  neutron lifetime, and $G_F$ from the measurement of muon lifetime, we can calculate the upper left element $V_{ud}$ of CKM matrix. Since $V_{ud}$ has the highest contribution to Eq.\,\eqref{eq:1}, a high precision measurement of $V_{ud}$ could help us test the unitarity of CKM matrix.
\subsection{The Fierz Interference Term $b$}
\label{C1S4S3}
\noindent
Since the electroweak interaction part of the SM has the $(V-A)\otimes(V-A)$ form, the Fierz interference term $b=0$ at leading order.

The electron spectrum could be retrieved by integrating out $\Omega_e$ and $\Omega_{\nu}$ from Eq.\,\eqref{eq:5}, note that only the Fierz interference term survives:
\begin{equation}
\Gamma (E_e) = \frac{d\omega}{dE_e} \propto F(Z,E_{e}) p_{e}E_{e}(E_{0}-E_{e})^{2}(1+b\frac{m_e}{E_e})\label{espec}\,.
\end{equation}
The Fierz interference term $b$ could be obtained from the precise measurement of the beta decay electron spectrum. Any deviation of its value from $b=0$ would indicate the existence of scalar and tensor interactions from physics beyond the SM.

\begin{figure}[h!]
\centering
\includegraphics[scale=0.8]{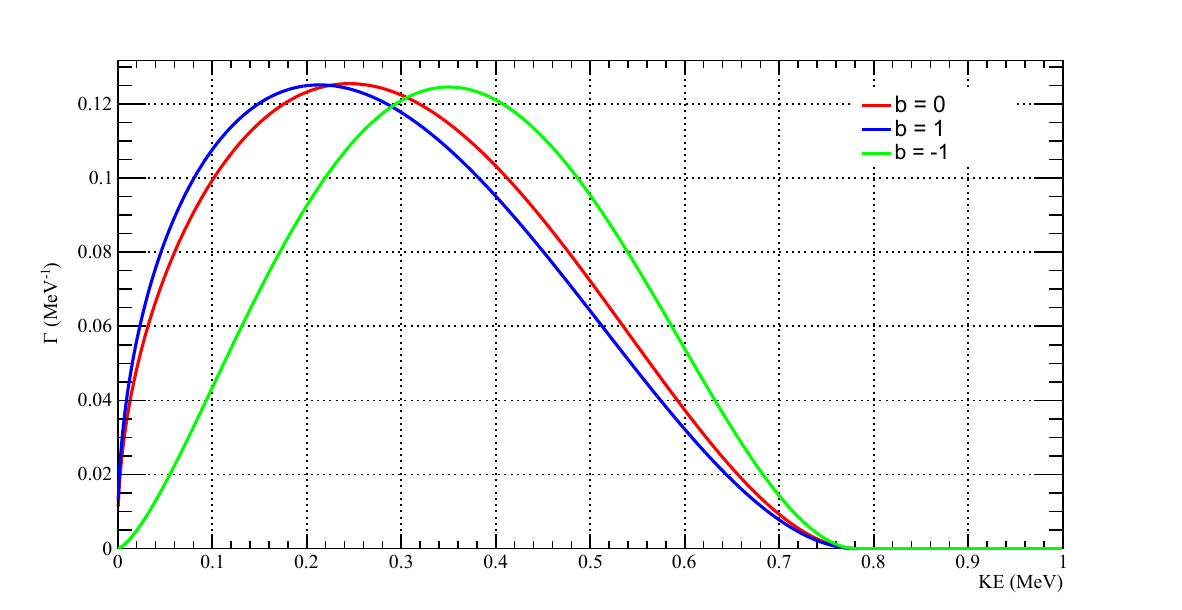}\label{fig_espec}
 \caption{The normalized beta decay spectrum with different $b$ values.}
 \label{fig:espec}
\end{figure}

As discussed above, the Fierz interference term is absent in the SM. The best determination of the Fierz term, consistent with 0, comes from the analysis of superallowed beta decays. Eq.\,\eqref{absoupparas} shows that superallowed beta decays (for which $M_{GT}=0$) are only affected by vector and scalar interactions. We expect vector interactions to be present, and the agreement between various parent nuclei allows us to put limit on scalar interactions (that is, a linear combination of $C_S$ and $C'_S$). Our measurements in neutron beta decay, combined with those from superallowed Fermi decays, allow us to find or constrain tensor interactions (that is, a linear combination of $C_T$ and $C'_T$) in the weak interaction.

\chapter{The Nab Experiment}
\label{C2}

\section{Theory and Method}
\label{C2S1}
\noindent
The Nab collaboration intends to measure the electron-antineutrino correlation coefficient $a$ to $\Delta a / a \sim 10^{-3}$, and the Fierz interference term at the level of $\Delta b \leq 3 \cross 10^{-3}$ in free neutron beta decay\cite{Fry:2018kvq}. As shown in Eq.\,\eqref{eq:6}, this will also give a result of the hadronic mixing ratio $\lambda$ to $\Delta \lambda / \lambda \sim 3 \cross 10^{-4}$\cite{10.1093/ptep/ptaa104}, which is competitive compared to the current experiment result $\lambda = -1.2756(13)$ provided by the Particle Data Group (PDG20)\cite{10.1093/ptep/ptaa104}. The Nab experiment detects decay electrons and protons from neutron beta decay. It imposes coincidence conditions to ensure that both particles are from the same decay event. For the electron, the total kinetic energy $E_e$ is meausured, and for the proton, the squared proton momentum $p_{p}^2$ is meausured. The measured data will be fitted according to Eq.\,\eqref{eq:5} to extract the coefficients of the electron-antineutrino correlation coefficient $a$ and the Fierz interference term $b$.

\subsection{The Extraction of the Electron-Antineutrino Correlation Coefficient $a$}
\label{C2S1S1}
\noindent
In Eq.\,\eqref{eq:5}, we obtain 
\begin{equation}
\begin{split}
\frac{dw}{dE_{e}d\Omega_{e}d\Omega_{\nu}} & \propto F(Z,E_{e})  p_{e}E_{e}(E_{0}-E_{e})^{2}(1+a\frac{\bm{p_{e}} \cdot \bm{p_{\nu}}}{E_{e}E_{\nu}}+b\frac{m_e}{E_e})\\
&= \rho(E_e) (1+a\frac{p_{e}}{E_{e}}\text{cos}\theta_{e\nu}+b\frac{m_e}{E_e})\,,
\end{split}
\label{decayrate}
\end{equation}
here $\theta_{e\nu}$ is the angle between the outgoing electron and the anti-neutrino.

\begin{figure}[H]
\centering
\includegraphics[scale= 0.8]{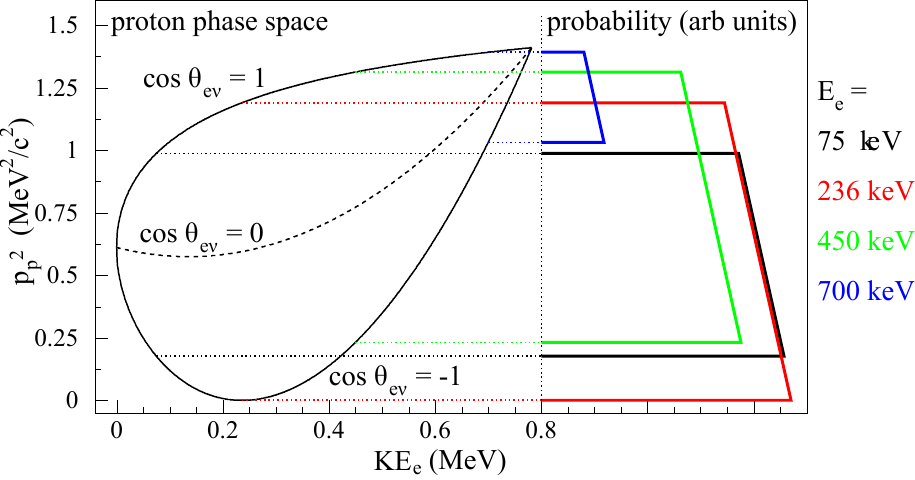}
 \caption[The teardrop plot for electron kinetic energy (KE$_e$) and proton energy ($p^2_p$).]{The teardrop plot for electron kinetic energy (KE$_e$) and proton energy ($p^2_p$). As shown on the right side of the plot, for any fixed KE$_e = E_e - m_ec^2$ , the distribution $P(p^2_p)$ of allowed $p^2_p$ will be linear in $p^2_p$, with a slope $\propto a$.}
 \label{teardrop}
\end{figure}

For free neutron beta decay, the momentum conservation is in the center-of-momentum frame: 
\begin{equation}
    \bm{p_{e}} + \bm{p_{\nu}}= -\bm{p_{p}}\,.
\end{equation}
We square this equation, solve for $\text{cos}(\theta_{e\nu})$, obtaining:
\begin{equation}
    \text{cos}(\theta_{e\nu}) = \frac{p^2_p-p^2_e-p^2_\nu}{2p_ep_\nu}\,.
    \label{momentumconserv}
\end{equation}
From the Eq.\,\ref{momentumconserv}, we find that for fixed $E_e$, cos($\theta_{e\nu}$) is linearly related to $p^2_p$ for fixed $E_e$ since the antineutrino energy $E_\nu$ could be determined from $E_\nu + E_e = E_0$. Thus a measurement of $E_e$ and $p^2_p$ can determine cos($\theta_{e\nu}$). Assume $b=0$ from the SM, we can find that the decay rate in Eq.\,\eqref{decayrate} is linearly dependent on $p^2_p$ within the allowed range, as shown in Fig.\,\ref{teardrop}. The value of $a$ could be extracted by fitting the decay rate to this two-dimensional histogram.

The direct measurement of the proton energy or momentum is difficult given that the proton has very a low kinetic energy ($\sim 10^2$\,eV level). Instead of directly measuring the proton energy, the Nab experiment uses an asymmetric spectrometer to measure the time of flight for proton ($t_p$), and find $p^2_p$ from $1/t^2_p$\cite{Alarcon2010FundingPF}.

\begin{figure}[h!]
\centering
\includegraphics[scale=1]{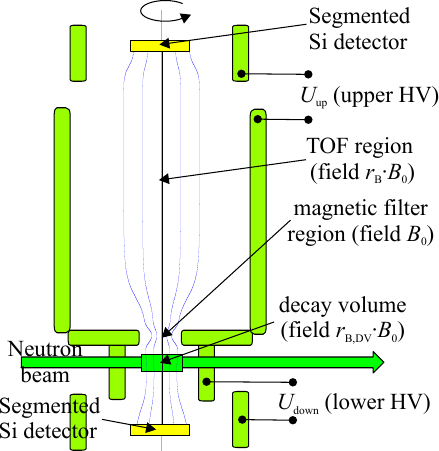}
 \caption[The Nab spectrometer.]{The Nab spectrometer\cite{Alarcon2010FundingPF}.}
\end{figure}

\begin{figure}[H]
    \centering
    \includegraphics{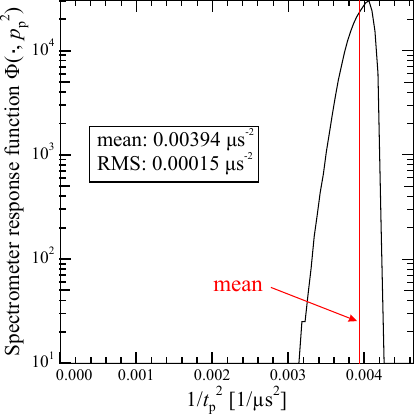}
    \caption[Detector response for a proton with $p^2_p$ = 0.95\,MeV/c.]{Detector response for a proton with $p^2_p$ = 0.95\,MeV/c\cite{Alarcon2010FundingPF}.}
    \label{phi}
\end{figure}

The Nab experiment will use a 7 meters high spectrometer, with 5 meters of time of flight (TOF) region. On each side of the spectrometer there's a silicon detector. The upper silicon detector will be located 5 meters above the decay region and the lower detector will be located 1.2 meters below the decay region. As discussed in section \ref{C2S2}, protons with $\bm{p_{p}}$ satisfying certain criteria will pass the filter region and be guided by the magnetic field line to the upper detector, the time of flight ($t_p$) for these protons will be measured. To relate $t^2_p$ with $p^2_p$, we have\cite{Alarcon2010FundingPF}:

\begin{equation}
P_t(1/t^2_p) = \int P_p(p^2_p) \Phi (1/t^2_p, p^2_p) dp^2_p\,,
\end{equation}
with $P_p(p^2_p)$ to be the $p^2_p$ distribution of outgoing protons, and $\Phi (1/t^2_p, p^2_p)$ to be the proton TOF response function, which maps $p^2_p$ to $1/t^2_p$ by averaging out the other unobserved quantities: the decay position, and the angle between the proton momentum and the magnetic field line at the decay position. For the Nab spectrometer, the width of $\Phi (1/t^2_p, p^2_p)$ is relatively small, Fig.\,\ref{phi} shows the plot of $\Phi (1/t^2_p, p^2_p)$.

\subsection{The Extraction of Fierz Interference Term $b$}
\label{C2S1S1}
\noindent
As shown in Eq.\,\eqref{espec}, the spectrum of neutron beta decay will be detected and fitted to extract the Fierz interference term $b$.

\begin{figure}[h!]
\centering
\includegraphics[scale=0.7]{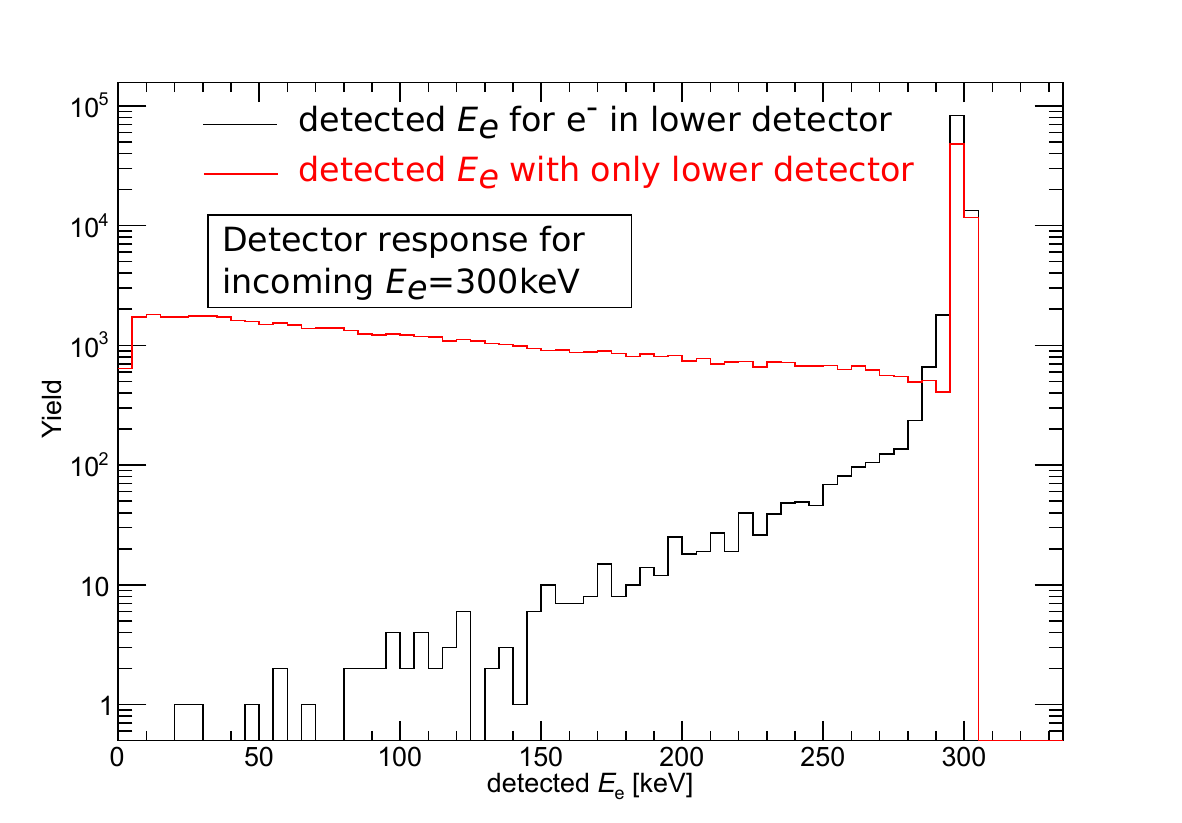}
 \caption[The energy distribution with only lower detector versus with both detectors.]{The energy distribution with only lower detector versus with both detectors. This figure shows the simulated distribution of detected energies for the case that only the lower detector is present (red line) or with both detectors (black line). With only the lower detector, we have a large low-energy tail, with both detectors, the tail of the detected $E_e$ is greatly suppressed. The threshold is 10 keV for the upper detector, and 10 keV for the first hit in the lower detector, and 40 keV for later hits in the lower detector. The energy is reconstructed by summing up all the hits above threshold. Due to the detector high voltage, an additional 30 keV or 1 keV is added to the total energy based on the last hit detector.}
 \label{backscattering}
\end{figure}

Given the electron energy endpoint of 0.782\,MeV, simulation shows that about 20\% of the electrons will be backscattered at least once, depositing only part of their energy in the detector. The Nab experiment will make use of both upper and lower detectors to collect electron hits. With this setup, a backscattered electron may hit the opposite detector. It may also reflect from the high magnetic field in the filter as discussed in chapter\,\ref{C3}, and come back to the first detector. If only one detector was used, the energy carried away by a backscattered electron going to the second detector would not be detected. The distribution of detected energies for monoenergetic electrons would have a tail to the low energy end because the detected electron energy is less than the true electron energy. With two detectors, the energy carried away by the backscattered electron could be recovered, and the electron kinetic energy could be reconstructed from all hits in both detectors. This helps to suppress the low energy tail of the detected electron energy distribution. As shown in Fig.\,\ref{backscattering}, if we sum up the deposited energy in both detectors, and add 30 keV or 1 keV to the total detected energy to make up for the detector voltage setting, we get the reconstructed energy shown as the black line.

\section{Experimental Setup}
\label{C2S2}

\begin{figure}[h!]
\centering
\includegraphics[scale=1]{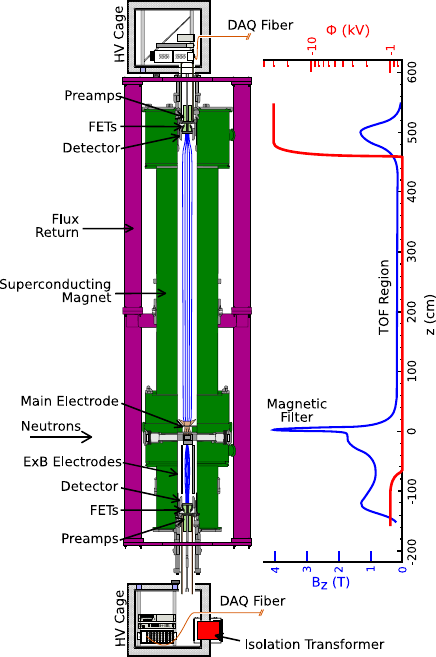}
\caption[The Nab spectrometer setup for Nab-$a$ configuration.]{The Nab spectrometer setup for Nab-$a$ configuration\cite{Fry:2018kvq}.}
\label{spectrometerdetail}
\end{figure}

\noindent
In the Nab experiment, according to the adiabatic approximation, electrons and protons born from the neutron beta decay will be guided by magnetic field lines. However, due to the difference of the initial momentum, the electron and proton from the same decay event usually have different guiding center\footnote{In the presence of the magnetic field and electric field together, for slow varying magnetic field, the charged particles' motion could be treated as spiral around the magnetic field line, the gyration center is called guiding center. In the next order approximation, the guiding center will be slowly drifting\cite{Jackson:1998nia}.}. The decay region has a characteristic magnetic field strength of 1.7\,T. Above the decay region, there is a strong magnetic field filter region with a characteristic magnetic field strength of 4\,T. Due the effect of the filter region, only part of the protons will be able to go to the top detector. We discuss this below for the Nab-$a$ configuration. 

For a proton moving in the magnetic field, assuming the adiabatic approximation, the magnetic flux through the particle's gyration orbit is constant:
\begin{equation}
    \pi r^2B = \pi r^2_0B_0\,,
      \label{adiabaticr}
\end{equation}
here $r$ is the gyration radius, $B$ is the magnetic field strength, and the subscript ``0'' means it is the value of the parameter at birth. Denoting the angle between the momentum of the proton and the magnetic field direction at its guiding center to be $\theta$, we obtain:
\begin{equation}
    r = \frac{\gamma m v }{Be}\text{sin}(\theta)\,,
    \label{mvbq}
\end{equation}
where $v$, $e$, $m$ are the velocity, charge, and mass of the proton, respectively, and $\gamma = \sqrt{(1-(v/c)^2)}$. The protons from neutron beta decay have kinetic energy end point lower than 750\,eV, and can be treated as nonrelativistic. Combining Eq.\,\eqref{adiabaticr} and Eq.\,\eqref{mvbq}, we have:
\begin{equation}
    \frac{v^2_\bot}{B} = \frac{v^2\text{sin}^2(\theta)}{B} = \frac{v^2_{\bot,0}}{B_0}\,,
    \label{vperp}
\end{equation}
where $v_\bot$ is the velocity in the direction perpendicular to the magnetic field. For non-relativistic protons, the velocity satisfies:
\begin{equation}
    v^2 = v^2_0 - \frac{2e(U-U_0)}{m}\,,
\end{equation}
here $U$ is the electrical potential. The parallel component of the velocity ($v_\parallel$) is:
\begin{equation}
\begin{split}
    v^2_\parallel &= v^2 - v^2_\bot \\
    &= v^2_0 - \frac{2e(U-U_0)}{m} - \frac{v^2_{\bot,0}}{B_0}B\\
    &= v^2_0 - \frac{2e(U-U_0)}{m} - \frac{v^2_0\text{sin}^2(\theta_0)}{B_0}B\,.
\label{filter}
\end{split}
\end{equation}

The magnetic field strength in the filter region is greater than that in the decay volume. Assuming the electrical potential is constant throughout these regions ($U-U_0 = 0$), with $v^2_\parallel > 0$, we have:
\begin{equation}
    B < \frac{B_0}{ \text{sin}^2(\theta_0)}\,.
\end{equation}

The above result shows that in order for the proton to pass through the filter region, the magnetic field strength ($B$) at its guiding center along the field line needs to be always less than $\frac{B_0}{\text{sin}^2(\theta_0)}$. Assume the maximum magnetic field strength along the field line is $B_{\text{max}}$, the above requirement is identical to:
\begin{equation}
\nonumber
B_{\text{max}} < \frac{B_0}{\text{sin}^2(\theta_0)}\,,
\end{equation}
or
\begin{equation}
\text{sin}^2(\theta_0) <\frac{B_0}{B_{\text{max}}}\,.
\end{equation}
Plugging $B_{\text{max}} = $4\,T and $B_0 =$ 1.7\,T in the above equation, we have:
\begin{equation}
\nonumber
    \text{sin}^2(\theta_0) < 0.425\,,
\end{equation}
or
\begin{equation}
\text{cos} (\theta_0) > 0.758\,.
\label{creteria}
\end{equation}

\begin{figure}[h!]
\centering
\includegraphics[scale=0.7]{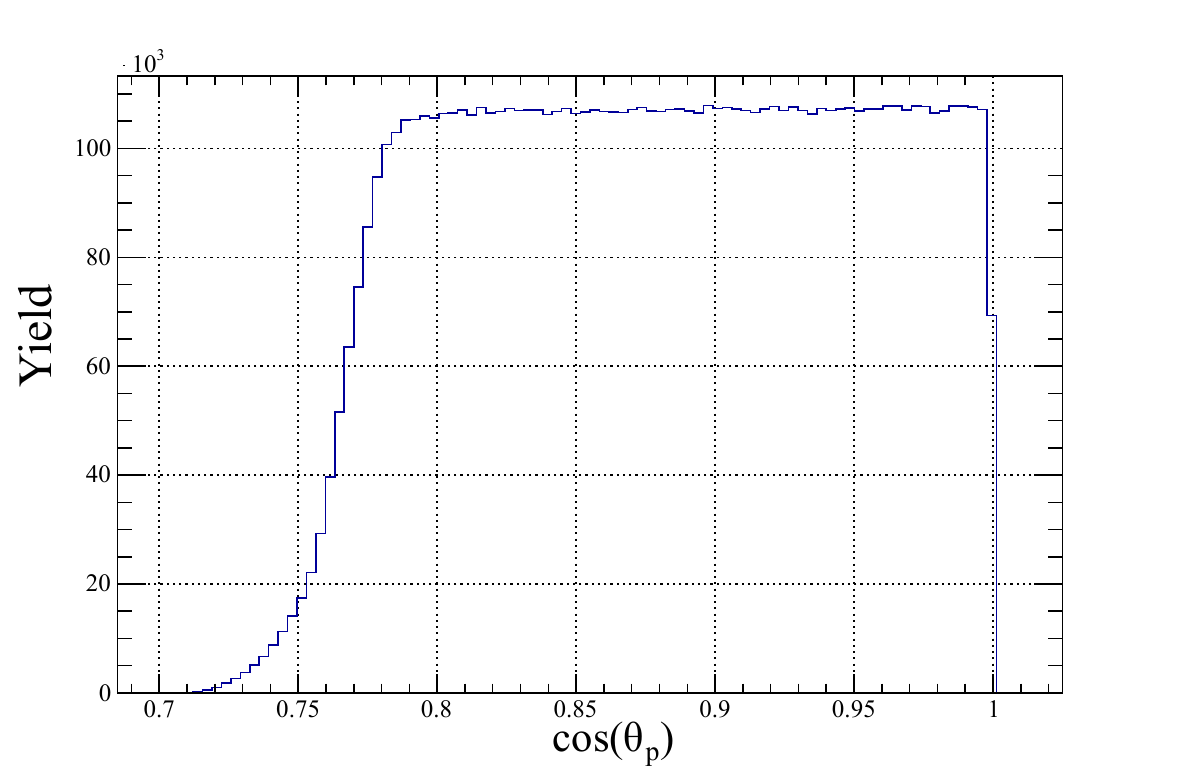}
\caption[The distribution of cos($\theta_p$) for protons that are able to pass the filter region.]{The distribution of cos($\theta_p$) for protons that are able to pass the filter region. The plot shows a soft cutoff cos($\theta_p$) around 0.75. This is different from the result in Eq.\,\eqref{creteria} because 1) $\theta_p$ here is the angle between proton momentum direction at birth and the $z$ direction, while $\theta_0$ in Eq.\,\eqref{creteria} is the angle between proton momentum direction and the magnetic field direction, 2) the actual values of $B_{\text{max}}$ and $B_0$ vary with field lines, the values used in the calculation are only the characteristic values. The second reason primarily accounts for the reason of the smearing since the maximum magnetic field strength on-axis is 4\;T but for off-axis the maximum strength is 4.23\;T.}
\label{dep_pPz}
\end{figure}

Fig.\,\ref{dep_pPz} shows the simulation results for the distribution of protons that are able to pass the through filter region. In this simulation, protons are born at the decay region with momentum direction uniformly in $4\pi$ solid angle.

The filtered protons then go through a 5 meter long TOF region with a characteristic magnetic field strength of 0.157\,T. The field line will guide the protons to the upper detector. The long TOF region volume allows for a precise measurement of the proton momentum from the proton TOF because protons with different momentum will be separate out during the long travel in the TOF region, with the travel time through the TOF region being roughly inversely proportional to the proton momentum.

On each side of the spectrometer, there is a silicon detector. For Nab-$a$ configuration, the top detector will be at a high voltage of $-$30\,kV to accelerate the protons at the end of TOF region, as shown in Fig.\,\ref{spectrometerdetail}. While protons are only detected on the upper detector, electrons will be detected in both detectors and the coincidence requirements in time and detected location will be applied to pair protons and electrons from the same decay event. The TOF of the proton will be estimated from the arrival time difference between the paired proton and the electron.

\begin{figure}
    \centering
    \includegraphics[scale = 0.6]{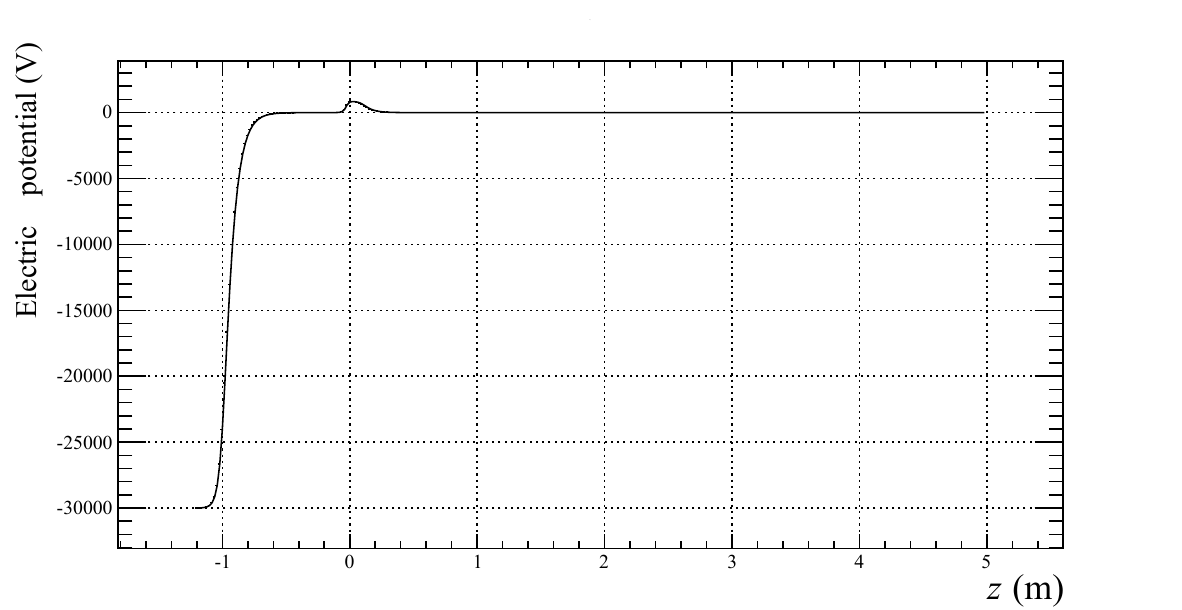}
    \caption{The electric potential along the $z$ direction for Nab-$b$ configuration.}
    \label{b_config_e}
\end{figure}

For Nab-$b$ measurement, we will use a different field setting for the measurement. While it's still possible to measure $b$ from the Nab-$a$ configuration, the selection of the forward protons for precision momentum measurement (as shown in Fig.\;\ref{dep_pPz}) will cut 7/8 for the decay events. To make the most out of all the decay events, Nab-$b$ configuration will have a electric potential configuration as shown in
Figure\;\ref{b_config_e}. In this configuration the filter region is at a voltage of 850\,V relative to the decay region. The decay protons that go upwards will be reflected by this voltage to the bottom detector, which is at a high voltage of $-$30\,kV, as shown in Fig.\,\ref{b_config_e}. The electrons will be collected on both detectors, as discussed in the previous section. The detected protons will be used to find electrons from true decay events, and separate them from background. This could help us suppress background through coincidence requirements.

\chapter{The Electric Field in the Neutron Decay Region}
\label{C3}
\noindent
In the Nab experiment, we detect the proton TOF and infer the squared proton momentum from it. As shown in Fig.\,\ref{spectrometerdetail}, free neutrons pass through the decay region and the charged decay particles (electrons and protons) are guided to the detectors by magnetic field lines.

\begin{table}
\begin{center}
\hspace*{-1.5cm}
\begin{tabular}{lcc}
\thickhline
 \textbf{Experimental parameter}& \textbf{Main specification} & $\bm{\Delta a/a}$\\
\thickhline
 \textbf{Magnetic field}   &    & \\
 \;\;  curvature at pinch&   $\Delta \gamma / \gamma = 2 \%$ with  $\gamma =\frac{d^2 B_z(z)}{B_z(0)dz^2}$ & $5.3\cross10^{-4}$  \\
\;\;  ratio $r_B = B_{TOF}/B_0$ &$(\Delta r_B) / r_B = 1 \%$ & $2.2\cross10^{-4}$\\
 \;\;  ratio $r_{B,DV} = B_{DV}/B_0$    &$(\Delta r_{B,DV}) / r_{B,DV} = 1 \%$ & $1.8\cross10^{-4}$\\
 \hline
  \textbf{Length of the TOF region}&     & none\\
 \hline
  \textbf{Electric potential inhomogeneity:}&   & \\
\;\; in decay volume / filter region & $\abs{U_F - U_{DV}} < 10$\,mV &  $5\cross10^{-4}$\\
 \;\; in TOF region & $\abs{U_F - U_{TOF}} < 200$\,mV &  $2.2\cross10^{-4}$\\
 \hline
  \textbf{Neutron beam:}&   & \\
 \;\;  position & $\Delta \overline{z_{DV}} < 2$\,mm &  $1.7\cross10^{-4}$\\
  \;\; profile (including edge effect)& Slope at edges < 10$\%$\,/cm &  $2.5\cross10^{-4}$\\
\;\;  Doppler effect & & small\\
 \;\;  Unwanted beam polarization & $\abs{\overline{P_{n}}} \ll 10^{-4}$ & $1\cross10^{-4}$\\
    \hline
   \textbf{Adiabaticity of proton motion:}&   & $1\cross10^{-4}$ \\
     \hline
   \textbf{Detector effects:}&   &  \\ 
\;\; Electron energy calibration & $\Delta E < 0.2$\,keV &  $2\cross10^{-4}$\\
\;\; Shape of electron energy response & fraction of events in tail to 1\% & $4.4\cross10^{-4}$\\
\;\; Proton trigger efficiency & $\epsilon_p < 100$\,ppm/keV & $3.4\cross10^{-4}$\\
\;\; TOF shift due to detector/electronics & $\Delta t_p < 0.3$\,ns & $3.9\cross10^{-4}$\\
   \hline
  \textbf{Electron TOF} & &small\\
   \hline
   \textbf{Residual gas}&  $p < 2 \cross 10^{-9}$\,torr   &  $3.8\cross10^{-4}$(prelim.) \\ 
    \hline
   \textbf{TOF in acceleration region} & $\Delta r_{\text{ground\;el.}} < 0.5$\,mm &  $3\cross10^{-4}$(prelim.)\\
\hline
   \textbf{Background / Accidental coincidences} & & small\\
\thickhline
      \textbf{Sum} & &  $1.2\cross10^{-3}$\\
\thickhline
\end{tabular}
\caption{\label{asystem}The systematic uncertainty budget for Nab-$a$ measurement.}
\end{center}
\end{table}

To achieve a precision of $\Delta a / a \sim 10^{-3}$, the Nab experiment needs to control the systematic uncertainties in different aspects of the experiment to the required level, as displayed in table\,\ref{asystem}\cite{Fry:2018kvq}.

One important requirement concerns the electrical potential inhomogeneity between the decay volume and the filter region. Due to this inhomogeneity, an unwanted electric field could be induced and contribute to uncertainty in the distribution of detected proton TOF\cite{kp:sean}; the cause of this field will be discussed in Sec.\,\ref{C3S2}. As shown in table\,\ref{asystem}, the potential difference caused by this unwanted field should be less than 10\,mV.

The specification for the potential difference came from parametric studies\cite{parametric}. In the parametric studies, the protons are assumed to have guiding center on the $z$ axis. The magnetic field from the decay volume to the upper detector on $z$ axis is modeled piecewise as:
\begin{equation}
    B=\begin{cases}
    B_0r_{B,DV}\,, & z < z'_2\,,\\
     k z + b\,, & z'_2 \leq z < -z_1 \,,\\
      B_0(1-\alpha^2z^2)\,, & -z_1 \leq z < z_1 \,,\\  
      -k z + b\,, & z_1 \leq z < z_2 \,,\\  
      B_0r_B\,, & z_2 \leq z < L\,,
      \end{cases}
      \label{bmodel}
\end{equation}
with the default setting of $z_1$ = 0.02\,m, $z_2$ = 0.1\,m, $\alpha$ =  15\,m$^{-1}$, $r_B$ = 0.04, $r_{B,DV}$ = 0.48, $B_0$ = 4\,T, and $L$ = 4.8\,m (this is the position of the upper detector). We also assume no electric field in this region. The values of $k$, $b$, $z'_2$ are computed such that the magnetic field from Eq.\,\eqref{bmodel} is continuous in $z$:
\begin{equation}
    \begin{split}
        k &= \frac{1-\alpha^2z^2_1-r_B}{z_2-z_1}B_0\,,\\
        b &= \frac{z_2(1-\alpha^2z^2_1)-r_Bz_1}{z_2-z_1}B_0\,,\\
        z'_2 &= \frac{(z_2-z_1)r_{B,DV}-z_2(1-\alpha^2z^2_1)+r_Bz_1}{1-\alpha^2z^2_1-r_B}\,.
    \end{split}
\end{equation}

\begin{figure}
    \centering
    \includegraphics[scale = 0.6]{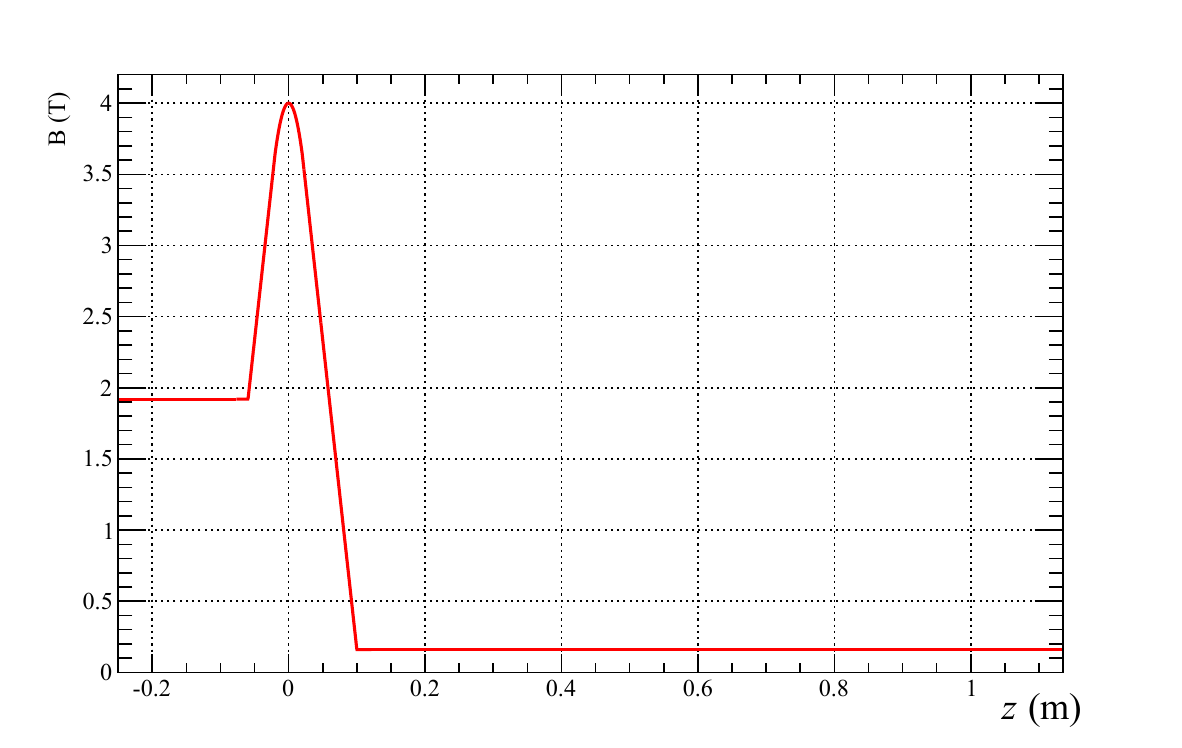}
    \caption[The model of magnetic field in the parametric studies.]{The model of magnetic field in the parametric studies\cite{parametric}. Note that the right constant part of the magnetic field extends to $z$ = 4.8 m.}
    \label{b_plot}
\end{figure}

As shown in Fig.\,\ref{b_plot}, in this model, the magnetic field has five pieces: the parabolic middle part $-z_1 \leq z < z_1$, the two linear parts with opposite slope $z'_2 \leq z < -z_1$ and $z_1 \leq z < z_2$, and two constant parts $z < z'_2$ and $z_2 \leq z < L$. Strictly speaking, $B(z)$ needs to have a continuous first order derivative with respect to $z$ to ensure adiabaticity, which is not the case at the connection points between different regions. However, the dependence of $t_p$ on unwanted distortions of the spectrometer, which is the subject of the parametric study, is the same as it would be for a similar but smoother field that we have in the Nab spectrometer, which has been designed to allow for adiabatic motion.

Under this magnetic field model, the TOF ($t_p$) for proton born with momentum\,$\bm{p}_p$ can be calculated as:
\begin{equation}
    t_p\left(p_p,\text{cos}(\theta_0)\right) = \frac{m_p}{p_p}\int^{L}_{z_0} \frac{dz}{\sqrt{1-\frac{B(z)}{B_0}\text{sin}^2(\theta_0)}}\,,
    \label{adnov}
\end{equation}
or
\begin{equation}
    \frac{1}{t^2_p\left(p^2_p,\text{cos}(\theta_0)\right)} = \frac{p^2_p}{(m_p\int^{L}_{z_0} \frac{dz}{\sqrt{1-\frac{B(z)}{B_0}\text{sin}^2(\theta_0)}})^2} = f\biggl(\text{cos}(\theta_0)), z_0\biggr) p^2_p\,.
    \label{t2p2theta}
\end{equation}

Here we used the adiabatic approximation from Eq.\,\eqref{filter} with no electric field, $z_0$ is the $z$ coordinate of the proton at birth. The angle $\theta_0$ is defined the same as in Eq.\,\eqref{filter}. Note that Eq.\,\eqref{t2p2theta} could be analytically calculated from the model Eq.\,\eqref{bmodel}, which is the point of using the field from Eq.\,\eqref{bmodel}.

In chapter\,\ref{C2}, we discussed the two-dimensional histogram of $E_e$ and $p^2_p$, and concluded that for fixed $E_e$, the probability distribution of $p^2_p$ is linear with $p^2_p$ inside the allowed region, with a slope proportional with the electron-antineutrino correlation coefficient $a$. Combine Eq.\,\eqref{decayrate} and Eq.\,\eqref{t2p2theta}, we can construct the decay rate as a function of $1/t^2_p$ (denoted as $P(t^2_p)$) and use it as the fitting function to extract $a$. As discussed in chapter\,\ref{C4}, the detector response for monoenergetic electrons is a Gaussian peak with an exponential tail. The decay rate $P(t^2_p)$ has been obtained by averaging over all true electron energies that can give the reconstructed $E_e$, and the height of the decay volume for $z_0$.

After obtaining the fitting function, the actual fitting was done in two steps. Step (1) fit the function with all the data and $a, b, L, E_{cal}$\footnote{$E_{cal}$ is the gain factor from the detector calibration, and is discussed in next chapter.} as free parameters. Step (2), fix the fitted $L$ and $E_{cal}$ from step (1), and fit with the inner 75\% of data\footnote{This is the inner 75\% percent of data on $1/t^2_p$.}

To obtain the fitting data, $1.09 \cross 10^{9}$\footnote{The number of data points is chosen such that relative statistical uncertainty $\Delta a /a = 10^{-3}$.} events are generated with 10\% background added. Recall that in Eq.\,\eqref{adnov}, we assumed that there is no electric field in the region in the fitting function. In this study, we assumed there is actually electrostatic potential difference between the decay region and the filter region (this potential difference is denoted as $U_F-U_{DV}$ with $U_F$ to be the potential of the filter region and $U_DV$ to be the potential of the decay region). Similar to Eq.\,\eqref{adnov}, we calculate the actual TOF ($t_p$) with the presence of the electrostatic potential difference as:
\begin{equation}
     t_p\left(p_p,\text{cos}(\theta_0)\right) = \frac{m_p}{p_p}\int^{L}_{z_0} \frac{dz}{\sqrt{1-\frac{2e(U_F-U_{DV})}{m_pv^2_0}-\frac{B(z)}{B_0}\text{sin}^2(\theta_0)}}\,,
     \label{adv}
\end{equation}
note that in this equation we used the non-relativistic approximation, $v_0$ is the velocity of the proton at birth.

The above relation is then applied together with the simulated information of the decayed particles at birth to create the two-dimensional histogram over $1/t^2_p$ and $E_e$ for all data points. This two-dimensional histogram is then fitted using the fitting function created above. Table\,\ref{dudv} shows the fitting result, this result is also plotted in Fig.\,\ref{dudv}. From the result, we find that for $\abs{U_F - U_{DV}}$ = 10\,mV, $\abs{\Delta a/a} = 5\cross 10^{-4}$.

\begin{table}[H]
\begin{center}
\begin{tabular}{|l|l|l|l|l|l|l|}
\hline
$U_{F} -U_{DV}$ (V) & fitted $L$ (m) & fitted Ecal & fitted $a$ & fitted $b$ & $\chi^2$ & dof \\ \hline
0             & 4.800047   & 1.000015      & -0.10294   & -0.00004   & 251           & 260          \\ \hline
0.05          & 4.800044   & 1.000019      & -0.10319   & 0.00006    &               & 260          \\ \hline
-0.05         & 4.800051   & 1.000011      & -0.10268   & -0.00013   & 250           & 260          \\ \hline
0.2           & 4.800034   & 1.000030      & -0.10395   & 0.00037    & 330           & 260          \\ \hline
-0.2          & 4.800060   & 0.999999      & -0.10193   & -0.00038   & 294           & 260          \\ \hline
\end{tabular}

\caption[The parametric study for the electrostatic potential in the decay region and filter region table.]{\label{dudv}The parametric study for electrostatic potential in the decay region and filter region table. $U_{F} -U_{DV}$ is the electrostatic potential difference between decay region and drift region, and dof stands for degree of freedom. As mentioned in the model, the input L for the data that is being fitted is 4.8\;m, and the input Ecal (gain factor) is 1 for this study.}
\end{center}
\end{table}

\begin{figure}[H]
    \centering
    \includegraphics[scale = 0.6]{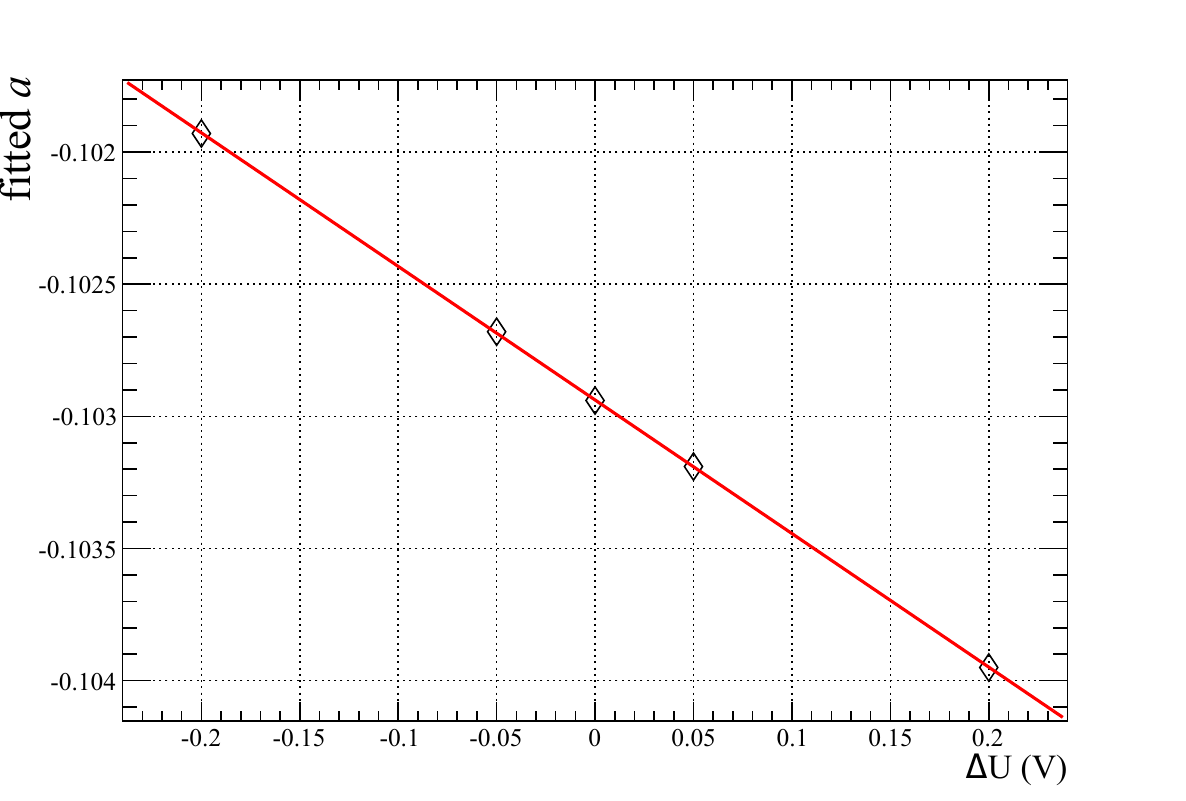}
    \caption[Plot of the fitted $a$ versus $\Delta U$, the potential difference in the decay region and filter region.]{Plot of the fitted $a$ versus $\Delta U$, the potential difference in the decay region and filter region.\footnotemark}
    \label{dudv}
\end{figure}

In the remainder of chapter, we will discuss our design of the decay volume and filter region for controlling the potential difference $\Delta U$, and discuss our solution to the problem to achieve the specification for the potential difference between the decay region and the filter region.

\footnotetext{The linearly fitted parameters for the blue line is $y =\beta_0 + \beta_1 x$ with $\beta_1 = -0.005$ and $\beta_0 = 0.1029$}

\section{The Design of the Decay Volume and the Filter Region}
\label{C3S1}
\begin{figure}[h!]
\centering
\includegraphics[scale=0.4]{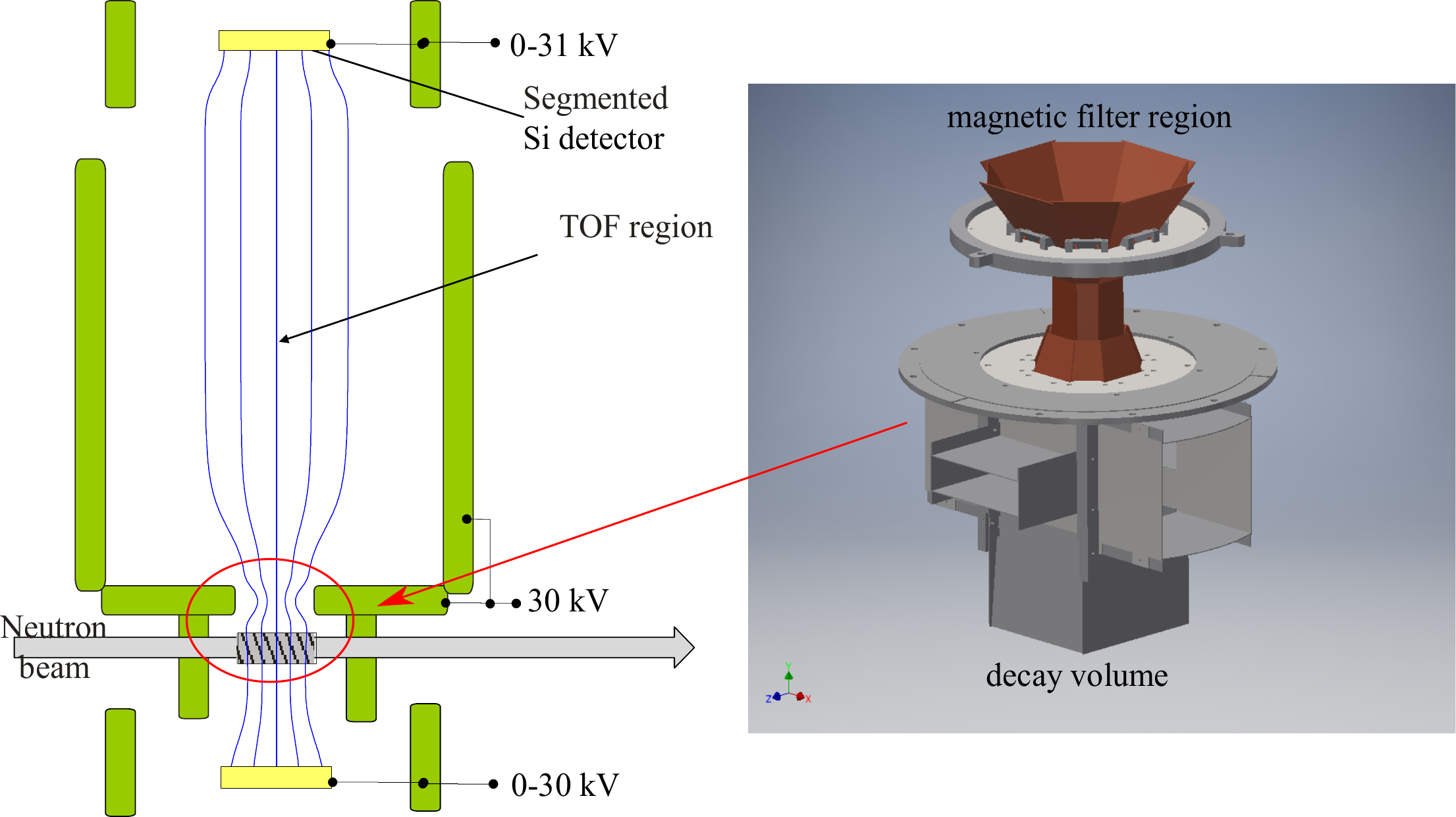}
\caption[The Nab electrode system for the magnetic filter region and the decay volume.]{The Nab electrode system for the magnetic filter region and the decay volume. Neutrons pass through the left and right side openings in the decay volume; the front and back openings are designed for the pumping and insertion of calibration source.}
\label{electrodes}
\end{figure}

\noindent
The Nab experiment uses an electrode system that surrounds the decay volume and the filter region, as shown in Fig.\,\ref{electrodes}. In the Nab-$a$ configuration, the electrode system and the magnetic field act as a filter, allowing only protons with a momentum direction sufficiently close to vertical to pass. As discussed later in this chapter, the electrode work function is taken into account. The work function is characterized at a level that allows to satisfy the specification proposed above\footnote{In the next chapter we will see that the work function difference of the material will create an electrical field.}. For Nab-$b$, the electrode system for the filter region will have an electric potential of 850\,V relative to the decay volume. As discussed in the chapter\,\ref{C2}, this voltage is used to reflect the protons so that we can collect all the protons in the lower detector. The detected protons will be used as a trigger to distinguish electron hits from background. In Nab-$b$, the small fluctuations of the electrostatic potential given by the workfunction are not important.

The top part of the electrode system is at the top of the filter region. It consists of eight pieces with four larger pieces and four smaller pieces. It is separated into two halves, left and right, that are electrically isolated. In the Nab experiment, the two halves will be on slightly different voltage to create a horizontal electric field. The purpose of the electric field is to clear out trapped particles. The bottom part of the electrode system covers the decay volume and the bottom of the filter region. The lower filter region is also made of eight electrode pieces to match the bottom of the upper drift region. The decay volume is made of 17 pieces of electrodes in various shape, as seen in Fig.\,\ref{numbering}, the five main pieces (piece No. 1, No. 36, No. 3, No. 16, and No. 17) that make up the main decay box contribute the most to the electric field inside. All the electrode pieces are made of titanium, which is a good non-magnetic material compatible with ultra high vacuum. However, its work function is not homogeneous enough for our specification. In the next section, we will discuss why the work function is important. As shown in table\,\ref{asystem}, The Nab-$a$ measurement requires the electrical potential difference between the decay volume and the filter region to be less than 10\,mV.

\section{The Work Function}
\label{C3S2}
\begin{figure}[h!]
\centering
\includegraphics[scale=0.34]{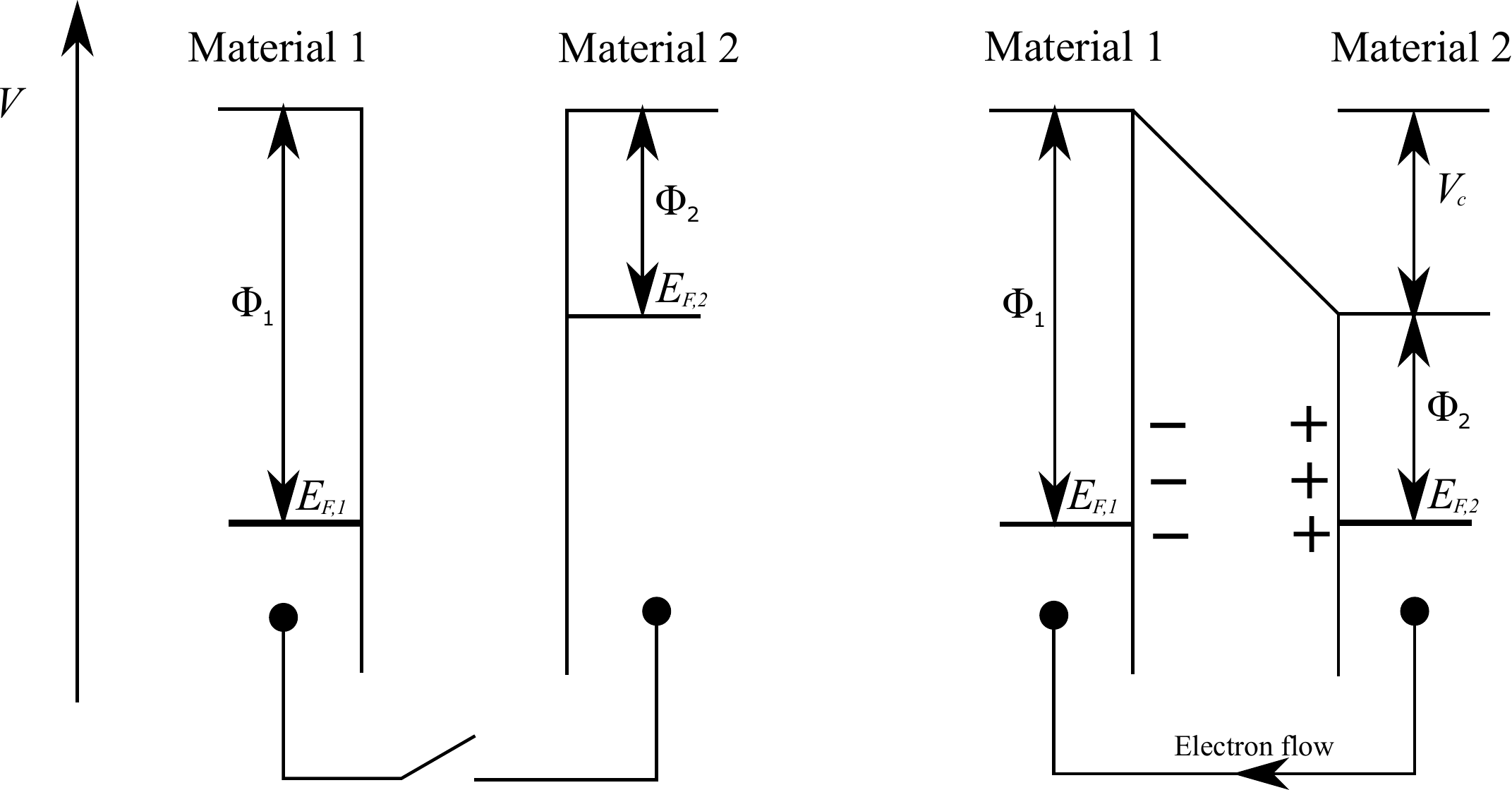}
\caption[Work function inhomogeneity causes an unwanted electric field.]{Work function inhomogeneity causes an unwanted electric field\cite{SKPmanual}.}
\label{wf_display}
\end{figure}

\noindent
The work function is defined as the minimum amount of work required to remove an electron from the surface of a solid. In the Nab experiment, the work function of the electrode material needs to be controlled to reduce the unwanted electrical field, Fig.\,\ref{wf_display} shows this effect. In this example, material 1 has a greater work function than material 2, once they are connected, the electrons will flow spontaneously from material 2 to material 1, making material 2 positively charged and material 1 negatively charged. This charge distribution will then create an unwanted electric field between the two materials. In the Nab experiment, the electrodes are made up of the same material and connected with each other. However, due to the impurities in the material, the work function at different parts of the electrode system is going to differ from each other.

The voltage difference ($V_{\text{cpd}}$) created by the work function difference ($\Phi_1 - \ \Phi_2$) is directly related by:

\begin{equation}
V_{\text{cpd}}= \frac{\Phi_1 - \ \Phi_2}{e}\,.
\label{wfvoltage}
\end{equation}

The Nab experiment requires the electrical potential difference between the decay volume and the filter region to be less than 10\,mV. To achieve this, the surface work function distribution within and across electrode pieces should have a controlled standard deviation of less than 10\,meV.

Table\;\ref{wfexample} shows typical values of work function for different materials. We can find that the 10\,meV deviation of work function requirement is harsh for most materials, including titanium which made up the electrode system.  Our solution to this problem is to coat all the inner surface of the electrode pieces, and test its work function with a device called the Kelvin probe.

\begin{table}[]
\scalebox{0.7}{
\begin{tabular}{llll|llll|llll}
\textbf{Element} & \textbf{Plane} & \textbf{$\Phi$/eV} & \textbf{Method} & \textbf{Element} & \textbf{Plane} & \textbf{$\Phi$/eV} & \textbf{Method} & \textbf{Element} & \textbf{Plane} & \textbf{$\Phi$/eV} & \textbf{Method} \\
Ag               & 100            & 4.64          & PE              &                  & 210            & 5.00          & PE              & Ru               & polycr         & 4.71          & PE              \\
                 & 110            & 4.52          & PE              & K                & polycr         & 2.29          & PE              & Sb               & amorp          & 4.55          &                 \\
                 & 111            & 4.74          & PE              & La               & polycr         & 3.5           & PE              &                  & 100            & 4.77          &                 \\
Al               & 100            & 4.20          & PE              & Li               & polycr         & 2.93          & FE              & Sc               & polycr         & 3.5           & PE              \\
                 & 110            & 4.06          & PE              & Lu               & polycr         & (3.3)         & CPD             & Se               & polycr         & 5.9           & PE              \\
                 & 111            & 4.26          & PE              & Mg               & polycr         & 3.66          & PE              & Si               & n              & 4.85          & CPD             \\
As               & polycr         & (3.75)        & PE              & Mn               & polycr         & 4.1           & PE              &                  & p 100          & (4.91)        & CPD             \\
Au               & 100            & 5.47          & PE              & Mo               & 100            & 4.53          & PE              &                  & p 111          & 4.60          & PE              \\
                 & 110            & 5.37          & PE              &                  & 110            & 4.95          & PE              & Sm               & polycr         & 2.7           & PE              \\
                 & 111            & 5.31          & PE              &                  & 111            & 4.55          & PE              & Sn               & polycr         & 4.42          & CPD             \\
B                & polycr         & (4.45)        & TH              &                  & 112            & 4.36          & PE              & Sr               & polycr         & (2.59)        & TH              \\
Ba               & polycr         & 2.52          & TH              &                  & 114            & 4.50          & PE              & Ta               & polycr         & 4.25          & TH              \\
Be               & polycr         & 4.98          & PE              &                  & 332            & 4.55          & PE              &                  & 100            & 4.15          & TH              \\
Bi               & polycr         & 4.34          & PE              & Na               & polycr         & 2.36          & PE              &                  & 110            & 4.80          & TH              \\
C                & polycr         & (5.0)         & CPD             & Nb               & 001            & 4.02          & TH              &                  & 111            & 4.00          & TH              \\
Ca               & polycr         & 2.87          & PE              &                  & 110            & 4.87          & TH              & Tb               & polycr         & 3.0           & PE              \\
Cd               & polycr         & 4.08          & CPD             &                  & 111            & 4.36          & TH              & Te               & polycr         & 4.95          & PE              \\
Ce               & polycr         & 2.9           & PE              &                  & 112            & 4.63          & TH              & Th               & polycr         & 3.4           & TH              \\
Co               & polycr         & 5.0           & PE              &                  & 113            & 4.29          & TH              & Ti               & polycr         & 4.33          & PE              \\
Cr               & polycr         & 4.5           & PE              &                  & 116            & 3.95          & TH              & Tl               & polycr         & (3.84)        & CPD             \\
Cs               & polycr         & 1.95          & PE              &                  & 310            & 4.18          & TH              & U                & polycr         & 3.63          & PE              \\
Cu               & 100            & 5.10          & PE              & Nd               & polycr         & 3.2           & PE              &                  & 100            & 3.73          & PE              \\
                 & 110            & 4.48          & PE              & Ni               & 100            & 5.22          & PE              &                  & 110            & 3.90          & PE              \\
                 & 111            & 4.94          & PE              &                  & 110            & 5.04          & PE              &                  & 113            & 3.67          & PE              \\
                 & 112            & 4.53          & PE              &                  & 111            & 5.35          & PE              & V                & polycr         & 4.3           & PE              \\
Eu               & polycr         & 2.5           & PE              & Os               & polycr         & 5.93          & PE              & W                & polycr         & 4.55          & CPD             \\
Fe               & 100            & 4.67          & PE              & Pb               & polycr         & 4.25          & PE              &                  & 100            & 4.63          & FE              \\
                 & 111            & 4.81          & PE              & Pd               & polycr         & 5.22          & PE              &                  & 110            & 5.22          & FE              \\
Ga               & polycr         & 4.32          & PE              &                  & 111            & 5.6           & PE              &                  & 111            & 4.45          & FE              \\
Gd               & polycr         & 2.90          & CPD             & Pt               & polycr         & 5.64          & PE              &                  & 113            & 4.46          & FE              \\
Ge               & polycr         & 5.0           & CPD             &                  & 110            & 5.84          & FE              &                  & 116            & 4.32          & TH              \\
Hf               & polycr         & 3.9           & PE              &                  & 111            & 5.93          & FE              & Y                & polycr         & 3.1           & PE              \\
Hg               & liquid         & 4.475         & PE              &                  & 320            & 5.22          & FE              & Zn               & polycr         & 3.63          & PE              \\
In               & polycr         & 4.09          & PE              &                  & 331            & 5.12          & FE              &                  & polycr         & (4.9)         & CPD             \\
Ir               & 100            & 5.67          & PE              & Rb               & polycr         & 2.261         & PE              & Zr               & polycr         & 4.05          & PE              \\
                 & 110            & 5.42          & PE              & Re               & polycr         & 4.72          & TE              &                  &                &               &                 \\
                 & 111            & 5.76          & PE              & Rh               & polycr         & 4.98          & PE              &                  &                &               &                
\end{tabular}}
\caption[Electron work function of elements.]{Electron work function of elements. The measurement method is indicated for each value. The following abbreviations appear: thermionic emission (TE), photoelectric effect (PE), field emission (FE), contact potential difference (CPD). The following abbreviations are not for measurement methods: polycrystalline sample (polycr) and amorphous sample (amorp). Values in parentheses are only approximate.\cite{1979STMP...85....1H,riviere1969work,michaelson1977work}}
\label{wfexample}
\end{table}

\section{The Kelvin Probe}
\label{C3S3}
\begin{figure}[h!]
\centering
\includegraphics[scale=1]{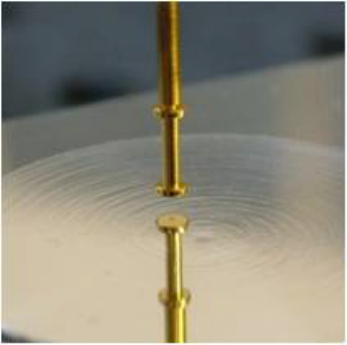}
\caption{The gold tip of the Kelvin probe, above a circular disk which is the sample in this picture.}
\label{kptip}
\end{figure}

\noindent
The Kelvin Method is an indirect technique to measure the difference of the work function between two different materials. As shown in Fig.\,\ref{kptip}, the Kelvin probe uses a small (2 mm diameter) vibrating gold tip to form a capacitor with the test sample.

Taking the distance between the tip and the sample to be $d$, and approximating the geometry as the one of a parallel plate capacitor, we get for the capacitance\footnote{Here we uses a simplified model proposed in \cite{SKPmanual}, for more complete calculation, please refer to \cite{completekp}}:
\begin{equation}
C(t) = \frac{\epsilon A}{d(t)}\,,
\label{parac}
\end{equation}
where $A$ is the area of the tip and $\epsilon$ is the absolute permittivity of air. The tip oscillates harmonically over the sample surface. The distance can be expressed as:
\begin{equation}
d(t) = d_0 + d_1 \text{sin}(\omega t) = d_0(1+e\text{sin}(\omega t))\,,
\label{paralellc}
\end{equation}
where $\omega$ denotes $2\pi$ times the vibrating frequency. The average distance is denoted by $d_0$ and the oscillation amplitude is denoted by $d_1$, and $e = d_1/d_0$.

To measure the work function, a voltage $V_b$ is applied through a circuit connecting the tip and the sample. The capacitor will be charged due to the applied voltage, and the amount of charge $Q$ is:
\begin{equation}
Q(t) =C(t)(V_b + V_{\text{cpd}}) = \frac{\epsilon A (V_b + V_{\text{cpd}})}{d_0(1+e \text{sin}(\omega t)}\,,
\label{paraq}
\end{equation}
here $V_{\text{cpd}}$ is the voltage difference induced by the difference of work function in Eq.\,\eqref{wfvoltage}. Due to the vibration of the tip, the current in the circuit will be:
\begin{equation}
I(t) = \frac{dQ(t)}{dt}= \frac{\epsilon A}{d_0}(V_b + V_{\text{cpd}})\frac{-e \omega \text{cos}(\omega t)}{[1+e \text{sin}(\omega t)]^2}\,.
\end{equation}

The following circuit diagram shows the detection circuit of the Kelvin probe. The tip current is converted to $V_{\text{out}}$ in two conversion stages. In the first stage, the I/V converter converts the current to a voltage, which is, the output voltage $V_1 = I_1R_f$. In the second stage, the preamplifier amplifies the voltage to obtain $V_{\text{out}}$, where $R_1$ and $R_2$ set the voltage gain ($V_{\text{out}} = -\frac{R_2}{R_1}V_1$). The output signal $V_{\text{out}}$ finally passes via a low pass filter and a analog-to-digit converter to the DAQ system. (b) is the simplified diagram for first stage. The tip parasitic capacitance is denoted by $C_{\text{p}}$, for low frequencies (<1000\,Hz), the signal loss to the parasitic capacitance $C_{\text{p}}$ is negligible.\cite{SKPmanual}
\begin{figure}[h!]
\centering
\includegraphics[scale=0.6]{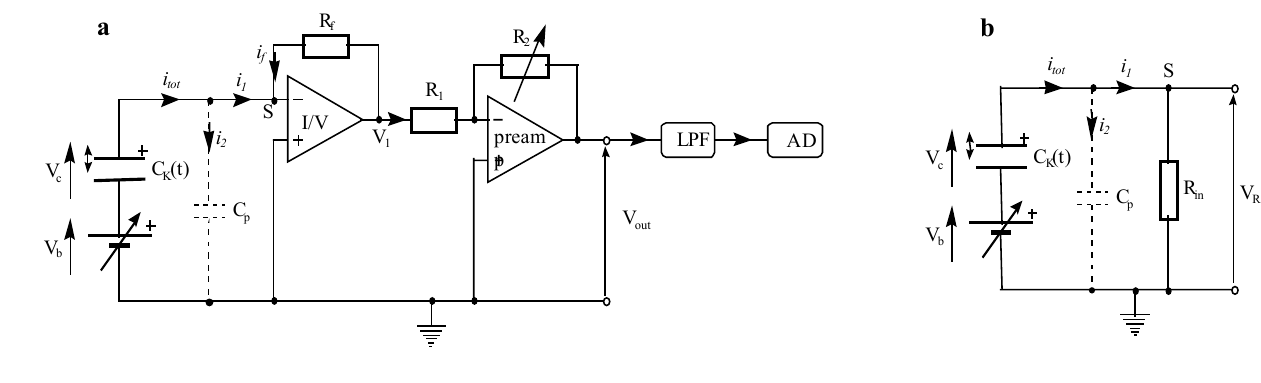}
\caption{The circuit diagram of Kelvin probe.}
\label{kpcircuit}
\end{figure}

Combining all this, the output voltage $V_{\text{out}}$ is:
\begin{equation}
V_{{\text{out}}}(t) =  \frac{\epsilon A}{d_0}(V_b + V_{\text{cpd}})R_f \frac{R_2}{R_1} \frac{e \omega \text{cos}(\omega t)}{[1+e \text{sin}(\omega t)]^2}\,.
\label{kpeq}
\end{equation}

The graph below shows the output signal of the Kelvin probe as a function of time. For all the measurement, we set $d_1 = 50$\,a.u.\footnote{The value of $d_0$ corresponds to a distance of about 30\,$\mu$m, and the value of $d_1$ corresponds to a distance of about 5\;$\mu$m} and $\omega$ = $2\pi \times$79.0\,rad/sec. 
\begin{figure}[h!]
\centering
\includegraphics[scale=1]{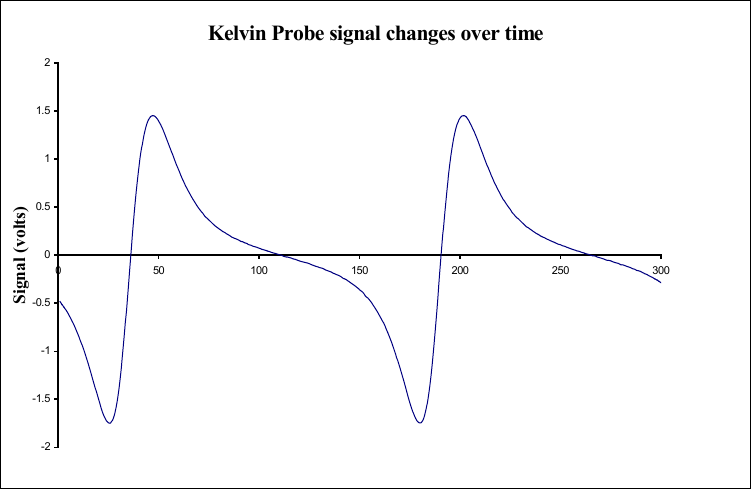}
\caption[The Kelvin probe output signal graph.]{The Kelvin probe output signal graph from \cite{SKPmanual}.}
\label{kpsignal}
\end{figure}

\begin{figure}[h!]
\centering
\includegraphics[scale=1]{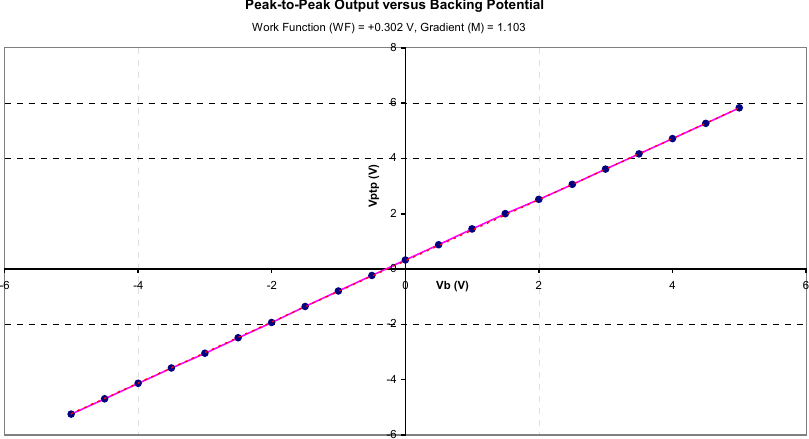}
\caption[The graph of $V_{\text{ptp}}$ versus $V_b$.]{The graph of $V_{\text{ptp}}$ versus $V_b$\cite{SKPmanual}. In the usual measurement protocol, only two points on the graph are used, this is done by flipping the direction of the voltage during the measurement.}
\label{vptp_vb}
\end{figure}

To extract the work function from the signal, the peak to peak value of this output signal is measured. In Eq.\,\eqref{kpeq}, the peak to peak voltage $V_{\text{ptp}}$ is linear in $V_b$ with other parameters to be fixed:
\begin{equation}
V_{\text{ptp}} = V_{\text{out}}(t)|_{\text{max}} - V_{\text{out}}(t)|_{\text{min}} =K(d_0) (V_b + V_{\text{cpd}})\,,
\label{ptpeq}
\end{equation}
with
\begin{equation}
    K(d_0) \propto  \frac{\epsilon A}{d_0}R_f \frac{R_2}{R_1}\,,
    \label{k0def}
\end{equation}
and the $x$ intercept is $-V_{\text{cpd}}$ as shown in Fig.\,\ref{vptp_vb}.

The analysis above is essentially the one that has been given to us by the manufacture. Since we believed that the approximation of the needle-sample capacitor as a parallel plate capacitor is quite rough, we alternatively assume the tip to be a small charged sphere with a radius $r$ that is much smaller than $d$, and the sample surface to be a infinite large plate. Applying the image charge method, the potential difference between small sphere and plate is:
\begin{equation}
\begin{split}
    V &= \frac{1}{4\pi\epsilon}\left[\frac{Q(t)}{r}-\frac{Q(t)}{2d(t)}\right]\\
        &=\frac{Q(t)}{4\pi\epsilon}\frac{2d(t)-r}{2rd(t)}\,,
    \label{ballv}
\end{split}
\end{equation}
the amount of charge carried on the capacitor is then:
\begin{equation}
 Q(t) = 4\pi\epsilon \frac{2rd(t)}{2d(t)-r} V\,.
\label{ballq}
\end{equation}

Comparing Eq.\,\eqref{ballq} with Eq.\,\eqref{paraq}, we find that this results in a different capacitance. For a more general model, we have:
\begin{equation}
 Q(t) = C(d(t))V\,,
\label{ballq}
\end{equation}
with $C(d)$ to be the capacitance as a function of the distance $d$.
Following the same calculation we get:
\begin{equation}
    V_{{\text{out}}} = (V_b+V_{\text{cpd}}) R_f \frac{R_2}{R_1} \frac{dC}{dt}\,.
\end{equation}
Assume $d(t) = d_0(1+e\text{sin}(\omega t))$, $\frac{dC}{dt}$ is also periodic in time, its peak to peak value is a function of $d_0$, $e$ and $\omega$:
\begin{equation}
    \frac{dC}{dt}|_{\text{max}} -  \frac{dC}{dt}|_{\text{min}}= f(d_0,e,\omega)\,.
\end{equation}
Similar to Eq.\,\eqref{kpeq}, the output voltage is then:
\begin{equation}
V_{\text{ptp}} =(V_b + V_{\text{cpd}})R_f \frac{R_2}{R_1} f(d_0,e,\omega)\,,
\end{equation}
with $d_0, e$, and $\omega$ to be fixed during the measurement, $V_{\text{ptp}}$ is always linear with $V_b +V_{\text{cpd}}$, as shown in Fig.\,\ref{vptp_vb}.

The Kelvin probe makes use of two points on the line by flipping the direction of the applied voltage $V_b \xrightarrow{} -V_b$, and calculating $V_{\text{cpd}}$ from the $x$ intercept. Then we can use Eq.\,\eqref{wfvoltage} to find the work function of the material. In the Nab experiment, the relative variation of the work function at different positions of the surface is of the most important, therefore for the following content of this chapter, the term ``work function'' refers to the relative work function with respect to the tip material (gold) of the Kelvin probe.

In the above calculation, we approximated the sample to be an infinite plate with a uniform work function. In reality, experiments showed that the measured work function is dependent on $d_0$ in the large scale. This is because that the Kelvin probe is most sensitive to a region of the sample surface right below the tip with a radius of $d_0$, due to the nonuniformity of the sample, the measurement is actually the average work function of that region, which depends on $d_0$. Furthermore, possible false effects due to the external charges may result in a signal output that depends on $d_0$. However, for a small range of $d_0$, the measured work function is weakly dependent on $d_0$. To make consistent measurements, the average distance $d_0$ needs to be the controlled. From Eq.\,\eqref{ptpeq} and Eq.\,\eqref{kpeq}, we can find that the slope ($K(d_0)$) of the line $V_{\text{ptp}}$ vs.\ $V_b$ is inversely proportional to $d_0$, this makes it possible for us to control the average distance $d_0$ by tracking the slope of the line (gradient) in Fig.\,\ref{vptp_vb}. To make consistent measurements, we always adjust $d_0$ until the gradient is within a certain range\footnote{The gradient is controlled by the device program within the range from \numrange{240}{260} a.u. The gradient is given in an internal unit of the Kelvin probe readout software. A greater value in gradient means a smaller value in actual distance $d_0$ between the tip of the Kelvin probe and sample. A calibration is not available.}.

\section{The Stability of the Kelvin probe}
\label{C3S4}
\noindent
As discussed in \cite{kp:sean}, there are two main considerations for the Kelvin probe: the noise and the stability. In this section we present the test results of the Kelvin probe. The Kelvin probe device used in this work is the scanning Kelvin probe (SKP5050) from KP Technology Ltd.

\begin{figure}[h!]
\centering
\includegraphics[scale=0.75]{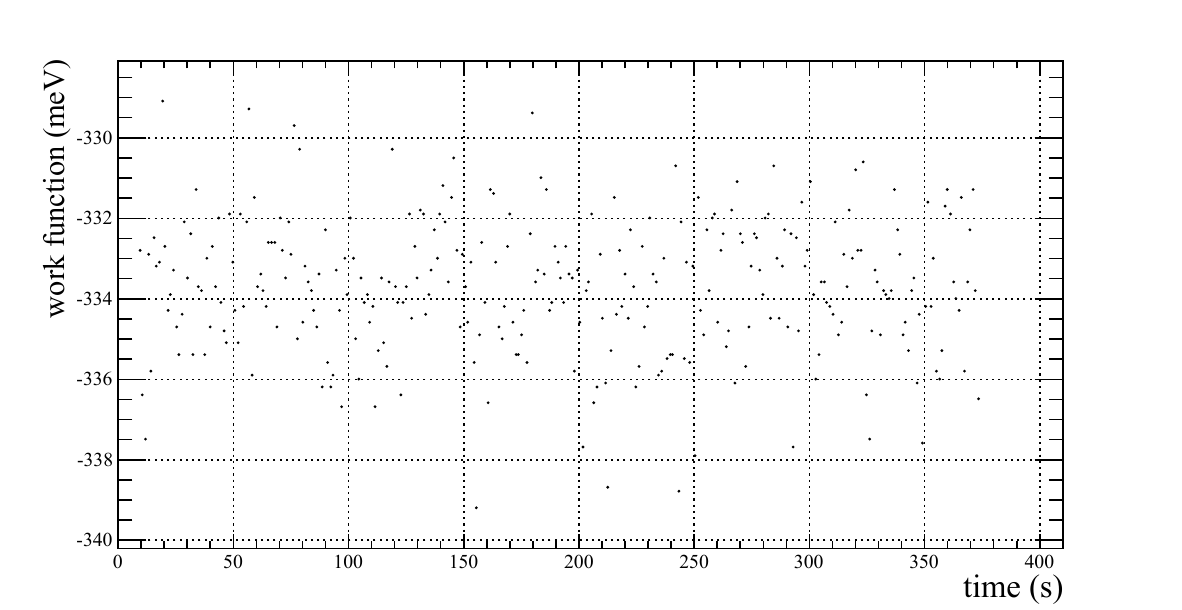}
\caption[The average work function of 40 measurements on the same point of the test sample over time.]{The average work function of 40 measurements on the same point of the test sample over time. The standard deviation of the work function is 1.6\,meV.}
\label{1000s}
\end{figure}

The noise of the Kelvin probe could be suppressed by averaging over multiple measurements, as shown in Fig.\,\ref{1000s}, with an average of 40 measurements, the measured work function noise over time on the same point of the testing sample is about 1.6\,meV, which is sufficient for our purpose.

The stability of the Kelvin probe for a longer time is a more complicated problem. For the same point on the testing sample, the measurement result from Kelvin probe drifts over time.

\begin{figure}[h!]
\centering
\includegraphics[scale=0.75]{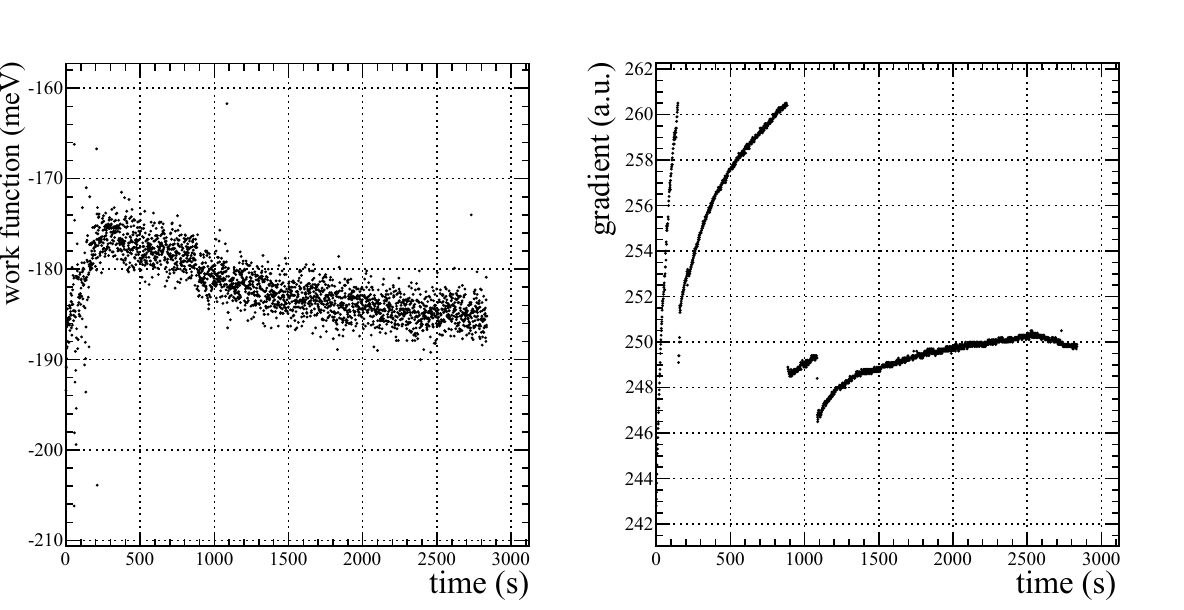}
\caption[The warming up of the Kelvin probe.]{The warming up of the Kelvin probe. The right plot shows how the gradient is drifting and gradually reaches a more stable state. The discontinuity point of gradient at $t \approx 150$\,s and $t \approx 900$\,s is caused by the self adjustment of the Kelvin probe, we can see that the gradient is about to exceed 260 a.u., and the device will automatically adjust $d_0$ to bring the gradient back to the range of \numrange{240}{260}\,a.u. The discontinuity point at $t \approx 1100$\,s is the result of an accidental perturbation from outside.}
\label{kpwarmup}
\end{figure}

The physical adjustments of the Kelvin probe, such as the replacement of the tip or the position-adjusting of the Kelvin probe can put the Kelvin probe into a unstable state. For it to reach its equilibrium state, we need to run the device for a while before taking any measurement. This warm up process usually takes few hours, with the pattern of a unidirectionally drifting gradient, as shown in Fig.\,\ref{kpwarmup}. The warm up process is also needed if the Kelvin probe is turned off for too long, or there is an accidental physical or electrical perturbation to the device. Note that after the Kelvin probe reaches a relative stable status, the drifting of the work function still exists, but with a much slower drift rate. Note also that we are primarily interested in stability of the work function, although it is typically not stable if the gradient is not stable.

\begin{figure}[h!]
\centering
\includegraphics[scale=0.75]{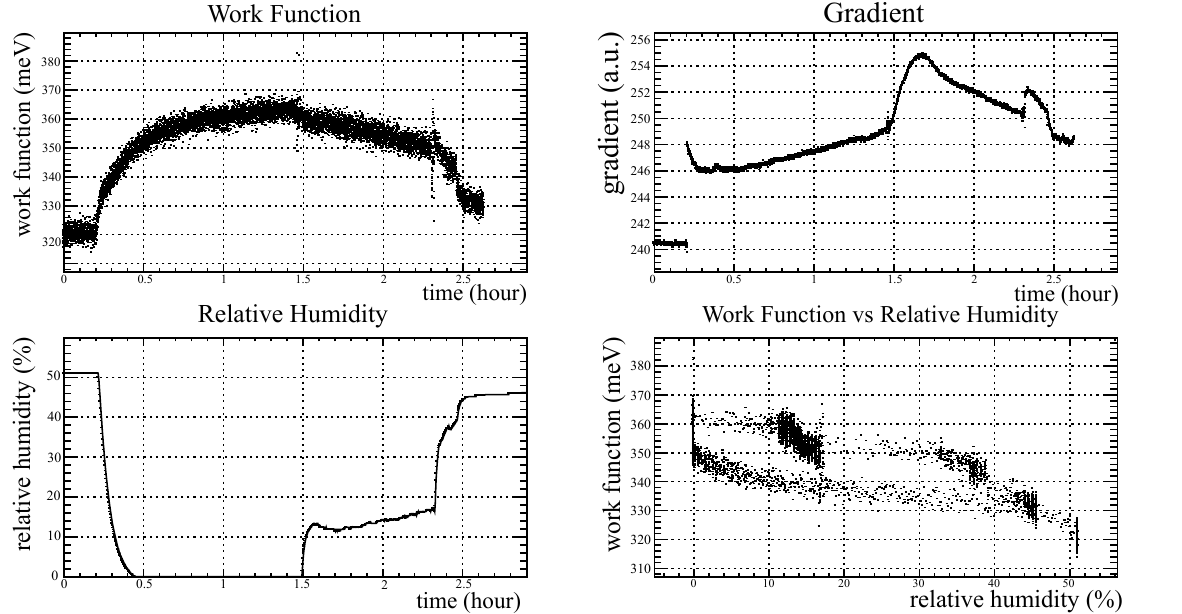}
\caption[The Kelvin probe measurement with the change of the relative humidity.]{The Kelvin probe measurement with the change of the relative humidity. The nitrogen gas flow was turn on at $t \approx 0.25$\,h, turned off at $t \approx$ 1.5\,h, turned on again at $t \approx$ 1.6\,h with a smaller flow rate, and turned off again at $t \approx$ 2.3\,h. The upper right graph indicates that the gas flow is affecting the gradient at $t$ = 0.25\,h, $t$ = 1.5\,h and $t$ = 2.3\,h, the bottom right graph shows measured work function depends on the relative humidity.}
\label{WF_H}
\end{figure}

The relative humidity plays a role in the measurement of the work function. To control the relative humidity, the Kelvin probe and the sample under test are placed in a sealed metal box, filled with dry nitrogen. The gas flow is controlled by a gas valve. Fig.\,\ref{WF_H} shows how the relative humidity is affecting the Kelvin probe measurement. Due to the condensation and the evaporation of the water vapor on the surface, the measured work function depends on the relative humidity level.

However, as shown in Fig.\,\ref{WF_H}, the measured gradient is also affected by the nitrogen gas flow, introducing a possible drift to the measured work function value. In the experiment, we don't find an one-to-one relationship between the relative humidity and the gradient. During the time of zero relative humidity, the gradient and the work function are still drifting. This drift could be caused by the ions introduced from the nitrogen gas or the small tremor introduced by the gas flow passing through the guiding tube.

On the other hand, replacing the test sample requires the sealed box to be opened, during this process, and we need to fill the box again with dry nitrogen after the replacement. Therefore, it is very difficult to measure and compare the work function across different samples at controlled relative humidity within a short period of time before the drift effect dominates. Furthermore, waiting many hours for each measurement to stabilize is unacceptable due to the large number of measurements that needed to be taken. As discussed previously, the electric potential variation is only sensitive to work function difference between varies parts of the electrode system, that means in order to get consistent work function measurement for different samples, we did use the reference sample method. With this method, a fixed reference sample is measured before or after the measurement of a series of samples. Based on the measured values of the reference sample, measurements of different samples at different times is compared via subtracting off the drift, as proposed in \cite{kp:sean}.

\section{The Coating of the Electrodes and the Measured Results}
\label{C3S6}
\begin{figure}[h!]
\centering
\includegraphics[scale=0.3]{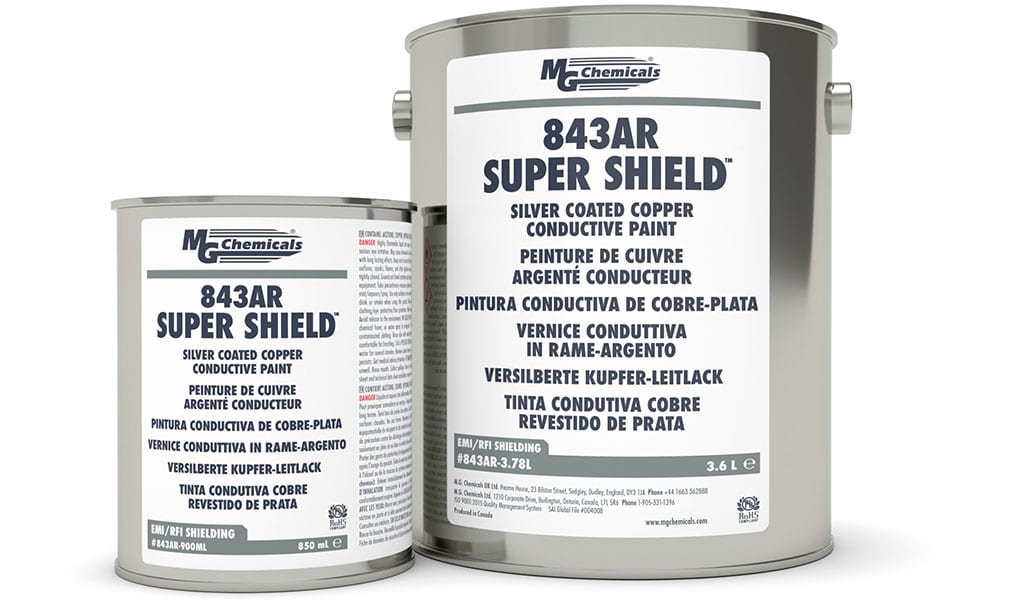}
\caption{Super Shield silver-coated copper conductive paint (product number 843AR) from MG Chemicals.}
\label{coating}
\end{figure}

\begin{figure}[h!]
\centering
\includegraphics[scale=0.5]{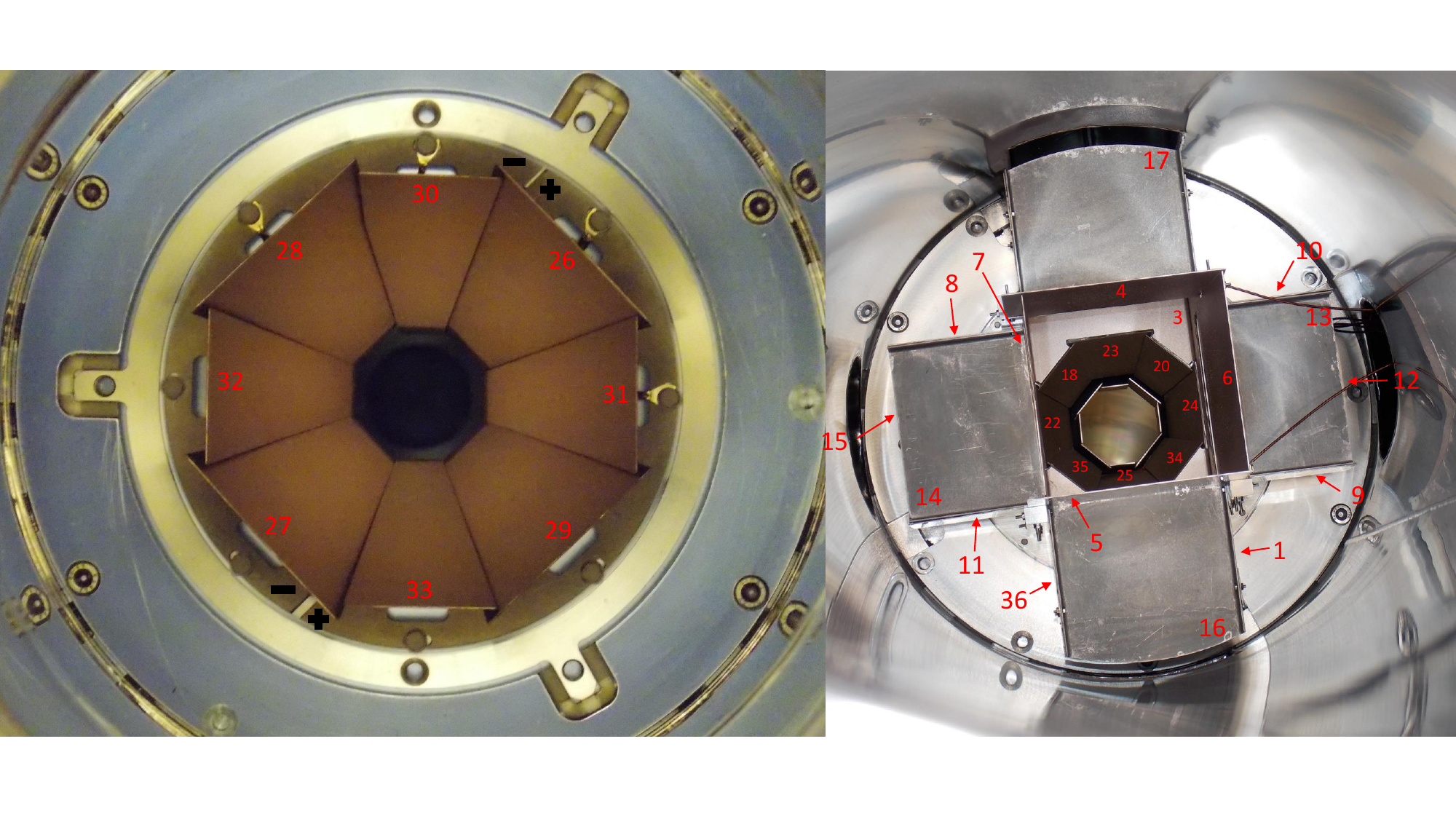}
\caption[The electrodes system numbering.]{The electrodes system numbering. The plus/minus sign on the left plot is engraved on the two holding plates. In both figures, neutron beam goes from bottom to top through the decay region.}
\label{numbering}
\end{figure}

\noindent
To achieve the 10\,meV requirement from table\,\ref{asystem}, we coated all the inner surfaces of the electrode pieces, which is directly affecting the electric field through the decay volume and the filter region. The coating is done with copper conductive coating, which consists of a solvent-based acrylic lacquer, pigmented with a highly conductive silver-coated copper flake\cite{mgc}, as shown in Fig.\,\ref{coating}. The surfaces of the electrodes are prepared with acetone and alcohol before coating. To make the drying process faster, the coated pieces are baked immediately after coating with a temperature of 50\,\textdegree C in an oven for 5 minutes\footnote{We never managed to get good uniformity in the work function across different coating batches, and therefore we coated all electrodes in one batch}. Each piece is labeled by a number, which is engraved at the corner of the non-coated side. Fig.\,\ref{numbering} shows the detailed numbering of all the pieces.

To make it possible to measure the work function and compare the results across different pieces, we repeatedly measure the work function of a reference sample, and correct the work function measurements by the reference sample.

Due to the limited scanning ability of Kelvin probe, it is very difficult to scan the whole area of the electrode pieces.  To get the work function, the measurements are done on distinct sampling areas or points on the pieces, depending on the shape of the pieces. For the large pieces that are making the main part of the decay region (pieces number 1, 3, 36), two areas from both sides are scanned. For smaller pieces that are making the upper and lower electrodes and the bottom wall of the decay box (pieces number 16-18, 20, 22-33, 34-35), only one area is scanned, for other pieces that 
are not as important, four points are measured for each piece. Table\,\ref{measureresult} shows the measured results. For all the measurements, piece number 26 is selected as the reference piece and all the other values are relative difference to this piece. The detailed scanning results are presented in appendix\,\ref{A1}.

\begin{table}[]
\centering
\scalebox{0.9}{
\begin{tabular}{ccccc}
\thickhline
piece no. & scan 1 (corrected)& scan 2 (corrected) & scan 1 (uncorrected)&  part name \\ \hline\hline
1         & -4.7   & -11.8  &-58.1   \\
36        & -20.6  & 11.0   &-73.9 & decay         \\
3         & -3.2   & -1.4  &-189.3  &  box        \\
16        & -7.3   &        & -52.4
 &    \\
17        & -1.7   &        &  -73.7 &                                \\ \hline
26*         & 0   &        &   -200.2&                 \\
27        & 5.0    &        &   -195.2 &                               \\
28        & -7.6   &        &  -207.7   &    upper                           \\
29        & -13.1  &        &    -213.3 &    drift                           \\
30        & -18.9  &        &   -211.2  &    region                          \\
31        & -24.8  &        &  -217.0   &                               \\
32        & -9.4   &        &   -201.6 &                               \\
33        & 2.8    &        &    -189.4 &                               \\ \hline
18        & -15.0  &        &   -215.2  &                \\
20        & -23.2  &        &  -95.2 &                                 \\
22        & -7.3   &        & -52.5  &lower                            \\
23        & 6.8    &        & -65.2  &  drift                              \\
24        & -13.0  &        &  -193.2 &  region                                 \\
25        & -17.4  &        & -62.6 &                                 \\
34        & -21.9  &        &  -222.1  &                                \\
35        & -18.2  &        &   -63.3 &                                \\ \thickhline
\end{tabular}
}
\bigskip
\footnotesize{*Piece no. 26 is the reference piece}
\caption[The scanning measurement results of the major electrode pieces in meV.]{The scanning measurement results of the major electrode pieces in meV. The work function measurement results without reference sample correction for scan 1 is also listed.}
\label{measureresult}
\end{table}

\begin{table}[]
\centering
\begin{tabular}{cccccc}
\thickhline
piece no. & point 1 & point 2 & point 3 & point 4 & part name \\ \hline\hline
4         & -2.6     & 3.6      & 4.8      & -10.6    &   \\
5         & 4.6      & 13.7     & 2.6      & 3.8      &           \\
6         & -4.4     & -9.1     & 42.3     & -9.7     &           \\
7         & 3.9      & -7.7     & -2.6     & -4.4     &           \\
8         & -3.2     & -15.9    & 8.7      & 2.1      &           \\
9         & 13.6     & -7.4     & 2.6      & -1.3     &   decay        \\
10        & -3.9     & 8.4      & -31.2    & 1.0      &    box       \\
11        & -6.7     & -7.0     & -2.3     & -17.5    &           \\
12        & -8.6     & -17.3    & -3.4     & -3.3     &           \\
13        & 12.3     & 1.3      & 0.4      & 6.9      &           \\
14        & 29.9     & 4.7      & 29.5     & -14.2    &           \\
15        & -17.4    & -16.1    & -7.3     & -11.3    &           \\ \thickhline
copper        &  377.8  &     &      &    &    other       \\ 
acrylic paint with graphite        &   406.7    &    &   &    &  material        \\
cleaned steel & 131.0   &    &   &    &\\ \thickhline

\end{tabular}
\caption[The point measurement results of other electrode pieces in meV.]{The point measurement results of other electrode pieces in meV. Here we only list the work function measurement values with reference sample correction.}
\label{wftable2}
\end{table}

As discussed in section\,\ref{C3S2}, the work function difference in table\,\ref{measureresult} and table\,\ref{wftable2} will introduce a electric field in across the decay region and the filter region. To study the electric potential distribution, the resulting electrostatic potential is simulated using COMSOL multiphysics.

\begin{figure}
    \centering
    \includegraphics[scale = 0.5]{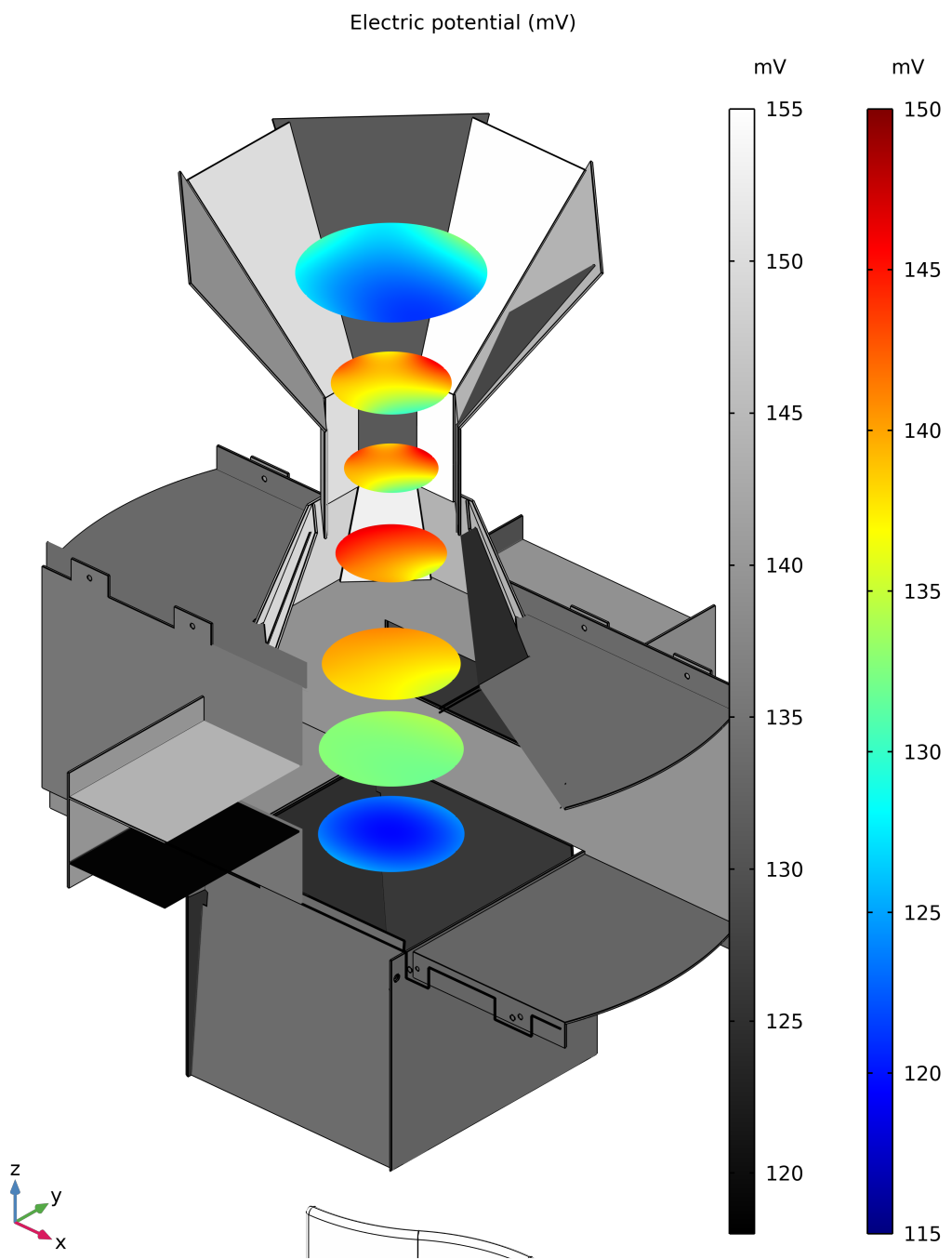}
    \caption[The voltage distribution in the decay and filter region from COMSOL simulation.]{The voltage distribution in the decay and filter region from COMSOL simulation\cite{electrodeab}. The voltage in the flux tube is plotted relative to the red to blue color legend, and the voltage on each electrode piece is plotted relative to the grayscale color legend.}
    \label{electrodes_potential}
\end{figure}

As discussed in \cite{electrodeab}, a COMSOL simulation is used to determine the voltage difference in the decay and filter region. The simulation takes the average value for the work function for each electrode, and models this by setting an external voltage on that electrode, the simulation shows that the 10\,mV difference between the decay region (second from bottom surface) and the filter region (third from top surface) is achieved. Only in the most extreme case is this limit approached.

\chapter{The Geant4 Simulation}
\label{C3_5}
\noindent
The Nab collaboration uses Geant4 (Geometry and Tracking) to simulate and study aspects of the Nab experiment. Geant4 is a toolkit for the simulation of the passage of particles through matter\cite{geant4}, which is based on Monte Carlo method and uses a probabilistic approach to handle the interaction of particle with matter. In Geant4, decay particles are generated and subsequently tracked. Tracking is done in multiple steps. Each step consists of two operations: (1)\,the computation of the particle position after the step through numerical calculation, which is based on the equations of motion for a particle in the electromagnetic field; (2)\,evaluation of interactions between the particle and the surrounding matter using the probability of occurrence for each physics process, which is calculated from the differential cross section of the interaction. The result from the second operation could lead to particle scattering, energy deposition, secondary particle generation, stopping of the particle, etc. The tracking of each particle ends once the particle loses all its kinetic energy, or when it leaves the experiment.

In this chapter, we will discuss the structure of the Nab (Geant4) simulation created by David McLaughlin\cite{NabSim} and further improved by Eric Stevens and Jason Fry. In particular, we will also talk about a situation that will make Geant4 unexpectedly stop and kill the particle event, and propose our solution.

\section{The Nab Simulation Geometry}
\label{C3_5S1}
\noindent
The Nab simulation models the Nab experiment, or at least the parts of it which are relevant for decay particle transport and detection. All relevant parts of the Nab experiment that are included obey cylindrical symmetry around the vertical axis, and the Nab simulation makes use of that in the electromagnetic field calculation\footnote{However the decay region and the filter region, as discussed in chapter\,\ref{C3}, is not cylindrically symmetric and are approximately treated to be cylindrically symmetric.}. We have defined the material for each component of the Nab setup. Table\,\ref{geomaterials} shows the current material type for major components of the geometry in the Nab simulation.
\begin{table}[]
\setlength\tabcolsep{10pt}
\def\arraystretch{1.2}
\centering
\begin{tabular}{cc}
\thickhline
\textbf{Component} & \textbf{Material} \\ \hline\hline
world                         & galactic          \\
electrode                     & titanium                \\
coil                          & iron                \\
grid                          & aluminium                 \\
dead layer                    & silicon                 \\
detector                      & silicon                \\ \thickhline
\end{tabular}
\caption[The material definition for major components of Nab simulation.]{The material definition for major components of Nab simulation. Since Geant4 does not support absolute vacuum, the world material is defined as ``galactic'', which is made up by hydrogen atom in the universe temperature, density and pressure.}
\label{geomaterials}
\end{table}

The Nab simulation calculates the electromagnetic field from the input source file using one of two ways: (1)\,a polynomial expansion of the exact solution of the field equations using elliptic integrals, developed by Ferenc Gl\"uck\cite{Glck2011AXISYMMETRICEF, Glck2011AXISYMMETRICMF}; and (2)\,a field map-based interpolation method that is evaluated faster\footnote{The interpolation method is started in \cite{NabSim}, and later revised by Eric Stevens and Jason Fry.}. The second method is an approximation method that takes the field value at fixed grid points from the first method, and interpolates the field value within the grid volume. Despite the slow initialization process at the beginning of the simulation when it reads a large file with field values used for the interpolation, the second method is much faster than the first method for subsequent use. To save computation time, the interpolation method is applied in most of the regions that are predefined in the simulation, and the first method is only used at regions when the field is changing sharply, as in this case the interpolation method is not precise enough.

\section{Generating Particles}
\label{C3_5S2}
In the Nab simulation, we generate events by simulating electrons and protons\footnote{Only electrons and protons are ran as outgoing particles in the simulation.} according to the decay rate from Eq.\,\eqref{decayrate}. The Coulomb correction term $F(Z,E)$ is approximated as\cite{WILKINSON1982474}:
\begin{equation}
    F(Z,E) = \frac{2\pi\eta}{1-e^{-2\pi\eta}},
\end{equation}
with
\begin{equation}
    \eta = \frac{\alpha E}{p_e c}.
\end{equation}

In the Nab simulation, we first sample the electron kinetic energy from a uniform distribution in the range from 0\,keV to beta decay energy endpoint\footnote{This endpoint energy is calculated as the energy different between the rest energy of a neutron and the sum of the rest energies of proton and a electron.}, then we uniformly sample electron outgoing angle and anti-neutrino outgoing angle within $4\pi$ solid angle. The energy of the anti-neutrino is then calculated by subtracting the electron energy from the beta decay endpoint energy\footnote{Here the kinetic energy of the proton is ignored since it is 2 to 3 orders of magnitude lower than the energy of electron and anti-neutrino.}. After we have all the information of electron and anti-neutrino, we calculate the proton momentum and its outgoing angle using momentum conservation\footnote{We also assume the sum of the momentum is zero.}, and then find the non-relativistic proton kinetic energy as $E_p = \frac{p_p^2}{2m_p}$.

To make the decay particle distribution agree with the decay rate from Eq.\,\eqref{decayrate}, we accept each event sampled using the above method with a probability proportional to the decay rate in Eq.\,\eqref{decayrate}. As a result, the decay particle distribution will be the same as given in Eq.\,\eqref{decayrate}. Energy and momentum are conserved for a neutron that decays at rest.

\begin{figure}
    \centering
    \includegraphics[scale = 0.6]{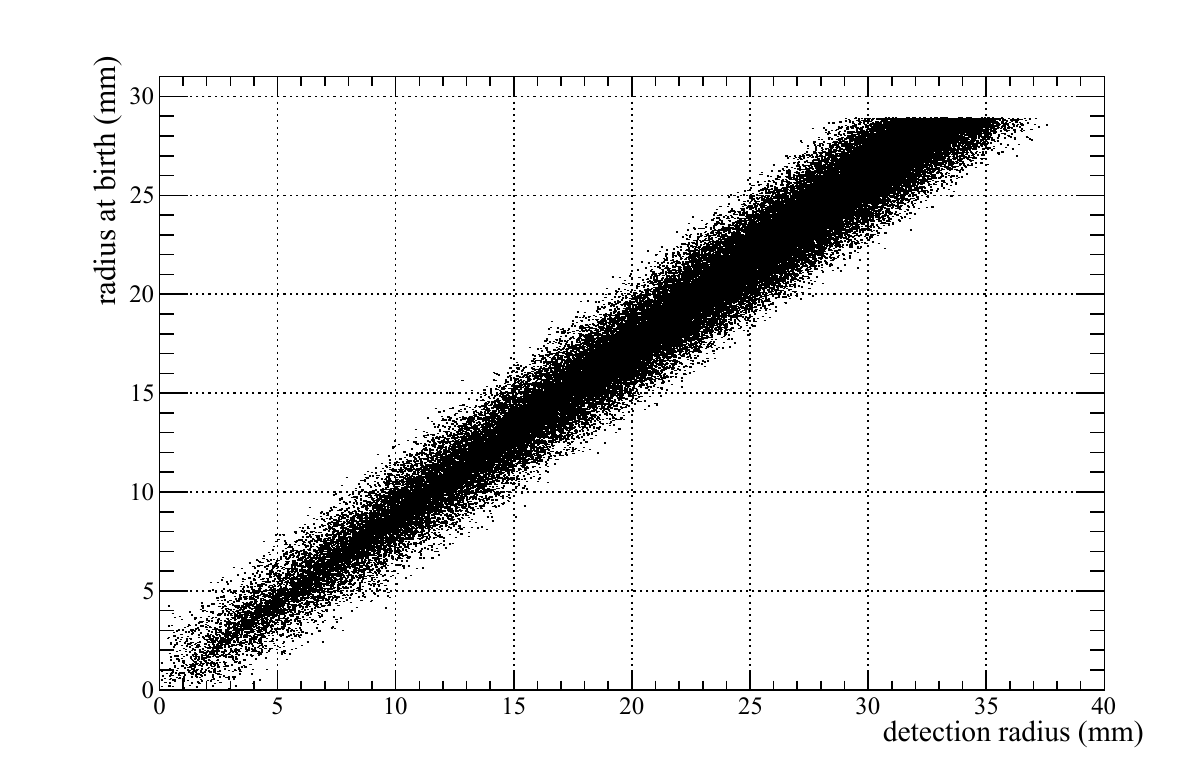}
    \caption{The relation between radial coordinate at birth vs.\ detection radial coordinate for electrons in Nab-$b$ configuration.}
    \label{detR-R}
\end{figure}

\begin{figure}
    \centering
    \includegraphics{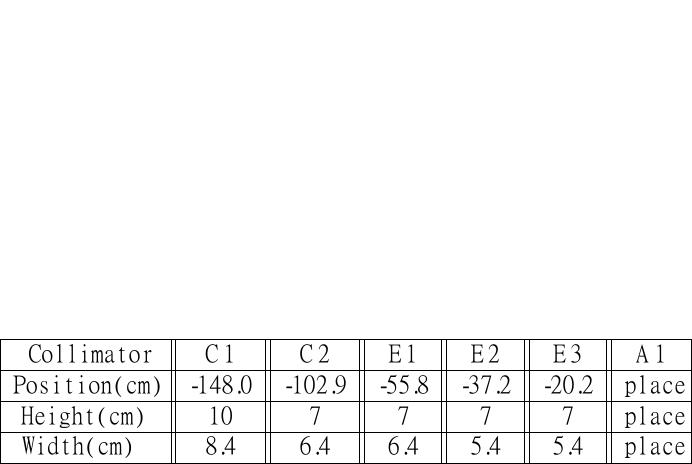}
    \caption[The Nab collimation design.]{The Nab collimation design\cite{maescott}.}
    \label{elizbeth}
\end{figure}

After the decay particle energies and outgoing angles are determined, we uniformly random sample a point from the decay volume, and set the particles starting from that point. As discussed in section \ref{C3_5S1}, the Nab simulation has a cylindrically symmetric geometry, with the decay volume also defined as a cylinder\footnote{The decay volume is a cylinder with vertical axis centered at $-13.189$\,cm, with a height of 8\,cm and radius of 2.89\,cm.}. Fig.\,\ref{detR-R} shows the relation between the detection radial coordinate of the electron and the radial coordinate of the electron at birth. As discussed in chapter\,\ref{C4}, the radius of the decay region is chosen such that it is enough to include the region of interest, but not too big that the particles would run into the electrodes. Fig.\,\ref{elizbeth} shows the Nab collimation system. The height of the decay volume cylinder is chosen to be similar to the true height of the neutron beam from the MC simulation study by Mae Scott\cite{maescott}.

\section{Particle Interactions, Energy Deposition, and the Early Termination of Protons}
\label{C3_5S3}
As described previously, Geant4 handles each step of the simulation with two operations. In the second operation, various physics processes are involved. Table\,\ref{g4interactions}\cite{NabSim} summarized the involved physics processes for the Nab simulation.
\begin{table}[]
\centering
\setlength\tabcolsep{15pt}
\def\arraystretch{1.2}
\begin{tabular}{ccc}
\thickhline
\textbf{Particle Type} & \textbf{Process Name} & \multicolumn{1}{l}{\textbf{Geant4 designation}} \\ \hline\hline
all charged particles       & multiple scattering   & msc                                             \\ 
                        &secondary electron emission & see                                      \\     \hline
electrons           & ionization            & eIoni                                           \\
                       & bremsstrahlung            & eBrems                                          \\ \hline
protons        & ionization            & hIoni                                           \\
                       & bremsstrahlung        & hBrems                                          \\ \hline
                       & compton scattering     & compt                                           \\
photons                & pair production       & conv                                            \\
                       & photoelectric effect  & phot                                            \\ \thickhline
\end{tabular}
\caption[Geant4 physics processes.]{Geant4 physics processes. The ionization process of the electrons has \texttt{dRoverRange} = 0.1 and \texttt{finalRange} = 100\,$\mu$m, the ionization process of the photons has \texttt{dRoverRange} = 0.1 and \texttt{finalRange} = 20\,$\mu$m. The variable \texttt{dRoverRange} determines the maximum fraction of the stopping range that can be travelled by an electron in a step, and the parameter \texttt{finalRange} fixes the minimum step size below which the electron is absorbed locally\cite{2007}.}
\label{g4interactions}
\end{table}

The Nab simulation defines the detector material as silicon in accordance with the Nab experiment\footnote{The detector has a radius of 6.3\,cm and 2.0\,mm thickness, and are centered at 5.0\,m and $-1.3$\,m respectively. Each detector has a 100\,nm of silicon deadlayer in front of its surface.}. The detectors are defined as sensitive detectors\footnote{This implemented as \texttt{NabSensitiveDetector} class, and this class is inherited from \texttt{G4VSensitiveDetector} class of Geant4.}\cite{NabSim}, which collects (1)\,``Track identifie'', (2)\,``Energy deposited'', (3)\,``Time of step'', (4)\,``Position vector'' (5)\,``Particle type'', (6)\,``Pixel number and position'' for any particle step in its volume. In particular, ``Energy deposited'', ``Time of step'', and ``Pixel number and position'' could be used to calculate the total deposited energy and position of the particle in the detector, and be used to study the detector response and detector waveform signal.

\begin{figure}
    \centering
    \includegraphics[scale=0.6]{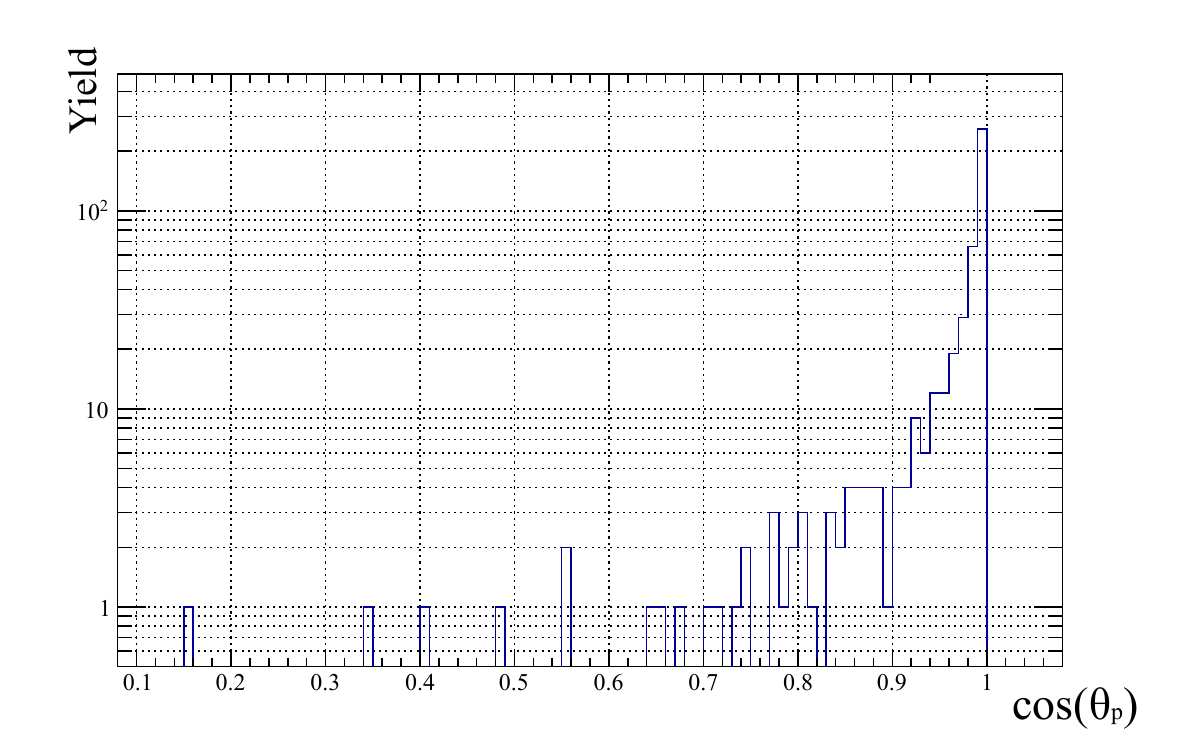}
    \caption[The cosine angle distribution early terminated protons.]{The cosine angle distribution early terminated protons. The $x$ axis is the cosine of the angle between proton born momentum and the $z$ direction.}
    \label{lost_protons}
\end{figure}

As described in table\,\ref{geomaterials}, the world material is defined as ``galactic'', which could be regarded as very low-density gas that is made of hydrogen atoms. This residual gas, although very thin, still interact with the decay particles with a low probability. The Nab-$b$ configuration, as described in Fig.\,\ref{b_config_e}, set the filter region to be at the voltage of 850\,V relative to the decay volume, in order to reflect all the decay protons to the bottom detector. At the reflecting point, the momentum component along the magnetic field line of the proton is zero. In addition, if the proton was born with its momentum direction close enough to the guiding center magnetic field line direction, this proton will lose almost all its kinetic energy at the reflecting point. In this scenario, the simulated ionization process between the proton and the residual gas, as shown in table\,\ref{g4interactions}, will dissipate all the kinetic energy of the proton within one step, and the proton track will be stopped and terminated by Geant4 due to zero kinetic energy. In real experiment, the proton will gain kinetic energy again from the electric field potential and turn back. The simulation, however, cannot handle this specific situation correctly. This caused the issue of the early termination of protons by Geant4. Fig.\,\ref{lost_protons} shows the angle distribution of the early terminated protons. From the figure we can clearly see that the protons are more likely to be terminated with $\text{cos}(\theta_{\text{p}})$ close to 1.

\begin{figure}
    \centering
    \includegraphics{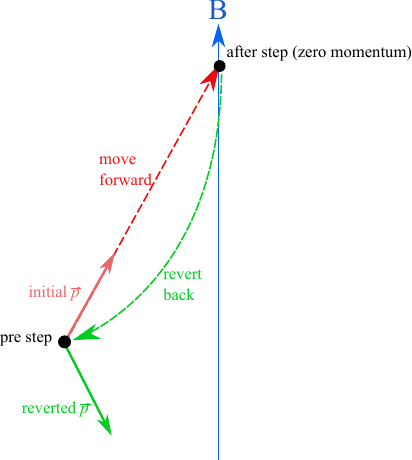}
    \caption[Illustration of a proton reverting its vertical motion.]{Illustration of a proton reverting its vertical motion. In this step, the proton moves from the pre-step point to the after-step point, and loses all its kinetic energy. This proton is then reverted back to the pre-step point and is given back its pre-step energy.}
    \label{revert_graph}
\end{figure}

As these protons are not stopped in reality, this early termination imposes a systematic error in the Nab simulation. To resolve the problem, we revive every proton when it is been terminated due to zero energy in the filter region. As illustrated in Fig.\,\ref{revert_graph}, the revived proton is brought back to the pre-step point, and is given back its pre-step kinetic energy to conserve energy. In order to make the proton moving along the same magnetic field guiding center, the momentum component in parallel to its guiding center magnetic field is reverted, while the momentum component vertical to its guiding center magnetic field remains unchanged.

On the flip side of the coin, there are protons born with very low kinetic energy. For these protons, it is expected for Geant4 to terminate their tracks because (1)\,they have very deficient kinetic energy at birth, and will spend a very long time before they travel to the detector, thus will not be counted as decay events\footnote{As discussed in chapter\,\ref{C5}, only protons arrive within a certain time window will be counted.}; (2)\,tracking these protons is very time-consuming and will increase the simulation run time. To avoid reviving these protons repeatedly, we revive each proton at most once. The simulation shows that only 0.0038\% of proton tracks are terminated under this situation (a fraction small enough that we can neglect it), and all these protons have kinetic energy < 1\,eV.

\begin{figure}
    \centering
    \includegraphics[scale = 0.6]{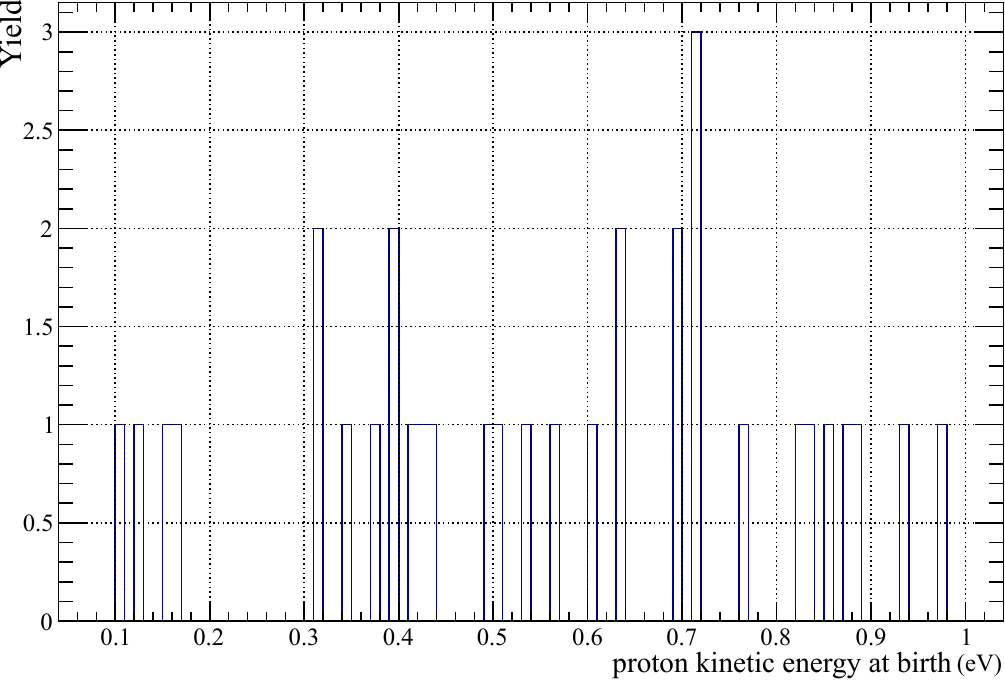}
    \caption[The born kinetic energy distribution of protons that are killed again after the revival process.]{The born kinetic energy distribution of protons that are killed again after the revival process. There is no event with the born kinetic energy of proton to be above 1\,eV.}
    \label{rever2}
\end{figure}

To confirm that we only revive the targeted protons, Fig.\,\ref{rever2} shows the born kinetic energy distribution of the protons that are revived but get killed by Geant4 again. From the figure, we find that among all the proton tracks, the terminated proton tracks have kinetic energy ($E_{\text{p}}$<1\,eV). This is expected from the revival mechanism.

\chapter{The Systematic Uncertainties of Fierz Interference Term $b$ }
\label{C4}
\noindent
As introduced in chapter\,\ref{C1} and chapter\,\ref{C2}, one important goal of the Nab experiment is to measure the Fierz interference term $b$ to the precision of $\Delta b \sim 3\cross 10^{-3}$. The Fierz interference term is predicted to be 0 due to the $V-A$ structure of the weak interaction from the Standard Model. A nonzero $b$ value will indicate the existence of other types of interaction, and will provide information on physics beyond the Standard Model.

To achieve the goal for the $b$ measurement, the detector calibration and detector response needs to be characterized to a certain level. In this chapter we will discuss the uncertainty budget in the Nab-$b$ configuration.

\section{The Fitting of the Electron Energy Spectrum}
\label{C4S1}
\noindent
To extract $b$, the beta decay electron spectrum will be fitted with Eq.\,\eqref{espec}, with $b$ as a free parameter. As Eq.\,\eqref{espec} states:
\begin{equation}
\nonumber
\Gamma (E_e) = \frac{d\omega}{dE_e} \propto F(Z,E_{e}) p_{e}E_{e}(E_{0}-E_{e})^{2}(1+b\frac{m_e}{E_e})\,,
\end{equation}
the proportionality factor should also be left as a free parameter. In the next section, we will also talk about other parameters that may be varied in the fitting procedure.

The fit range is also important. For the electron spectrum, a wider range of fitting could provide us with more statistics and lower error, but including too low energy (energy < 150\,keV) of the spectrum could introduce noise and other systematic bias to the results. Table\,\ref{numberstats} shows the fitted error for $b$ with respect to the starting point of the fitting range.

\begin{table}[]
\centering
\renewcommand{\arraystretch}{1.5}
\begin{tabular}{c|ccccc}
\thickhline
Starting energy (keV) & 0   & 100& 160  & 200  & 300  \\
\hline\hline
fitted $b$ uncertainty     & $\frac{7.5}{\sqrt{N}}$ &  $\frac{10.1}{\sqrt{N}} $ &  $\frac{12.9}{\sqrt{N}} $&  $\frac{15.6}{\sqrt{N}} $ &  $\frac{26.4}{\sqrt{N}} $\\\hline
fitted $b$ uncertainty (free gain)     & $\frac{9.1}{\sqrt{N}}$ &  $\frac{12.7}{\sqrt{N}} $ &  $\frac{16.8}{\sqrt{N}} $&  $\frac{20.4}{\sqrt{N}} $ &  $\frac{36}{\sqrt{N}} $\\
\thickhline
\end{tabular}
\caption[The fitted uncertainty in $b$ and the starting fitting energy relation.]{The fitted uncertainty in $b$ and the starting fitting energy relation, $N$ is the total number of decay events, including the ones which don't make it through proton TOF and electron energy cut.}
\label{numberstats}
\end{table}

\begin{figure}[h!]
\centering
\includegraphics[scale=0.6]{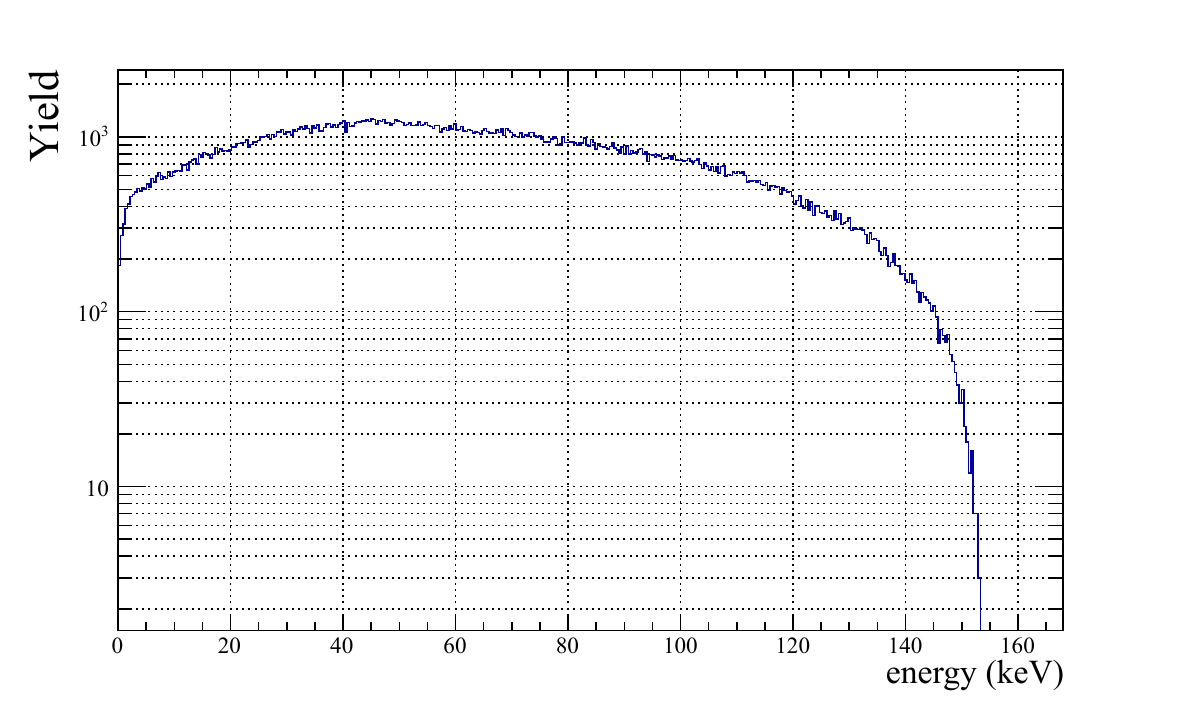}
\caption{The energy spectrum of electrons that can not make it to the detector.}
\label{lostelectrons}
\end{figure}

In the Nab-$b$ setting, the bottom detector will be on a voltage of $-$30\,keV to accelerate the protons. Due to the same reason as discussed in Chapter\;\ref{C2}, the magnetic mirror will reflect electrons born with insufficient longitudinal momentum. Due to the high voltage of the bottom detector, part of the reflected electrons will be reflected again, and been trapped between the bottom detector and the filter region. Fig.\,\ref{lostelectrons} shows the trapped electron spectrum.

According to Geant4 simulation results for the Nab-$b$ configuration, the electron energy fitting starting point should be at $\sim$160\,keV to avoid trapped protons. From table\,\ref{numberstats}, in order to get the fitting uncertainty of $b$ to be at $10^{-3}$ level, the Nab experiment needs $10^8$ or more decay events\footnote{To get enough statistics, $10^8$ events were ran with the simulation with integer random seeds from 0 to 999 inclusive.}. In the remaining content of this chapter, all the fitting will be done in the range of 160\,keV to 780\,keV, with a bin width of 3.33\,keV.

\section{The Detector Calibration and Response}
\label{C4S2}
\noindent
In the Nab experiment, the Fierz interference term is measured directly from the beta spectrum of the electron. The Nab experiment uses two 127 hexagonal pixels silicon detectors with an area of 70\,mm$^2$ each and outer rings of partial pixels.

\begin{figure}[h!]
\centering
\includegraphics[scale=1]{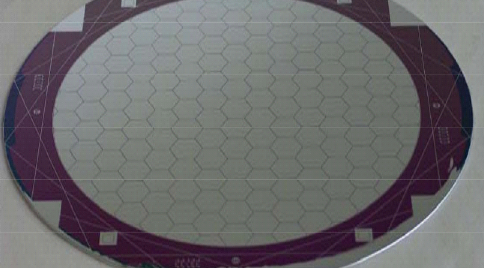}
\caption{The Nab experiment silicon detector (back side).}
\label{detector}
\end{figure}

\begin{figure}[h!]
\centering
\includegraphics[scale=1]{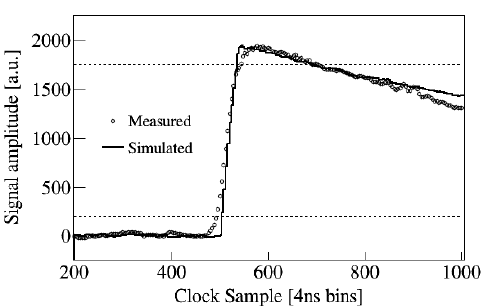}
\caption[Waveform of a 363.8\,keV conversion electron from $^{113}$Sn with simulated detector signal waveform.]{Waveform of a 363.8\,keV conversion electron from $^{113}$Sn with simulated detector signal waveform\cite{BROUSSARD201783}. Dashed horizontal lines mark 10\% and 90\% of the maximum amplitude.}
\label{waveform}
\end{figure}

Once hit by a particle, the particle deposits energy in the form of electrons and hole pairs. The bias voltage across the silicon detector moves the electrons and holes, the electrons to the back, and the holes to the front. This process builds up charge on the readout electrode at the backside (one or more of the hexagonal pixels as shown in Fig.\,\ref{detector}), which is then detected by a preamplifier whose output signal is recorded in our DAQ system. Fig.\,\ref{waveform} shows a typical waveform.

The waveform is then digitized in analog to digital converters (ADCs), and the height of the waveform is attributed to an ADC channel (ch)\footnote{The attributed channel is proportional to the height of the waveform in principle, with the noise of the waveform being corrected.}, the particle hit energy and the channel have the general relationship:
\begin{equation}
    E = g \cross \text{ch} + \text{offset} + (\text{non-linearity term})\,,
\end{equation}
where $g$ represents the conversion gain factor.

This relation between detected energy ($E$) and output pulse height (ch) needs to be calibrated with calibration sources. As presented in \cite{BROUSSARD201783}, the Nab detectors calibration have achieved a 3\,keV FWHM and a noise threshold of 6\,keV with high linearity at typical pixels, this also indicates that the nonlinearity term is small.

The actual detector calibration is planned to provide an in-situ measurement of the gain factor ($g$) and offset. It allows to reconstruct the energy as:
\begin{equation}
    E' = g' \cross \text{ch} + \text{offset}'\,,
\label{calibresponse}
\end{equation}
the ($'$) marking means these values are measured and calculated from calibration.

Due to the imperfection of the calibration, the relation between measured energy ($E'$) and the true energy ($E$) could be formulated as following:
\begin{equation}
    E' = \alpha  + \beta E + (\text{non-linearity term})\,,
\label{erespons}
\end{equation}
and a perfect calibration will have $\alpha = 0, \beta = 1$, and non-linearity term = 0.

With this imperfect knowledge of the detector, our reconstructed electron energy spectrum $f'(E')$ will also be different from real spectrum:
\begin{equation}
    f'(E') = f(E)\frac{dE}{dE'}\,.
\label{imperfection}
\end{equation}
Since $f'(E')$ and $f(E)$ are different, the fitting will produce a $b'$ that is different from the true $b$ value. In this section, we review the level of knowledge about the detector calibration required to achieve the goal of Nab-$b$.

\subsection{The Gain Factor}
\label{C4S2S2}
\noindent
To get the electron spectrum, we need to find the gain factor in the detector calibration relation. Any deviation of the measured gain factor ($g'$) from the true gain factor ($g$) will affect the fitted $b$ value.

\begin{figure}
     \centering
     \includegraphics[scale=0.7]{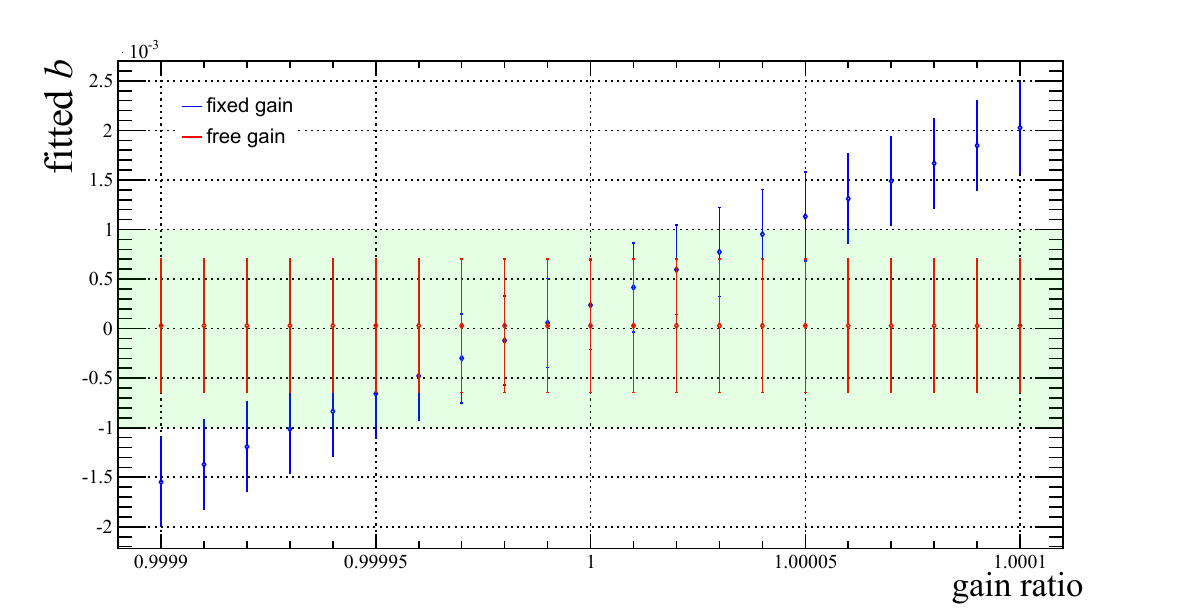}
     \caption[The fitted $b$ value versus the gain ratio $g/g'$.]{The fitted $b$ value versus the gain ratio $g/g'$. The blue points show the true result with $g'$ fixed in the fitting. The red points, however, show the fitted $b$ response when we allow $g'$ to be free again during fitting, in which case the fitted $b$ should not be affected by the gain error any more. The green area shows the region of $\abs{b} < 10^{-3}$. The linearly fitted parameters for the blue line is $y =\beta_0 + \beta_1 x$ with $\beta_1 = 17.9\pm 1.6$ and $\beta_0 = -17.9 \pm 1.6$.}
     \label{gain}
 \end{figure}

We calculated an electron energy histogram and introduced statistical noise consistent with the one expected for $10^9$ events, and we added noise consistent with  $10^8$ background events with uniform energy distribution. Assuming our calibrated gain factor ($g'$) is different from the true gain factor ($g$), the reconstructed histogram is also going to be different. In this study, we reconstructed many different histograms, with different $g'$ values, and fitted them to extract $b$. Fig.\,\ref{gain} showed the relation between gain ratio ($g/g'$) and the fitted b value.
 
 According to the graph, to control the error of fitted $b$ value within $\Delta b \sim 10^{-3}$, we need to know the gain ration to $\Delta g'/g < 5 \cross 10^{-5}$, which is very difficult to achieve. To solve this problem, this parameter could be left free in the final fitting process, at the cost of a slight increasing of the fitted $b$ error.

\subsection{The Offset}
\label{C4S2S1}
\begin{figure}
     \centering
     \includegraphics[scale=0.7]{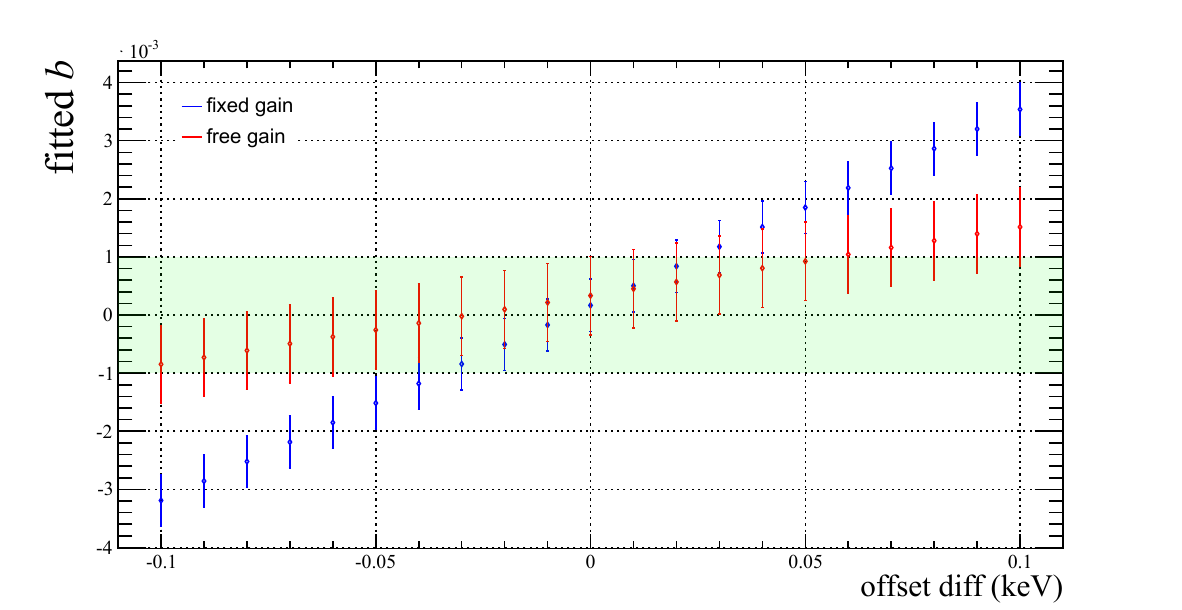}
     \caption[The fitted $b$ value versus the offset difference ($\text{offset} - \text{offset}'$).]{The fitted $b$ value versus the offset difference ($\text{offset} - \text{offset}')$. The blue points show the relation with fixed gain the red points show the relation with gain as a free fit parameter. The plot demonstrates that the free gain parameter also helps to reduce the error of the offset. The green area shows the region of $\abs{b} < 10^{-3}$. The linearly fitted parameters for the blue points is $y =\beta_0 + \beta_1 x$ with $\beta_1 = 0.0118 \pm 0.0024$ and $\beta_0 = 3.3\cross 10^{-4}  \pm 1.5 \cross 10^{-4}$, for the red points $\beta_1 = 0.0337 \pm 0.0016$ and $\beta_0 = 1.69\cross 10^{-4} \pm 9.8\cross 10^{-5}$, respectively.}
     \label{offset}
 \end{figure}

\noindent
To study the effect of the offset, we assume $\alpha$ and the non-linearity term vanish in Eq.\,\eqref{erespons}, and input different values of $\alpha$, the offset difference ($\text{offset} - \text{offset}'$), to evaluate $b$ from fitting. The same method is applied as in section\,\ref{C4S2S2}, Fig.\,\ref{offset} shows the fitted result with different offset difference.

According to the graph, to achieve our goal, we need to determine the offset to 0.03\,keV if we fix the gain factor in the fitting, or to 0.08\,keV if we keep gain factor as a free parameter.

\subsection{The Non-linearity}
\label{C4S2S3}
\noindent
The Nab detector is highly linear. To study the non-linearity term, we first model it to be an analytic quadratic term: non-linearity term of the energy of various $ \propto \text{ch}^2$, then we model the non-linearity to be a Gaussian perturbation on the calibration sources.

\begin{figure}
     \centering
     \includegraphics[scale=0.7]{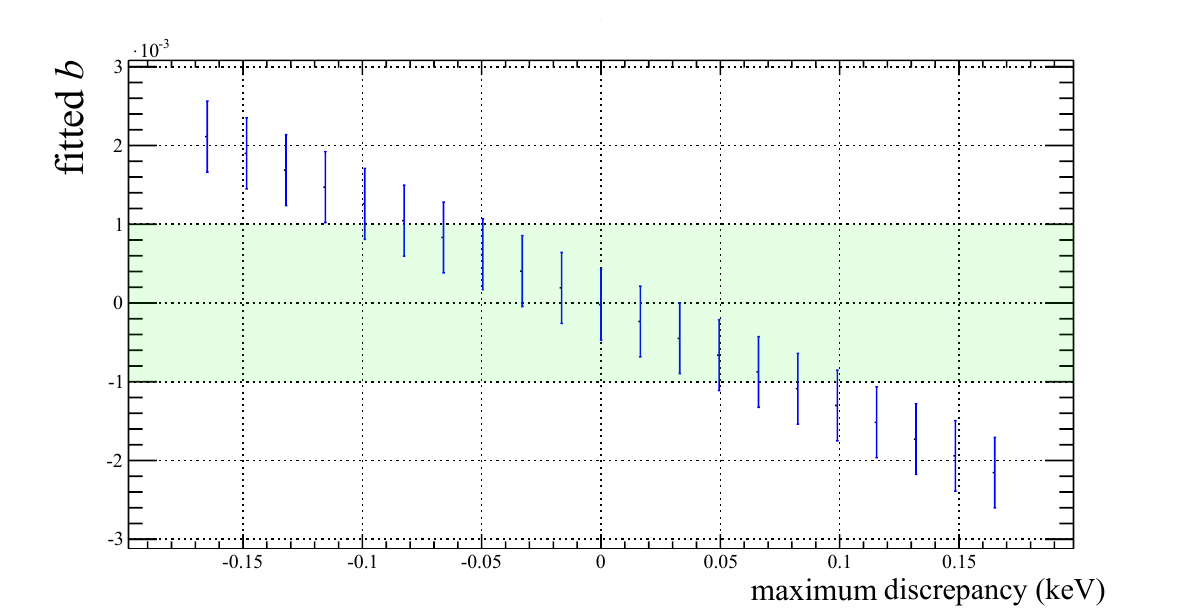}
     \caption[The fitted $b$ value versus the maximum discrepancy from calibration.]{The fitted $b$ value versus the maximum discrepancy from calibration. The green area shows the region of $\abs{b} < 10^{-3}$. The linearly fitted parameters for the blue points is $y = \beta_0 + \beta_1 x$ with $\beta_1 = -0.01293 \pm  0.00098$ and $\beta_0 =  -2.1 \cross 10^{-5} \pm 9.8 \cross 10^{-5}$.}
     \label{nonlinear_quadratic}
 \end{figure}

In the first model, we model non-linearity of the detector calibration to be quadratic: 
\begin{equation}
     E = g \cross \text{ch} + \text{offset} + k \cross \text{ch}^2\,.
\end{equation}
We calibrated this model using Eq.\,\eqref{calibresponse}, and used the calibrated $g'$ and offset$'$ to reconstruct the electron spectrum. We subsequently adjust the value of $k$ to find the effect of this non-linear term to our fitted $b$ value. To make the graph more meaningful, we use the maximum discrepancy in the calibration\footnote{The maximum discrepancy is defined as the maximum difference between the measured energy and the calibrated energy from the calibration sources.} to characterize the strength of the non-linear term. Fig.\,\ref{nonlinear_quadratic} shows the relation between maximum discrepancy and the fitted $b$ value.

According to the plot, to achieve our goal, we need to determine the maximum discrepancy to 0.07\,keV. Note that this energy is the the maximum distance of the quadratic line to the linear line in the energy range from 0\,keV to 1000\,keV. The actual required discrepancy is dependent on the energy of the calibration source.

The second model, we will assume propose a Gaussian model that is dependent on the actual calibration sources. Table\,\ref{calibsource} shows the energy of the calibration sources for the Nab detector.

\begin{table}[]
\centering
\begin{tabular}{l|ll}
\thickhline
calibration source & \multicolumn{2}{l}{energy (keV)} \\ \hline \hline
Ba-133              & 45.0\,,        & 76.6            \\
Cd-109              & 87.0\,,           &                 \\
Ce-139              & 136.6\,,          &                 \\
Sn-113              & 371.9\,,         &                 \\
Bi-207              & 504.5\,,          & 996.9           \\ \thickhline
\end{tabular}
\caption[The calibration sources for the Nab spectrometer.]{\label{calibsource}The calibration sources for the Nab spectrometer. As the table shows, Ba133 and Bi207 have two spectral lines.}
\end{table}

\begin{figure}
     \centering
     \includegraphics[scale=0.7]{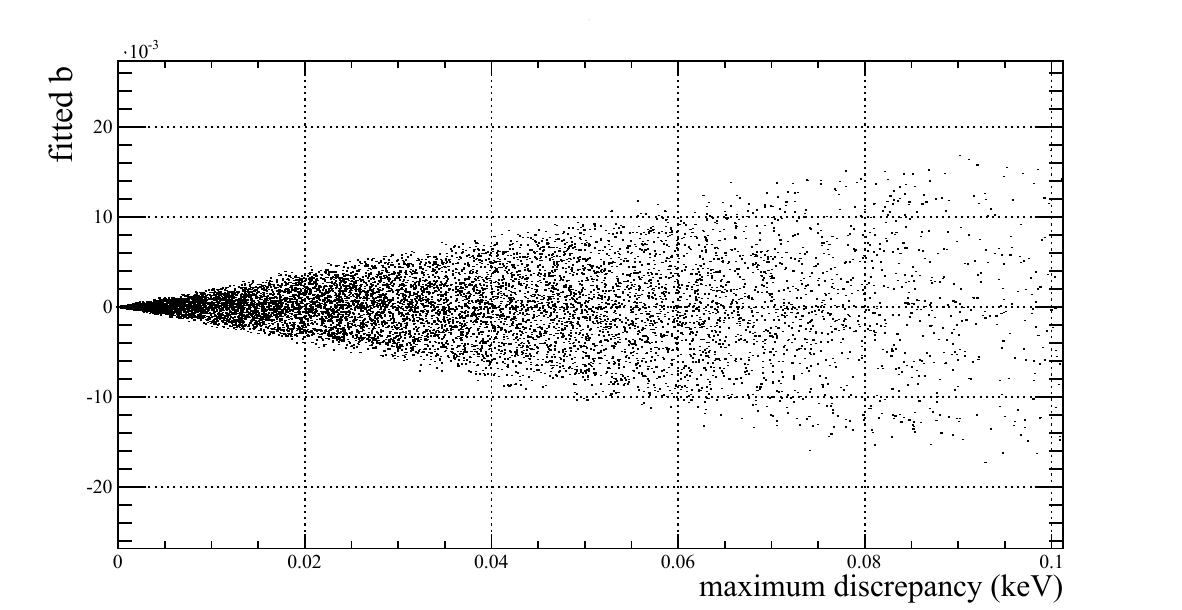}
     \caption{The fitted $b$ value versus the maximum discrepancy from calibration.}
     \label{nonlinear_calib}
 \end{figure}

We will add a random Gaussian distributed perturbation to each of the calibration energies. To construct the true detector calibration function, we will model the nonlinear term as a third order spline that connects the perturbed energy points. The detector calibration function Eq.\,\eqref{calibresponse} will be obtained by fitting only to the calibration source energy lines. Finally we reconstruct the spectrum with the true detector calibration function, and fit it to extract the $b$ value. In this study, due to symmetry, the absolute value of the maximum discrepancy from calibration energy lines is used to characterize the strength of the non-linearity. In this study, we sampled the Gaussian perturbation with different width, a total number of 10000 samples are created, Fig.\,\ref{nonlinear_calib} shows the distribution of these samples.

\begin{figure}
     \centering
     \includegraphics[scale=0.7]{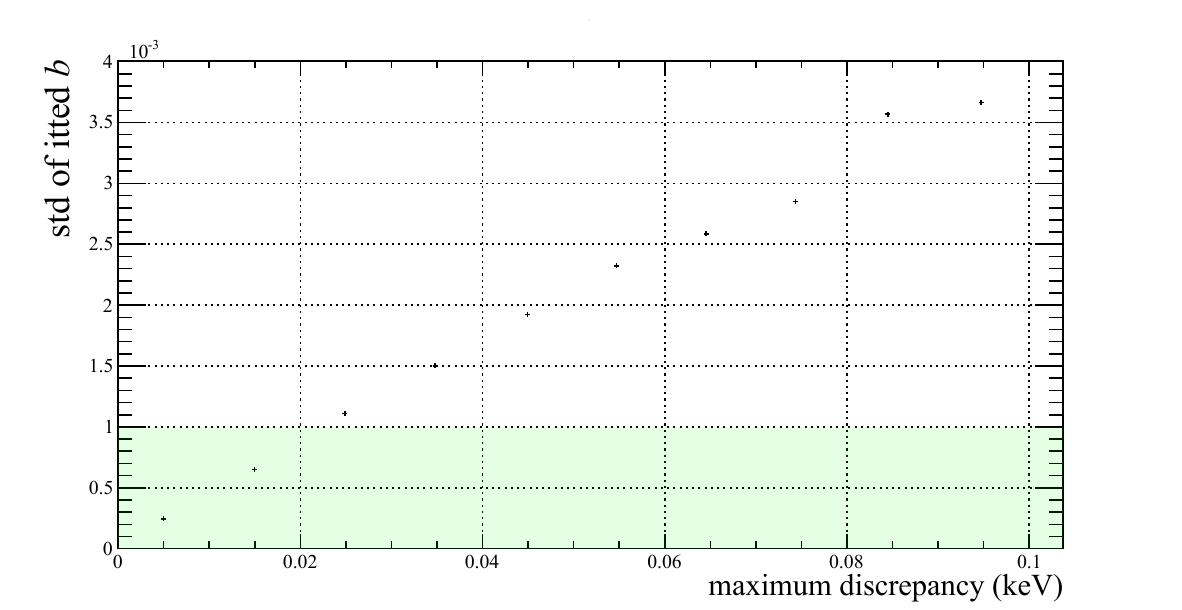}
     \caption[The relation of the standard deviation (std) of the fitted $b$ value to the maximum discrepancy energy.]{The relation of the standard deviation (std) of the fitted $b$ value to the maximum discrepancy energy. The green area shows the region of std of fitted $b < 10^{-3}$.}
     \label{std_nonlinear_calib}
 \end{figure}

This plot shows that in the random Gaussian model, the range of fitted $b$ values gets wider as the amplitude of the discrepancy increases. The envelope of the distribution shows the worst case of the detector non-linearity effect on $b$, and the standard deviation of the spread in the $y$ direction could reflect the average cases of the detector nonlinearity effect under this model. Fig.\,\ref{std_nonlinear_calib} shows the standard deviation (std) of the spread as a function of the maximum discrepancy\footnote{In this graph, points from Fig.\,\ref{nonlinear_calib} are binned in 0.01 keV bins, the std of fitted $b$ and the average of the maximum discrepancy energy in each bin are calculated and plotted.}.

However, this study inputs the non-linearity term in a random way, which could be wiggly and exaggerate the influence of the detector non-linearity. To connect this graph to the quadratic model in the previous subsection, we could view Fig.\,\ref{nonlinear_quadratic} as a specific subset of points from Fig.\,\ref{std_nonlinear_calib}.

As shown in the figure, assuming the non-linearity term to be a random Gaussian perturbation, we need to know the maximum discrepancy of the calibration to 0.02\,keV.

\subsection{The Width of the Detector Response}
\label{C4S2S4}
\noindent
The Nab detector has reached the energy resolution of 3\,keV full width half maximum (FWHM)\cite{BROUSSARD201783}. Due to the width of the detector response, for a given reconstructed $E_e$, the electrons could have a range of values as their initial energy. In this analysis, we will create a histogram with the beta spectrum Eq.\,\eqref{espec} convoluted with the following Gaussian function:
\begin{equation}
    G(E,E_e) = \frac{1}{\sqrt{2\pi}\sigma} e^{-\frac{(E-E_e)^2}{2\sigma^2}}\,,
\label{gauss}
\end{equation}
the parameter $\sigma$ controls the FWHM of the detector response peak: FWHM = $2.355 \sigma$. 

To analyse this effect, we will then use the beta spectrum convolved with Gaussian functions of different FWHM, as the fitting function, and extract $b$ from the fitting.

 \begin{figure}[H]
     \centering
     \includegraphics[scale=0.7]{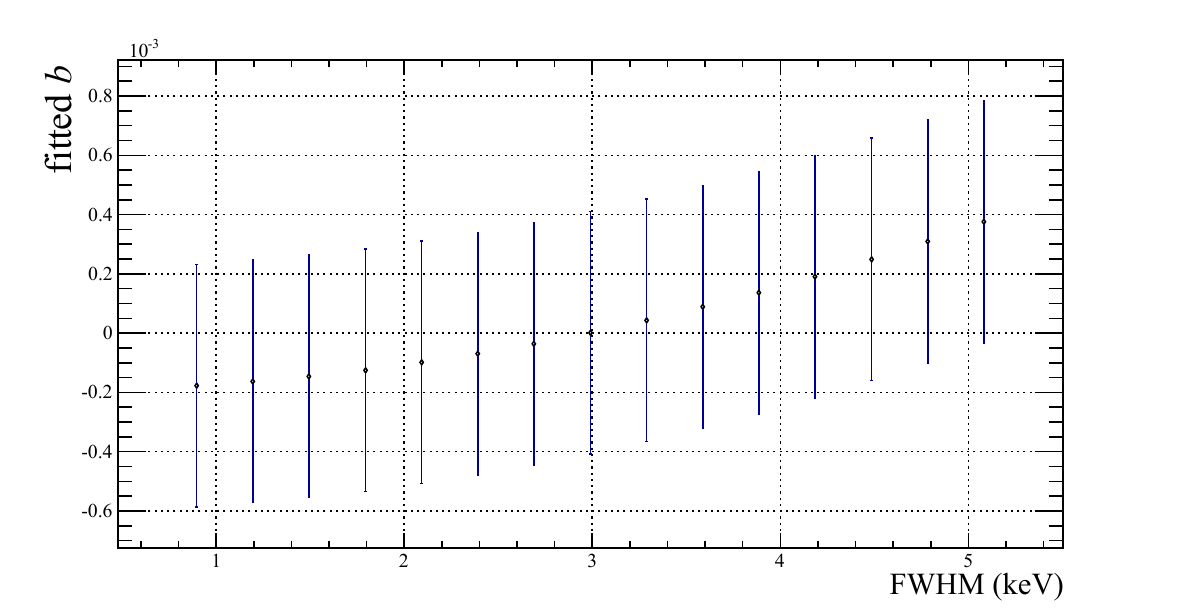}
     \caption[The relation of the fitted $b$ value to the FWHM of the detector response.]{The relation of the fitted $b$ value to the FWHM of the detector response.}
     \label{peakwidth_3keV_2}
 \end{figure}
 
The result if shown in Fig.\,\ref{peakwidth_3keV_2}, as we can see, the fitted $b$ result is very weakly sensitive to the width of the peak.

\subsection{The Tail of the Detector Response}
\label{C4S2S5}
\begin{figure}
     \centering
     \includegraphics[scale=0.7]{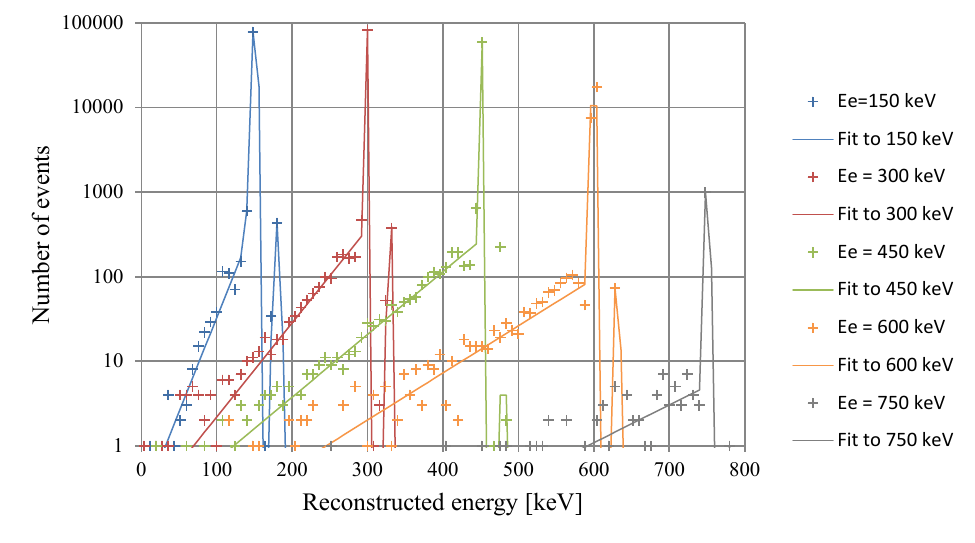}
         \caption[The detector response function for electrons with energies that are multiples of 150 keV.]{The detector response function for electrons with energies that are multiples of 150 keV. The data points are obtained from simulation, and the solid line in the graph is the proposed analytical model.}
     \label{detector_response}
 \end{figure}

\noindent
Due to the bremsstrahlung radiation of the electrons, the stopping of the detector dead layer and grid, some electrons lose a significant amount of their energy before they enter the active part of the detector. This would cause a low energy tail in the detector response function. The Nab collaboration has studied realistic Geant4 Monte Carlo simulations of the detector response function \cite{electronenergydetermination}. Fig.\,\ref{detector_response} shows the simulated detector response of electrons with a range of discrete energies generated in this study.

The detector response model is:
\begin{equation}
\begin{split}
    T(E, E_e)&= \begin{cases}
      t \frac{E_e}{E_fE_g} \text{exp}(\frac{E-E_e}{E_g})\,, & E < E_e\,,\\
      0\,, & E \geq E_e\,,\\
    \end{cases}   \\
    P(E,E_e) &= (1-t\frac{E_e}{E_f})\delta (E-E_e)\,,\\
\end{split}
\label{tail}
\end{equation}
with $E_f = 300$\,keV and $E_g = 0.13 E_e$, $t$ is the parameter to define the amount of tail.

\begin{figure}
     \centering
     \includegraphics[scale=0.5]{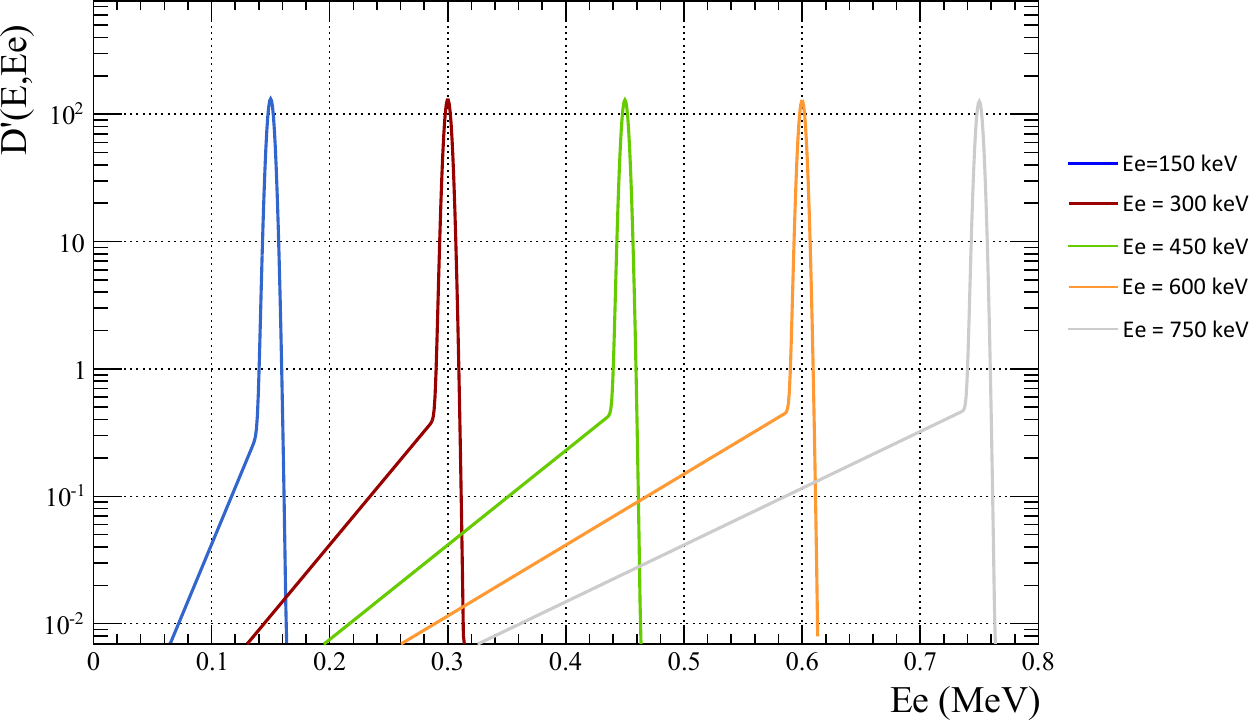}
         \caption[The function of $D(E,E_e)$ with different $E_e$.]{The function of $D(E,E_e)$ with different $E_e$. To compare this figure with Fig.\,\ref{detector_response}, note that each curve in this plot has the normalized area of 1 underneath.}
     \label{detector_response_function}
 \end{figure}

And the full detector response is:
\begin{equation}
    D(E,E_e)=T(E,E_e) + P(E,E_e)\,.
\label{tailmodel}
\end{equation}
To have a consistent tail model, model in Eq.\,\ref{tailmodel} is convolved with the Gaussian function in Eq.\,\eqref{gauss}:
\begin{equation}
\begin{aligned}
    D'(E,E_e) =& (D\ast G)(E,E_e)\\
    = &t\frac{E_e}{2E_f E_g}\text{exp}[\frac{2E_g(E-E_e)+\sigma^2}{2E_g^2}]\text{erfc}[\frac{E_g(E-E_e)+\sigma^2}{\sqrt{2}E_g\sigma}]\\
    &+(1-t\frac{E_e}{E_f})\frac{1}{\sqrt{2\pi}\sigma}\text{exp}[-\frac{(E-E_e)^2}{2\sigma^2}]\,,
\end{aligned}
\label{tailconv}
\end{equation}
in the above calculation, the convolution is done with respect to $E$, and $E_e$ or $E_g$ is treated as a constant. The first line of Eq.\,\eqref{tailconv} is the tail part and the second line is the peak part from Eq.\,\eqref{tail}, the purpose of the convolution is to (1) transfer the tail into a smooth function, and (2) make the model consistent with the model from Eq.\,\eqref{gauss} when $t$ = 0. Fig.\,\ref{detector_response_function} shows the plot of $D'(E,E_e)$ with different $E_e$, $t$ is set to 0.021.

\begin{figure}[H]
     \centering
     \includegraphics[scale=0.7]{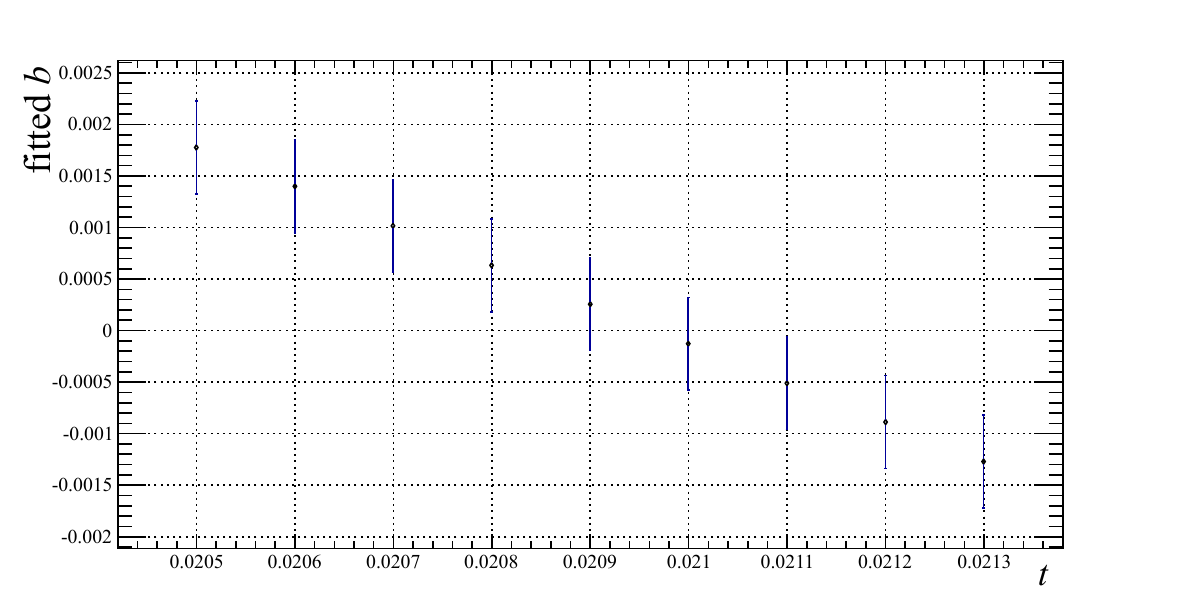}
     \caption[The relation of the fitted $b$ value to the tail percentage $t$ of the detector response.]{The relation of the fitted $b$ value to the tail percentage $t$ of the detector response. The green area shows the region of $\abs{b} < 10^{-3}$. The fitted parameters for the is $y = \beta_0 + \beta_1 x$ with $\beta_1 = -3.81 \pm  0.58$ and $\beta_0 =  0.080 \pm 0.012$.}
     \label{tail_percent}
 \end{figure}

With the above tail model, we study the effect of the tail, we input a 3\,keV FWHM for the peak and $t = 0.021$ for the tail as our detector response, and create and fit histograms with the same method from the previous subsection. In the fitting algorithm, we make the fitting function by the same method with different fixed $t$ value, and find the retrieved $b$ value as a function of $t$. Fig.\,\ref{tail_percent} shows the result of the fitted $b$ value with different input fitting parameter $t$. From the fitting result, we find that, for $\Delta b = 10^{-3}$ budget, we need to determine the tail percentage parameter to 1\%. As shown in table\,\ref{asystem}, this is also the uncertainty budget requirement for Nab-$a$ configuration.

\begin{figure}[H]
     \centering
     \includegraphics[scale=0.7]{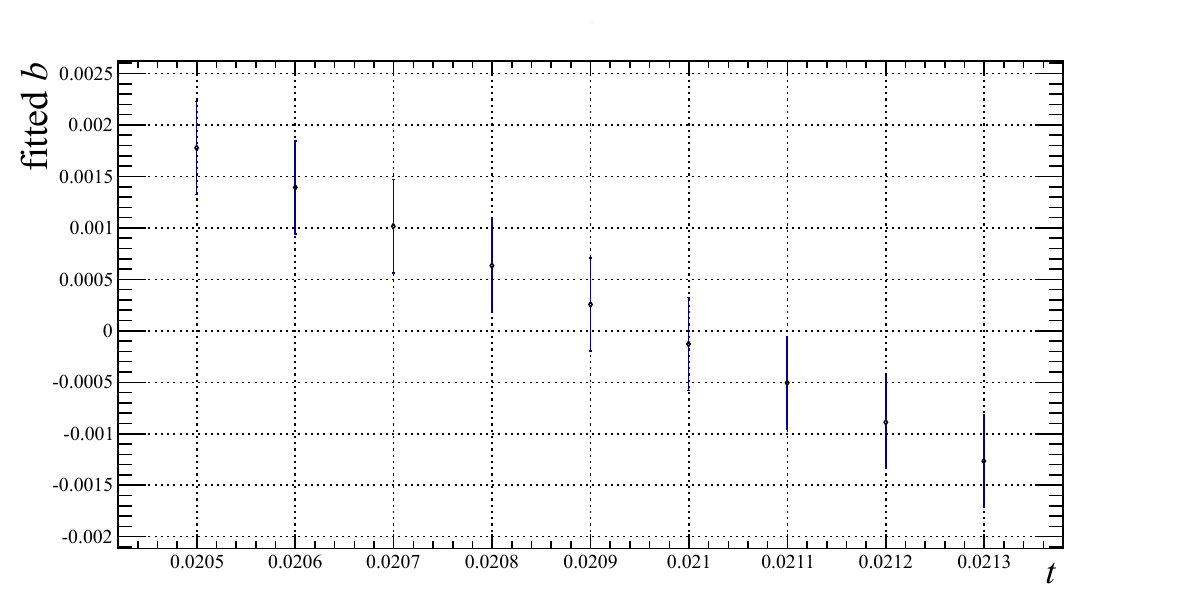}
     \caption[The relation of the fitted $b$ value to the raw tail model.]{The relation of the fitted $b$ value to the raw tail model of Eq.\,\ref{tail}. The green area shows the region of $\abs{b} < 10^{-3}$.}
     \label{tail_percent_WO_sigma}
 \end{figure}

For a consistent check, we also calculated the effect of the tail using the raw tail model Eq.\,\eqref{tail}, the result is similar, Fig.\ref{tail_percent_WO_sigma} shows the result.

\section{The TOF Cut of Proton Detection}
\label{C4S3}
\begin{figure}[H]
     \centering
     \includegraphics[scale=0.7]{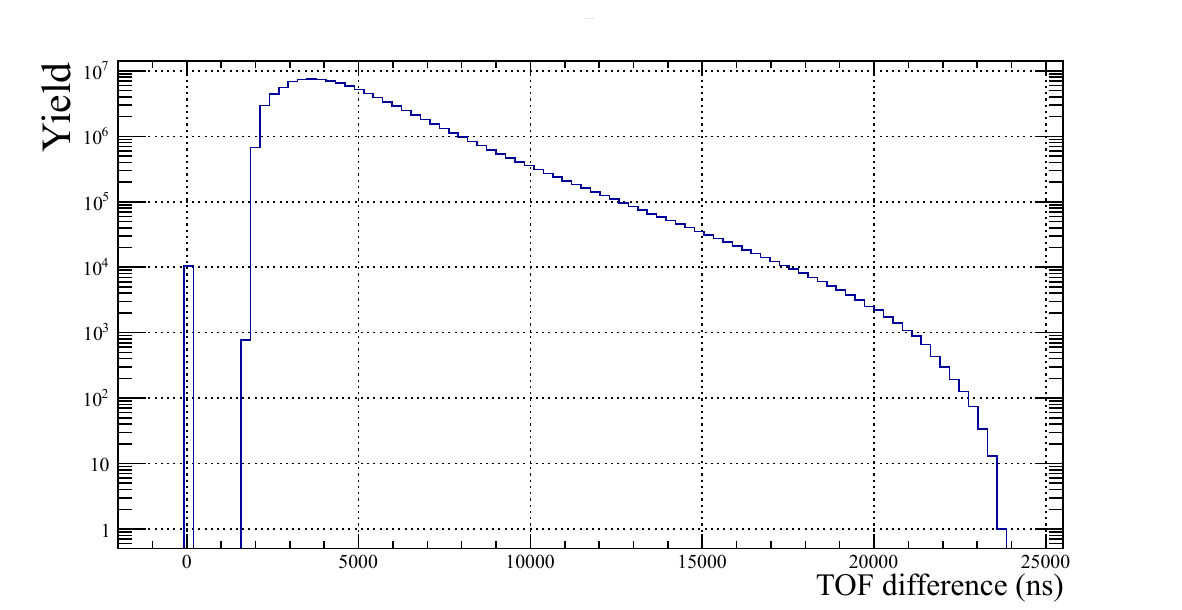}
     \caption[The TOF difference between protons and electrons.]{The TOF difference between protons and electrons. The peak at zero represents the protons or electrons that are not able reach to the detectors, this is because 1) electrons decelerated by the lower detector electric field and stopped in the detector dead layer, 2) electrons with too low energy bounced endlessly between the lower detector and the filter region, 3) protons with too low longitudinal momentum spent too long a time travelling in the spectrometer. In this study we omitted these events since they will not be recorded in the real experiment.}
     \label{TOF_diff}
 \end{figure}

\noindent
In the Nab experiment, we apply a coincidence requirement to find true decay events. In general, a coincidence time window will be applied, which means electron-proton pair candidates with TOF difference\footnote{TOF difference is defined as the TOF of the proton subtracted by the TOF of the coincidence electron} within a certain time window will be considered and accepted in into the histogram. Fig.\,\ref{TOF_diff} shows the simulation result for the TOF difference for coincident events. 

\begin{figure}[H]
     \centering
     \includegraphics[scale=0.7]{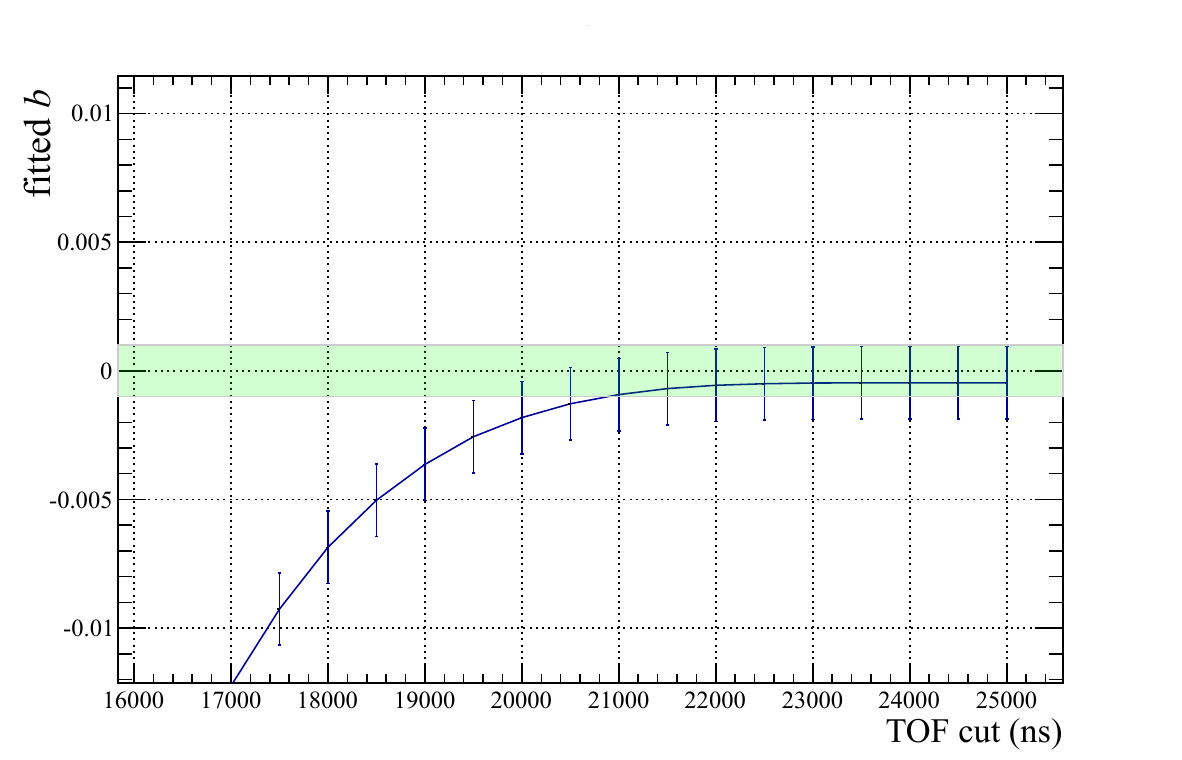}
     \caption[The relation of the fitted $b$ value to TOF cut of proton.]{The relation of the fitted $b$ value to TOF cut of proton. The green area shows the region of $\abs{b} < 10^{-3}$.}
     \label{TOF_cut}
 \end{figure}

The applied time window cut results in some events been excluded from the histogram, and therefore introduces a systematic bias to our fitted $b$ value. To study this effect, we simulated $10^8$ decay events in Geant\,4.10.2 with Nab-$b$ configuration ($b=0$) as introduced in section\,\ref{C2}. For each event, we calculate the TOF difference of proton and electron (the electron always arrives earlier than the proton), and impose a TOF difference cut on the electron energy histogram. Fig.\,\ref{TOF_cut} shows the fitted $b$ values as a function of TOF cut applied to the histogram.

From the graph we find that, the time window will make our fitted $b$ value less than expected. The fitted $b$ approaches to our input $b$ value as TOF cut increases, since a higher fraction of number of events is accepted. In order to make the systematic error of $b$ within $10^{-3}$, we should have the TOF window to be from 2\,$\mu$s to at least 21\,$\mu$s\footnote{In Fig.\,\ref{TOF_cut}, the fitted b value doesn't approach to 0 as TOF cut increases, this is due to the randomness of the simulation. Increasing the number of events of the simulation makes the asymptote of the curve closer to 0.}. 

\section{The Proton Beam Polarization}
\label{C4S54}
\begin{figure}[H]
     \centering
     \includegraphics[scale=0.7]{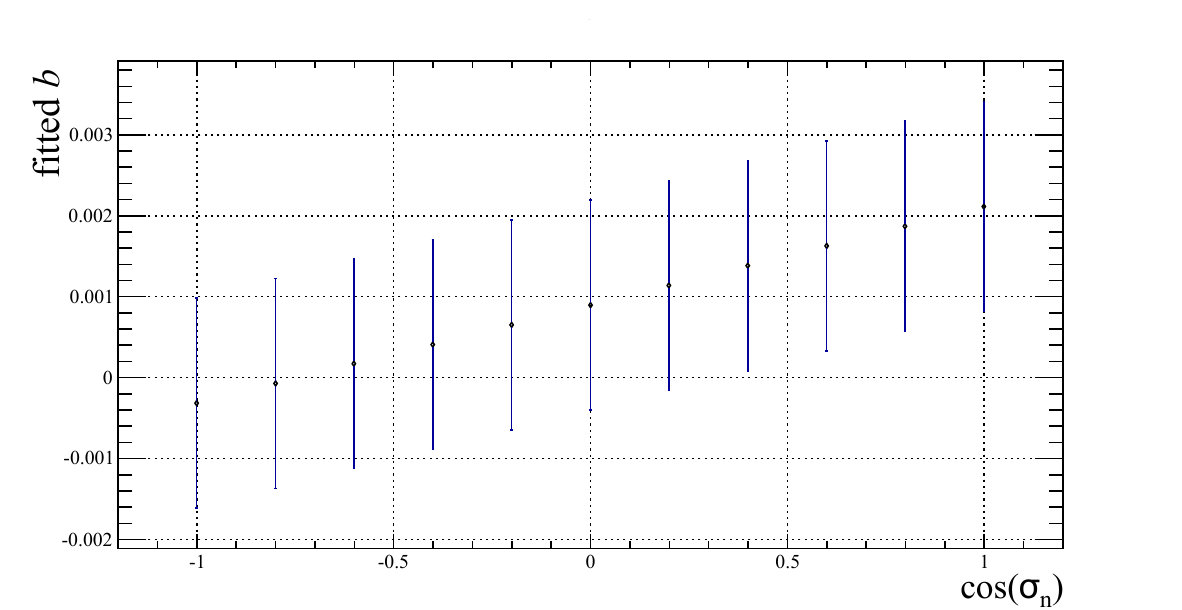}
     \caption[The relation of the fitted $b$ value to the polarization of the neutron.]{The relation of the fitted $b$ value to the polarization of the neutron. The fitted $b$ value is off from 0 but is still within the uncertainty at cos($\sigma_{n}$) = 0 due to the randomness of the events.}
     \label{Polarization}
 \end{figure}

\noindent
The neutron beam we used for the Nab experiment is considered to be unpolarized, for all the previous analysis we assumed $\langle \bm{\sigma_{n}} \rangle$ in Eq.\,\eqref{eq:3} to be 0. However, the average polarization of the neutrons cannot be perfectly 0, and any deviation can in principle affect the beta energy spectrum. Running Geant4 with $10^8$ events is really time consuming. In this study, we first created $10^8$ simulated events with 0 polarization, and introduced the polarization by re-weighting the existing histogram. To get a group of histograms with different neutron polarization, we re-weight each event using its profile as follows:

\begin{equation}
    w = \frac{1+a\frac{\bm{p_{e}} \cdot \bm{p_{\nu}}}{E_{e}E_{\nu}}+b\frac{m_{e}}{E_{e}}+ \langle \bm{\sigma_{n}} \rangle\cdot(A\frac{\bm{p_{e}}}{E_{e}}+B\frac{\bm{p_{\nu}}}{E_{\nu}})}{1+a\frac{\bm{p_{e}} \cdot \bm{p_{\nu}}}{E_{e}E_{\nu}}+b\frac{m_{e}}{E_{e}}}\,.
\end{equation}
The $\langle \bm{\sigma_{n}} \rangle$ value is set to be different values ranging from $-1$ to 1. The result is shown in Fig.\,\ref{Polarization}.

From the result, we can find that the fitted $b$ value is very weakly sensitive to the neutron polarization. From Eq.\,\eqref{eq:3} we can also find that the effect of $\langle \bm{\sigma_{n}} \rangle$ vanishes if we integrate the decay rate over 4$\pi$ solid angle. The weak sensitiveness shown in Fig.\,\ref{Polarization} is actually an artificial effect: since the data points are obtained by re-weighting the same simulation, different data points are actually correlated.

\section{The Proton Detection Efficiency}
\label{C4S5}
\begin{figure}[H]
     \centering
     \includegraphics[scale=0.7]{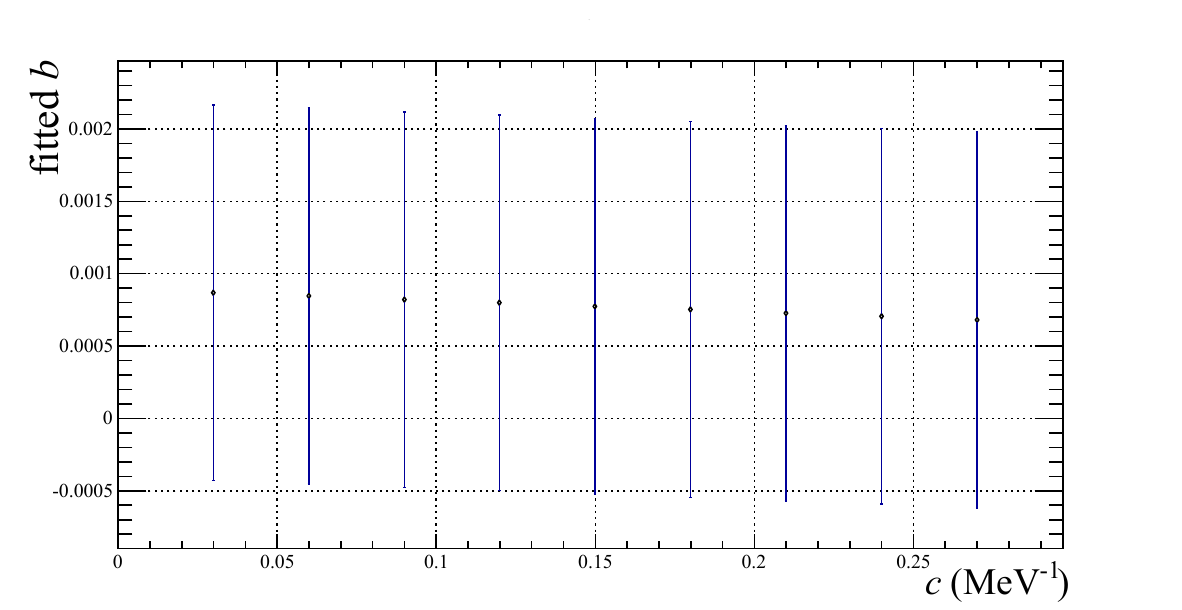}
     \caption[The relation of the fitted $b$ value to the proton detection efficiency.]{The relation of the fitted $b$ value to the proton detection efficiency. The linearly fitted parameters for the blue line is $y = \beta_0 + \beta_1 x$ with $\beta_1 = -0.00060 \pm  0.00017$ and $\beta_0 =  0.00056 \pm 0.00094$.}
     \label{proton_efficiency}
 \end{figure}

\noindent
The coincidence method requires a good detection efficiency for all protons. In the Nab-$b$ configuration, all protons are detected in the lower detector. Due to the 30\,kV accelerating voltage at the lower detector, the protons arriving at the detector surface will have kinetic energy ranging from $30$\,keV to $30.75$\,keV. The detector has different detection efficiency for protons with different energies. Here we model the detection efficiency as:

\begin{equation}
    \text{efficiency} \propto 1 + c E_p\,,
    \label{proton_eff}
\end{equation}
where $c$ is the parameter that characterizes the detection efficiency of the detector, and $E_p$ is the proton energy. In our Geant4 Monte Carlo simulation, we create histograms with different proton detection efficiency parameter ($c$) using the same reweighting method that described in the preceding subsection, and fit it with $c$ = 0\,MeV$^{-1}$. The result is shown as in Fig.\,\ref{proton_efficiency}.

\begin{figure}[H]
     \centering
     \includegraphics[scale=0.7]{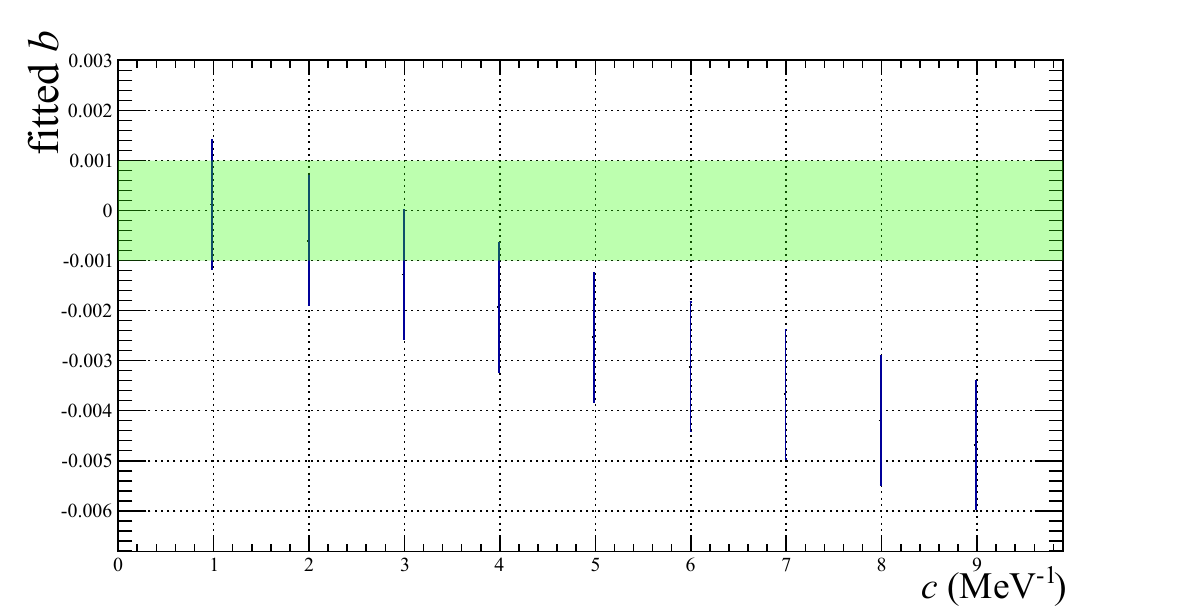}
     \caption[The relation of the fitted $b$ value to the proton detection efficiency over a larger scale.]{The relation of the fitted $b$ value to the proton detection efficiency over a larger scale. This plot shows how proton detection efficiency parameter $c$ affects the fitted $b$. Due to the limited range of proton kinetic energy, this effect is weakly sensitive.}
     \label{proton_efficiency_larger}
 \end{figure}

We can also find that the fitted $b$ value is not sensitive to the proton detection efficiency defined in Eq.\,\eqref{proton_eff}. To see this effect in a larger scale, Fig.\,\ref{proton_efficiency_larger} shows the fitted $b$ value versus a larger range of c values. From the graph we find that, to control our fitted $b$ error within $10^{-3}$, we need the proton efficiency parameter $c$ to be less than 2\,MeV$^{-1}$ = 2000\,ppm/keV, this is much less severe than the condition for the measure of the a parameter, as shown in table \ref{b_uncertainty}.

\section{The Edge Effect}
\label{C4S6}
\noindent
In the strong magnetic field of the Nab spectrometer, both electrons and protons gyrate around a guiding center which is essentially a magnetic field line. The gyration radius depends on the transverse momentum of the electron, which takes larger values for larger values of kinetic energy. A infinitely large detector capable of detecting all the electrons will not show an edge effect, but a detector with limited acceptance area will miss some of the electrons. Electrons with greater gyration radius are more likely to be missed, which introduces a distortion in the electron spectrum and, if uncorrected for, a bias in the extracted value for $b$. This is called the edge effect, and we will explain it after.

\begin{figure}[H]
    \centering
    \includegraphics[scale = 0.6]{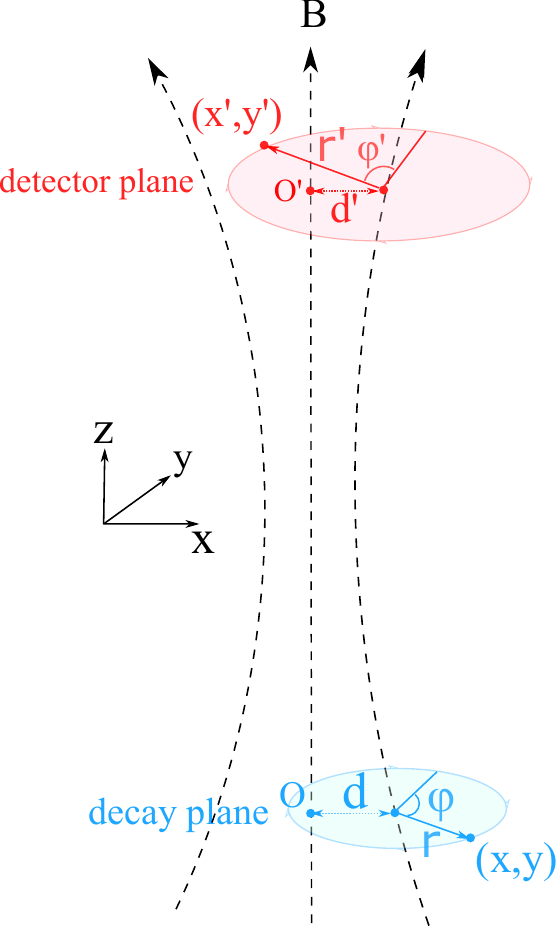}
    \caption[The schematic diagram of the electrons gyration from the decay volume to the detector.]{The schematic diagram of the electrons gyration from the decay volume to the detector. The origin of the coordinate systems is the center of the decay volume and the center of the detector plane, respectively.}
    \label{gyration}
\end{figure}

Fig.\,\ref{gyration} shows a schematic diagram of the gyration.
Assuming the neutron beam goes in the $y$ direction, our problem could be simplified to a one-dimensional problem in the $x$ direction. Consider an electron created at position $(x,y)$ in the decay region plane $O$, that hits the detector at $(x',y')$ in the detector plane $O'$. The magnetic field line guiding center of the electron is at a distance $d$ from the origin $O$. The gyration has a initial phase of $\phi$, with a gyration radius of $r$. We have:
\begin{equation}
\begin{split}
    x &= r \text{sin}(\phi) + d\,, \\
    x'&=  r' \text{sin}(\phi') + d'\,.
\end{split}
\label{gyraeq1}
\end{equation}
In adiabatic approximation\cite{Jackson:1998nia}, we have:
\begin{equation}
\begin{split}
    \frac{d'}{d} = \frac{r'}{r} = \sqrt{\frac{B}{B'}}\,.
\end{split}
\label{gyrarelation}
\end{equation}
Combine Eq.\,\eqref{gyraeq1} and Eq.\,\eqref{gyrarelation}, we have $x'$ as a function of $x$ as:
\begin{equation}
\begin{split}
    x' &= r' \text{sin}(\phi') + d'\\
    &=\sqrt{\frac{B}{B'}}r\text{sin}(\phi') + \sqrt{\frac{B}{B'}} d\\
    &=\sqrt{\frac{B}{B'}}r\text{sin}(\phi') + \sqrt{\frac{B}{B'}} [x - r \text{sin}(\phi)]\\
    &=\sqrt{\frac{B}{B'}}r[\text{sin}(\phi') - \text{sin}(\phi)] + \sqrt{\frac{B}{B'}} x\,,\\
\label{gyraeq}
\end{split}
\end{equation}
or:
\begin{equation}
\begin{split}
    \frac{1}{r}(\sqrt{\frac{B'}{B}}x' - x) = \text{sin}(\phi') - \text{sin}(\phi)\,.
\label{gyradis}
\end{split}
\end{equation}
In general, the gyration phase $\phi'$ at the detector depends on the phase at creation. However, due to the vertical extent of the decay volume that leads to a large variation of the flight path for neutrons decay with differing vertical coordinates, $\phi'$ and $\phi$ could be treated as uncorrelated variables, each with a uniform distribution in $(0, 2\pi]$. The probability distribution for $\eta =$ sin($\phi$) is then:
\begin{equation}
    p(\eta)= 2\frac{1}{2\pi} \frac{d\phi}{d\eta} \\ = \begin{cases}
     \frac{1}{\pi} \frac{1}{\sqrt{1-\eta^2}}\,, & \abs{\eta} \leq 1\,,\\
      0\,, & \abs{\eta} > 1\,.\\
    \end{cases}   \\
\end{equation}
Denote $\eta' =-\text{sin}(\phi')$, then $\eta'$ has the same distribution as $\eta$. Since $\phi$ and $\phi'$ are uncorrelated, $\eta$ and $\eta'$ are also uncorrelated, the distribution of the sum $\xi = \eta + \eta'$ is the convolution of the two individual probability distributions:
\begin{equation}
\begin{split}
q(\xi) &=  \int^{+\infty}_{-\infty} p(\eta) p(\xi-\eta)d\eta \\&= \begin{cases}
\frac{1}{\pi^2}\int^{\text{min}(1,\xi+1)}_{\text{max}(-1,\xi-1)} \frac{1}{\sqrt{1-\eta^2}} \frac{1}{\sqrt{1-(\xi-\eta)^2}}d\eta\,,& -2 < \xi < 2\,,\\
0\,, & \abs{\xi} \geq 2\,.\\
\end{cases}\\
\end{split}
\end{equation}

\begin{figure}[H]
    \centering
    \includegraphics[scale = 0.7]{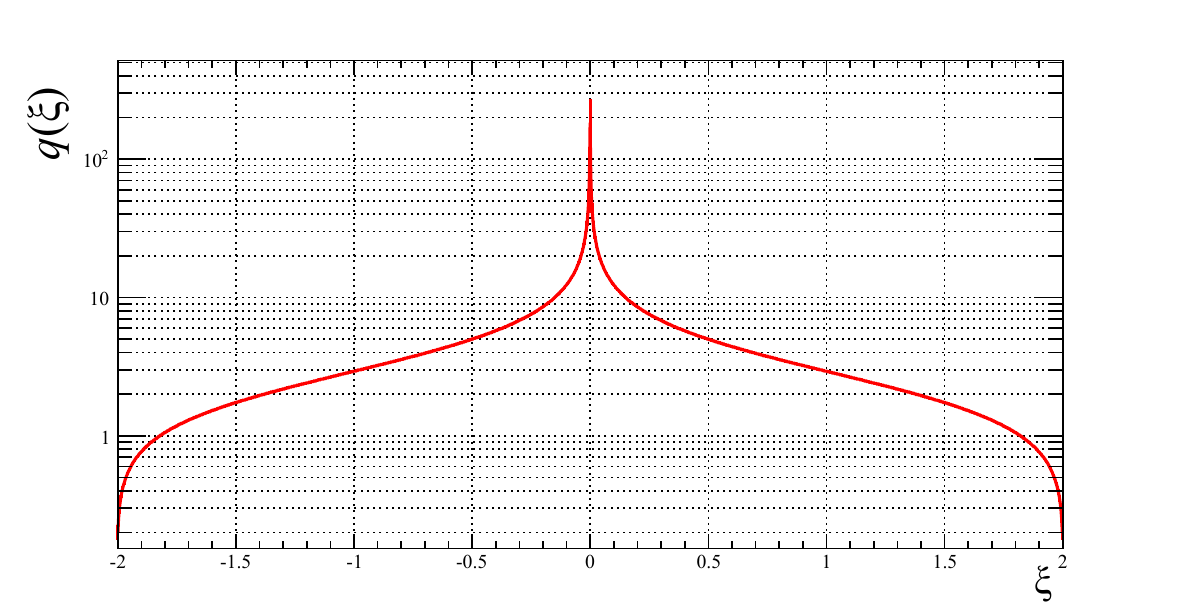}
    \caption{The plot of $q(\xi)$.}
     \label{T_x}
\end{figure}

According to Eq.\,\eqref{gyradis}, the distribution of $x' = \sqrt{\frac{B}{B'}} (r\xi + x)$ is then:
\begin{equation}
    p'(x') = q \biggl( \xi(x') \biggr) \frac{d\xi}{dx'} = \frac{1}{r}\sqrt{\frac{B'}{B}}q
    \left(\sqrt{\frac{B'}{B}}\frac{x'}{r} - \frac{x}{r}\right)\,.
\label{x'dist}
\end{equation}
The distribution $q(x)$ is plotted in Fig.\,\ref{T_x}. Together with Eq.\,\eqref{x'dist} we can see that the distribution of $x'$ is centered at $\sqrt{\frac{B}{B'}}x$ and symmetrically spread out to $2\sqrt{\frac{B}{B'}}r$ in both directions\footnote{A similar discussion is provided in \cite{2014}}. To understand this result, considering the distribution of $x'$ without gyration, the target point will be uniquely determined by mapping $x$ in the decay plane into $x'$ in the detector plane with the magnetic field, which gives a distribution of $p'(x') = \delta(x' - \sqrt{\frac{B}{B'}}x)$. With gyration, the distribution of the electron will spread out around the projection point $x' = \sqrt{\frac{B}{B'}}x$.

\begin{figure}[H]
    \centering
    \includegraphics[scale = 0.7]{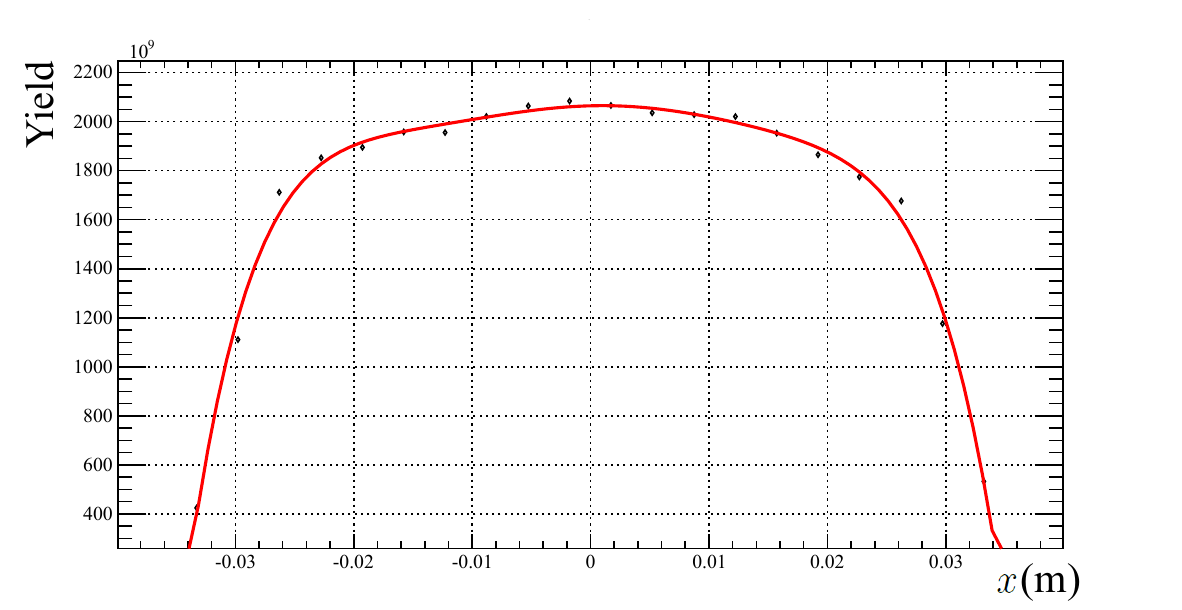}
    \caption[A neutron beam profile from MC simulation.]{A neutron beam profile from MC simulation\cite{maescott}. The markers show the simulation result and the red curve is the 6 order polynomial fitting curve of the simulation.}
     \label{neutron_profile}
\end{figure}

The gyration radius of an electron at birth is a function of its energy and momentum direction relative to the magnetic field. From the McStas simulation in \cite{maescott}, we find the neutron beam profile shown in Fig.\,\ref{neutron_profile}. The profile will be mapped onto the detector plane using Eq.\,\eqref{x'dist}. The distribution of the electrons at the detector is spread out due to the magnetic field ratio, which is not problematic. It is additionally spread out due to the gyration, which is the concern here. Due to the difference in gyration radius, electrons with higher energy are likely to spread out further, resulting a wider distribution on the detector plane. If the beam profile was uniform, this would not be a problem: due to the symmetry of the beam profile, for any acceptance region, the number of electrons going into the acceptance region from outside equals to the number of electrons going out of the acceptance region from inside. On the other hand, if the neutron beam profile is changing, those numbers will be different. Moreover, electrons with higher gyration radius will be affected more since it is more likely for these electrons to travel across the boundary of the acceptance region.

As a consequence, if we detect all the electrons, the collected energy spectrum will be a consistent representation of the neutron beta decay spectrum. But if we are only collecting electrons in a limited region, the collected spectrum will be distorted, especially close to the region where the neutron beam profile is changing rapidly, in our case, the neutron beam edge. This distortion is what we call the edge effect. If uncorrected, it leads to a bias in the fitted value for the $b$ coefficient. We note that there is no edge effect in the direction along the beam, as the decay density is constant in this direction. There is also no edge effect in the vertical direction, as we collect decays irrespective of the vertical coordinate of the starting point.

\begin{figure}[H]
    \centering
    \includegraphics[scale = 0.7]{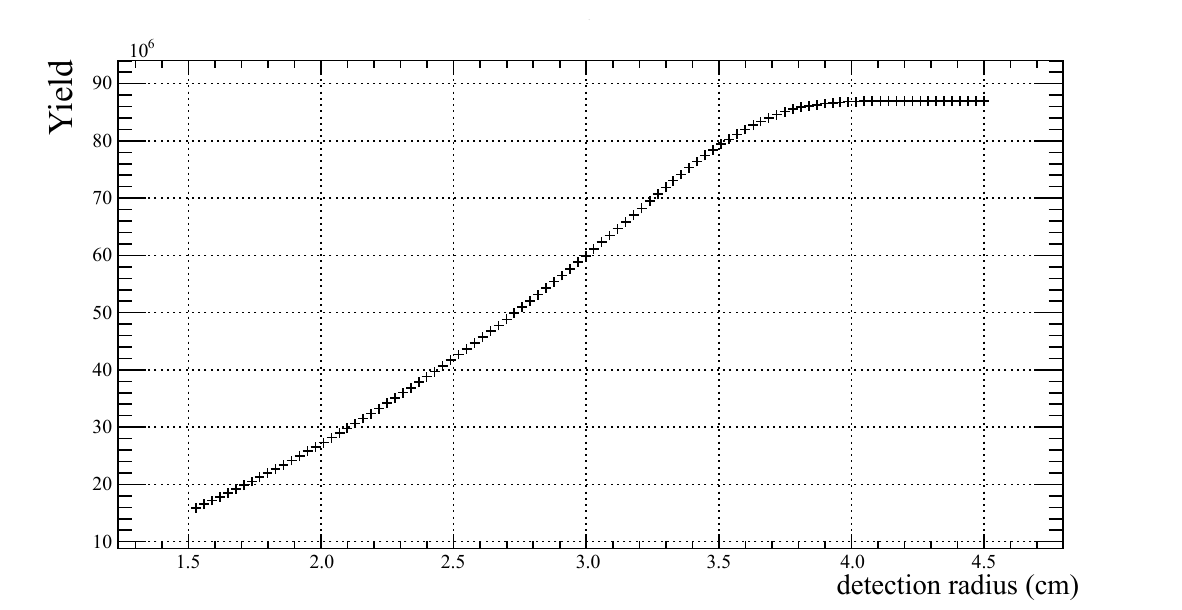}
    \caption[The number of electrons collected as a function of the cutting radius.]{The number of electrons collected as a function of the cutting radius. The detection area grows quadratically as a function of the cutting radius. As a consequence, the number of collected electrons also increases quadratically until the neutron beam edge is reached. Beyond that, the curve becomes flat.}
  \label{count_radius}
\end{figure}

To study the edge effect, we simulate decay events in Nab-$b$, and collect electrons on both detectors only within a given radius\footnote{The actual collecting shape is set to be a hexagon, to make it consistent to the Nab detector shape in Fig.\,\ref{detector}, radius here is defined as the distance from vertex to the center of the hexagon.}. In this study, the neutron beam profile is also applied to the decay region. Fig.\,\ref{count_radius} shows the number of electron counts as a function of the cut off radius. 

\begin{figure}
    \centering
    \includegraphics[scale = 0.7]{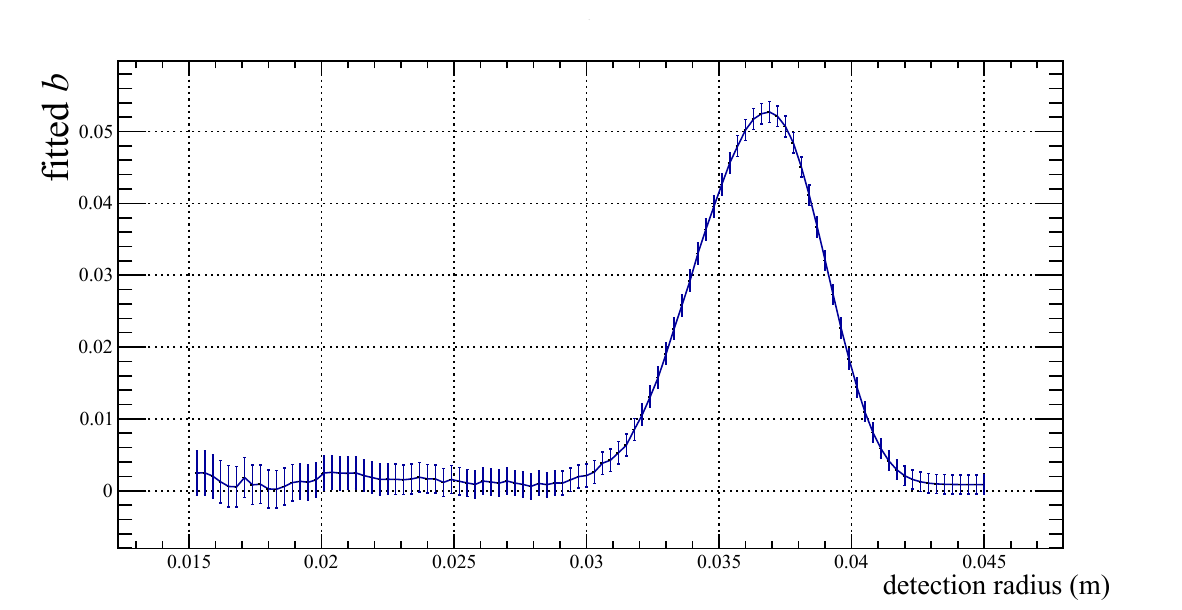}
    \caption[The fitted $b$ value as a function of the radius of the circular acceptance area on the detector.]{The fitted $b$ value as a function of the radius of the circular acceptance area on the detector. We also notice that the error bar of the fitted $b$ value increases as the radius decreases. this is because the reduction in counts. Note that we use the same simulation for all data points, and just adjust the acceptance area of the detector. The noise on the data points is highly correlated.}
 \label{fitted_b_cut_radius}
\end{figure}

Fig.\,\ref{fitted_b_cut_radius} shows the fitted $b$ value as a function of the radius cut-off. As we expected, the fitted $b$ value is biased when the cutting radius is at the range from 3.0\,cm to 4.2\,cm. As the cut off radius decreases, the fitted $b$ value goes back to the input value, this is because the neutron beam profile gets more uniform as $x$ decreases, so that the spreading effect caused by the gyration is less influential.

Note that due to the statistical randomness, the fitted $b$ value from the Nab simulation data is biased towards positive direction in comparison to the expected value of $b = 0$\footnote{This could be seen from the rightmost point of Fig.\,\ref{fitted_b_cut_radius}. If there was no bias, the fitted $b$ value would go back to $0$ as the detection radius went large enough to accept all the decay electrons.}. The actual edge effect is seen by subtracting this bias from the fitted $b$ value. To make our systematic uncertainty of fitted $b$ to be less than $10^{-3}$, we found that we need to restrict our detection radius to be less than 3\,cm, or greater than 4.5\,cm, at the cost of losing 1/2 of events. In Fig.\,\ref{fitted_b_cut_radius}, one can observe that the fitted $b$ value for all events (unaffected by an edge effect), that is, for a detection radius of $4.5$ cm, is not more different than $10^{-3}$ from the fitted $b$ value at $2.9$ cm. In the Nab experiment, we are expecting 1600 decay events per second. Assume we only record 1/2 of those, it will take less than two days to to reach a statistical uncertainty well below our systematic uncertainty goal.

\begin{figure}[H]
    \centering
    \includegraphics[scale = 0.7]{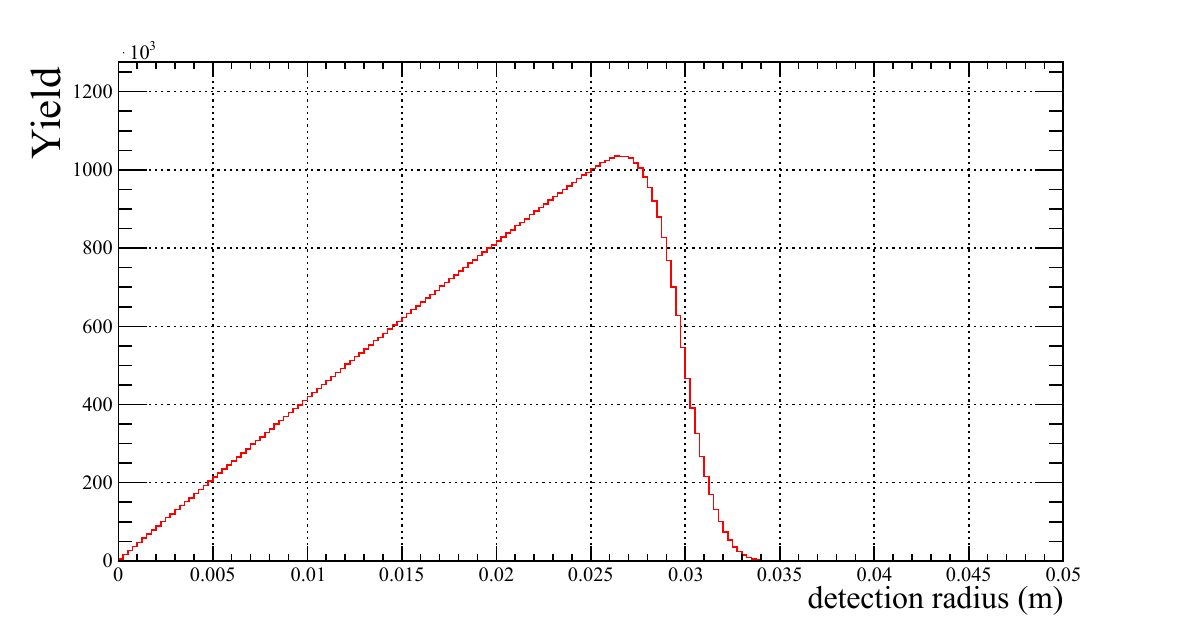}
    \caption[The detection radius for protons.]{The detection radius for protons. The red line shows the detection radius for coincidence protons, with the electron detection radius to be less than 3.0\,cm.}
  \label{proton_detection_r}
\end{figure}

In the above analysis, we imposed a detection radius cut to the decay electrons, and assumed that we detected all the coincidence protons to distinguish these decay electrons from background events. Similar to the electron detection radius, these coincidence protons also have a certain range of detection radius due to the guiding of the magnetic field. Fig.\,\ref{proton_detection_r} shows the detection radius for these coincidence protons.From the figure, we find that in order to collect the required coincidence protons, we need to detect all protons with $r<3.5$\,cm on the lower detector. 

 \section{The Summary Table For the Systematic Uncertainty of $b$}
\label{C4S7}
\noindent
As discussed in this chapter, the uncertainty table for the Fierz interference term $b$ is summarized in table\,\ref{b_uncertainty}.

\begin{table}
\begin{center}
\hspace*{-1.5cm}
\begin{tabular}{lcc}
\thickhline
 \textbf{Experimental parameter}& \textbf{Main specification} & $\Delta b$\\
\thickhline
 \textbf{Detector calibration and response}   &     \\
 \;\;  gain factor&   free parameter   & none\\ 
\;\;  offset & $\Delta$offset < 0.08\,keV&  $10^{-3}$\\
 \;\;  nonlinearity determination    & |maximum discrepancy| < 0.07\,keV&  $10^{-3}$\\
  \;\;  FWHM determination    & negligible &  none \\
   \;\;  Tail    &$\Delta$tail percentage < 1\% &  $10^{-3}$\\
 \hline
  \textbf{TOF cut}& Tof cut > 22\,$\mu$s  &  $10^{-3}$ \\
 \hline
  \textbf{Neutron beam polarization}& negligible  &  none\\
  \hline
  \textbf{Proton detection efficiency}& negligible & none  \\
  \hline
  \textbf{Edge effect}& detection radius < 2.9\,cm & $10^{-3}$ \\
\thickhline
\textbf{Sum} & & $2.2 \cross 10^{-3}$\\
\thickhline
\end{tabular}
\caption[The systematic uncertainty budget of $b$.]{\label{b_uncertainty}The systematic uncertainty budgets of $b$. For the detector calibration part, the results are based on using the gain as a free fit parameter as discussed in section\,\ref{C4S2S1}.}
\end{center}
\end{table}
\chapter{Conclusion}
\label{C5}
\noindent
The goal of the Nab collaboration is to measure the electron-antineutrino correlation parameter $a$ to $\Delta a / a \sim 10^{-3}$ and the Fierz interference term $b$ to $\Delta b \sim 3 \cross 10^{-3}$. These measurements are intended to lead to a competitive evaluation of the ratio of axial-vector to vector coupling ($\lambda$) in the weak interaction. Combing the measurement of $\lambda$ and the neutron lifetime we can also calculate the upper left element $V_{ud}$ of the CKM matrix, and test the unitarity of the CKM matrix by calculating $\abs{V_{ud}}^2+\abs{V_{us}}^2+\abs{V_{ub}}^2$. It is worth noting that the most recent result fails the test of unitarity at the 3$\sigma$ level\cite{10.1093/ptep/ptaa104}. The determination of $b$ will shed light on physics beyond the Standard Model, more specifically, the existence of scalar, and tensor type interactions in the weak interaction.

The measurement of the electron-antineutrino correlation coefficient $a$ is done by measuring the squared momentum of the proton and the energy of electron from neutron beta decay. Due to the low kinetic energy of the proton (<1\,keV), it is very difficult to directly measure the proton energy. The Nab experiment uses a 5 m long TOF region to utilize the longitude momentum of the proton, and connect the squared proton momentum with the TOF of the proton with the proton response function. The final extraction of $a$ is done by fitting the two-dimensional histogram of squared proton momentum vs.\ the electron energy plot (the tear drop plot). As shown in Eq.\,\eqref{propto_p}, for fixed electron energy, the slope of the distribution in squared proton momentum is constant and proportional to $a$.

To evaluate $b$, the electron energy spectrum is measured, and is fitted to the neutron beta decay rate in Eq.\,\eqref{espec}. In the Nab experiment, there are two silicon detectors on both side of the spectrometer. Electrons are guided by the magnetic field to fly in the spectrometer. The two detector setup enables nearly full recovery of the energy loss due to backscattering by summing up all the valid hit energies from coincidence pixels within a time window. The suppression of electron the backscattering is needed for both $a$ and $b$ measuremetns.

In the $a$ measurement, the proton TOF is measured by detecting protons in the upper detector and subtracting its detecting time by its coincidence electron detecting time.
To achieve our goal for measurement of $a$, we need to control the electrical potential difference between the decay volume and the filter region to certain degree. This is because the unwanted potential difference could accelerate the protons, and disturb the TOF measurement. The Nab experiment requires the electric potential difference between the filter region and the decay volume to be less than 10\,mV. To achieve this requirement, we used electrode pieces made by titanium to control the electric field environment in the decay region and the filter region. Due to the surface inhomogenity, the inner surface of the electrode pieces need to be treated, otherwise the inhomogenity of the work function could introduce unwanted electric field. We coated the inner surface of the electrode pieces and characterized their work function. As shown by COMSOL simulation, our coated electrodes could provide a low electrical field environment, which satisfies the experimental requirement.

The measurement of $b$ requires very good determination of the electron energy. To achieve our experimental goal, the detector response function for the electron energy needs to be characterized to a certain level. In this thesis, specifications are given for the necessary degree of knowledge of the electron energy response function, and also of other factors that could result in a systematic uncertainty of the $b$ measurement. A summary of the results could be found in table\,\ref{b_uncertainty}. The main challenge for the $b$ measurement is the calibration and characterization of the Nab detectors. As discussed in \cite{Fry:2018kvq}, these specifications are achievable.

\include{chap6}
\include{chap7}
\include{chap8}
\appendix

\chapter{The Measurement Result Of Electrode Pieces}
\label{A1}
\noindent
The detailed scanning results of the work function values from Table \ref{measureresult} are presented here. The 2 dimensional scan is done on 6 $\cross$ 6 points with a distance of 1.27 mm between points, the total scanning area for each piece is 40.32 mm$^2$. From large pieces, such as the piece that consists of the decay box, an area on the middle of each side is chosen to be scanned. For the upper drift region pieces, the middle area of the larger end is scanned. For other pieces, the scan is always done close to the middle region.

In the measurements of piece No.31 and the first scan of piece No.3, we could see that there are work function outliers, this could be caused by local bumps, damages or dust particles on the surface. \added[]{All average work function values are calculated based on all the data points}. The work function values have already been calibrated by the reference piece. The unit of all measurements is meV.

\begin{figure}[ht] 
 \centering 
 \begin{subfigure}{1\columnwidth} 
 \includegraphics[width=\linewidth] {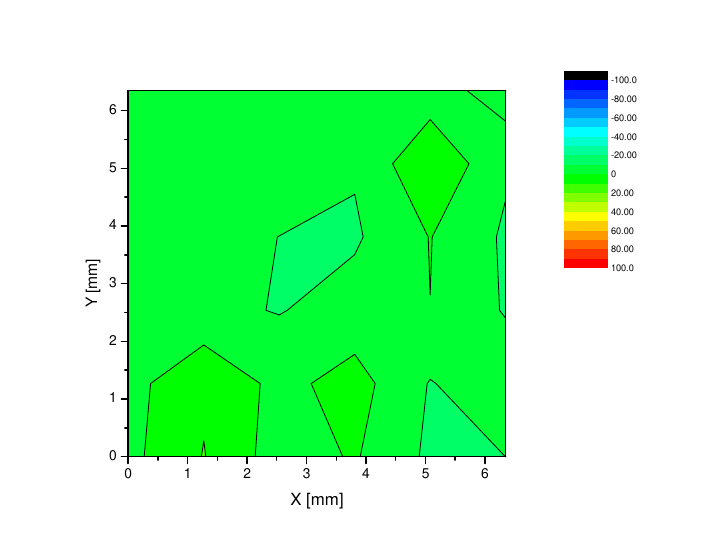} 
\caption{Work function contour plot.}    
 \end{subfigure}     
 \bigskip

 \begin{subfigure}{\columnwidth}     
 \centering     
 \renewcommand\tabularxcolumn[1]{m{#1}}     
 \renewcommand\arraystretch{1.3}     
 \setlength\tabcolsep{2pt}
\begin{tabular}{rrrrrr}
\hline
 -8.8 & -6.5 &  -7.5 &  -9.0 &  -5.9 &   6.2 \\
 -9.0 & -8.6 &  -4.1 &  -9.1 &   9.0 &  -8.6 \\
 -3.9 & -6.8 & -10.1 & -11.3 &   0.3 & -11.5 \\
 -5.3 & -7.7 & -10.5 &  -5.9 &  -0.1 & -10.9 \\
 -3.6 &  8.5 &  -2.9 &   3.9 & -10.6 &  -2.9 \\
 -2.8 & 10.4 &  -4.9 &   0.9 & -11.9 & -10.0 \\
\hline
\end{tabular}
\caption{Work function data table.}      
 \end{subfigure}     
 \caption{Piece no.1, the average work function is -4.7\,meV and its standard deviation is 6.0\,meV.}     
 \label{p1}     
 \end{figure}

\begin{figure}[ht] 
 \centering 
 \begin{subfigure}{1\columnwidth} 
 \includegraphics[width=\linewidth] {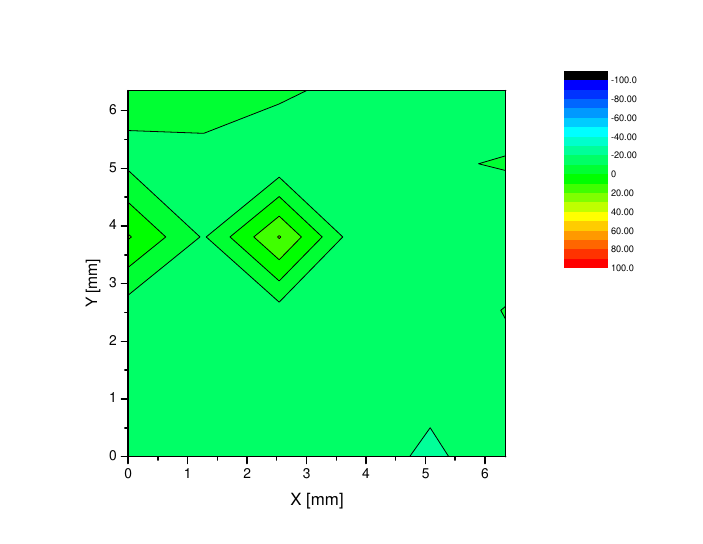} 
\caption{Work function contour plot.}    
 \end{subfigure}     
 \bigskip

 \begin{subfigure}{\columnwidth}     
 \centering     
 \renewcommand\tabularxcolumn[1]{m{#1}}     
 \renewcommand\arraystretch{1.3}     
 \setlength\tabcolsep{2pt}
\begin{tabular}{rrrrrr}
\hline
  -7.5 &  -4.4 &  -8.5 & -12.7 & -16.2 & -15.8 \\
 -12.1 & -14.0 & -16.9 & -11.5 & -11.3 &  -9.3 \\
  10.9 & -11.1 &  20.7 & -15.8 & -11.0 & -17.1 \\
 -15.5 & -12.9 & -13.8 & -11.0 & -15.8 &  -9.6 \\
 -13.0 & -13.9 & -10.7 & -15.9 & -17.6 & -12.9 \\
 -16.4 & -12.3 & -17.3 & -15.8 & -21.6 & -15.1 \\
\hline
\end{tabular}
\caption{Work function data table.}      
 \end{subfigure}     
 \caption{Piece no.1, the average work function is -11.8\,meV and its standard deviation is 7.5\,meV.}     
 \label{p1}     
 \end{figure}

\begin{figure}[ht] 
 \centering 
 \begin{subfigure}{1\columnwidth} 
 \includegraphics[width=\linewidth] {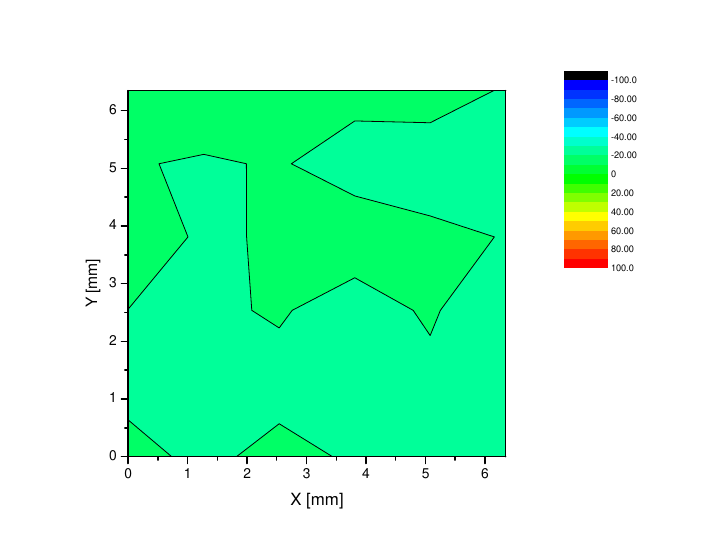} 
\caption{Work function contour plot.}    
 \end{subfigure}     
 \bigskip

 \begin{subfigure}{\columnwidth}     
 \centering     
 \renewcommand\tabularxcolumn[1]{m{#1}}     
 \renewcommand\arraystretch{1.3}     
 \setlength\tabcolsep{2pt}
\begin{tabular}{rrrrrr}
\hline
 -17.2 & -12.1 & -19.8 & -16.6 & -16.9 & -20.6 \\
 -19.2 & -21.2 & -19.1 & -24.8 & -24.0 & -21.9 \\
 -14.0 & -21.6 & -18.8 & -13.9 & -18.4 & -20.3 \\
 -20.1 & -21.8 & -19.0 & -24.9 & -18.6 & -29.1 \\
 -22.7 & -21.4 & -23.2 & -24.8 & -22.7 & -27.6 \\
 -17.3 & -22.0 & -17.4 & -21.1 & -21.8 & -26.8 \\
\hline
\end{tabular}
\caption{Work function data table.}      
 \end{subfigure}     
 \caption{Piece no.36, the average work function is -20.6\,meV and its standard deviation is 3.7\,meV.}     
 \label{p1}     
 \end{figure}

\begin{figure}[ht] 
 \centering 
 \begin{subfigure}{1\columnwidth} 
 \includegraphics[width=\linewidth] {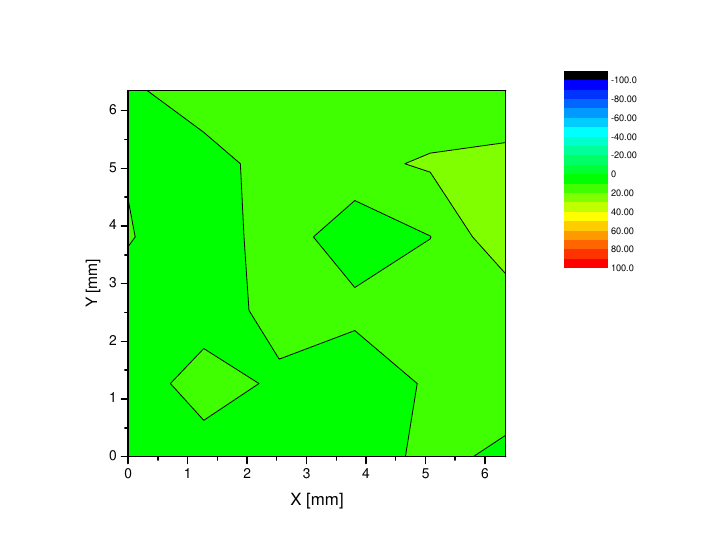} 
\caption{Work function contour plot.}    
 \end{subfigure}     
 \bigskip

 \begin{subfigure}{\columnwidth}     
 \centering     
 \renewcommand\tabularxcolumn[1]{m{#1}}     
 \renewcommand\arraystretch{1.3}     
 \setlength\tabcolsep{2pt}
\begin{tabular}{rrrrrr}
\hline
  7.4 & 17.7 & 15.5 & 10.3 & 12.2 & 15.3 \\
  9.3 &  4.2 & 16.1 & 17.4 & 21.3 & 21.9 \\
 10.7 &  3.0 & 16.1 &  2.6 &  9.8 & 28.0 \\
  5.2 &  1.3 & 15.8 & 13.3 & 16.6 & 12.1 \\
  0.0 & 17.9 &  7.1 &  1.3 & 11.8 & 14.0 \\
  8.1 &  2.2 &  3.9 &  5.5 & 12.2 &  8.3 \\
\hline
\end{tabular}
\caption{Work function data table.}      
 \end{subfigure}     
 \caption{Piece no.36, the average work function is 11.0\,meV and its standard deviation is 6.5\,meV.}     
 \label{p1}     
 \end{figure}

\begin{figure}[ht] 
 \centering 
 \begin{subfigure}{1\columnwidth} 
 \includegraphics[width=\linewidth] {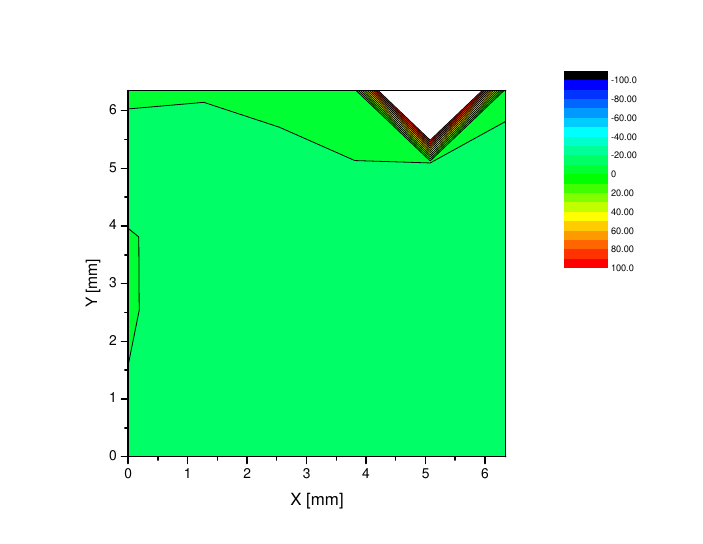} 
\caption{Work function contour plot.}    
 \end{subfigure}     
 \bigskip

 \begin{subfigure}{\columnwidth}     
 \centering     
 \renewcommand\tabularxcolumn[1]{m{#1}}     
 \renewcommand\arraystretch{1.3}     
 \setlength\tabcolsep{2pt}
\begin{tabular}{rrrrrr}
\hline
  -7.4 &  -8.9 &  -8.3 &  -6.3 & 337.8 &  -7.7 \\
 -17.9 & -15.4 & -11.8 & -10.2 & -13.6 & -13.1 \\
  -8.9 & -16.4 & -17.0 & -10.4 & -13.7 & -12.6 \\
  -9.0 & -16.1 & -10.7 & -14.7 & -12.3 & -17.5 \\
 -10.3 & -11.7 & -12.1 & -11.5 & -15.7 & -17.4 \\
 -11.7 & -14.1 & -17.4 & -15.8 & -16.3 & -17.8 \\
\hline
\end{tabular}
\caption{Work function data table.}      
 \end{subfigure}     
 \caption{Piece no.3, the average work function is -3.2\,meV and its standard deviation is 57.7\,meV.}     
 \label{p1}     
 \end{figure}

\begin{figure}[ht] 
 \centering 
 \begin{subfigure}{1\columnwidth} 
 \includegraphics[width=\linewidth] {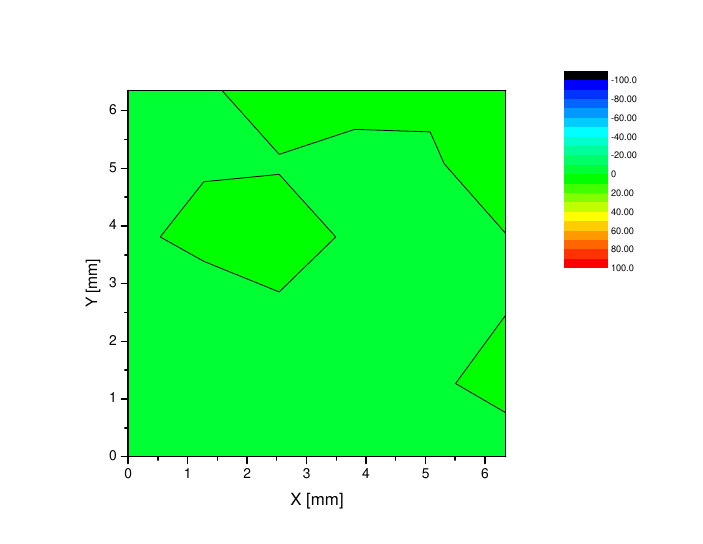} 
\caption{Work function contour plot.}    
 \end{subfigure}     
 \bigskip

 \begin{subfigure}{\columnwidth}     
 \centering     
 \renewcommand\tabularxcolumn[1]{m{#1}}     
 \renewcommand\arraystretch{1.3}     
 \setlength\tabcolsep{2pt}
\begin{tabular}{rrrrrr}
\hline
 -2.9 & -1.3 &  4.1 &  0.8 &  2.5 &  0.7 \\
 -2.8 & -1.2 & -0.6 & -0.7 & -1.9 &  8.4 \\
 -2.7 &  3.7 &  3.6 & -1.2 & -2.8 & -0.4 \\
 -2.5 & -7.4 & -1.2 & -4.7 & -7.4 & -0.1 \\
 -2.4 & -0.9 & -1.6 & -1.3 & -0.8 &  1.6 \\
 -4.8 & -5.3 & -7.0 & -2.7 & -3.2 & -2.4 \\
\hline
\end{tabular}
\caption{Work function data table.}      
 \end{subfigure}     
 \caption{Piece no.3, the average work function is -1.4\,meV and its standard deviation is 3.2\,meV.}     
 \label{p1}     
 \end{figure}

\begin{figure}[ht] 
 \centering 
 \begin{subfigure}{1\columnwidth} 
 \includegraphics[width=\linewidth] {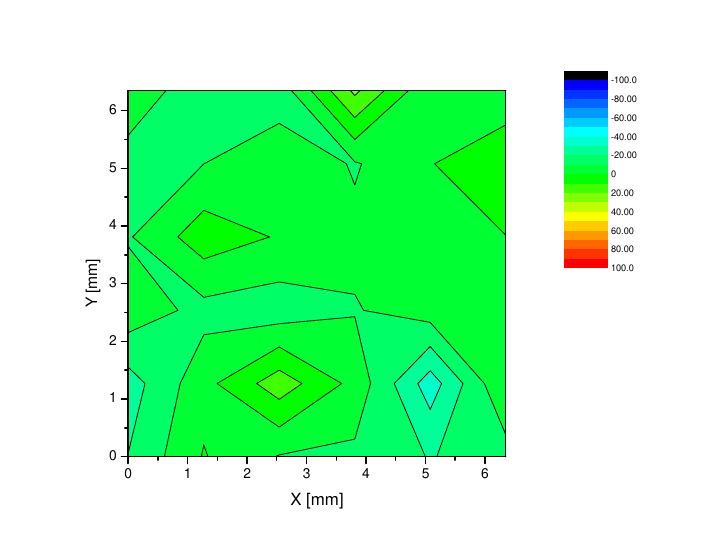} 
\caption{Work function contour plot.}    
 \end{subfigure}     
 \bigskip

 \begin{subfigure}{\columnwidth}     
 \centering     
 \renewcommand\tabularxcolumn[1]{m{#1}}     
 \renewcommand\arraystretch{1.3}     
 \setlength\tabcolsep{2pt}
\begin{tabular}{rrrrrr}
\hline
  -8.2 & -11.9 & -16.2 &  22.2 &  -9.0 &  -5.3 \\
 -11.2 & -10.1 &  -2.4 & -11.0 &  -0.4 &   5.9 \\
 -11.0 &   5.7 &  -0.9 &  -7.7 &  -6.6 &  -0.2 \\
  -3.6 & -13.4 & -15.8 & -10.7 &  -5.1 &  -9.1 \\
 -24.8 &  -3.4 &  15.7 &  -3.3 & -35.3 &  -0.1 \\
 -19.8 &   0.6 & -10.7 & -12.2 & -20.6 & -14.3 \\
\hline
\end{tabular}
\caption{Work function data table.}      
 \end{subfigure}     
 \caption{Piece no.16, the average work function is -7.3\,meV and its standard deviation is 10.4\,meV.}     
 \label{p1}     
 \end{figure}

\begin{figure}[ht] 
 \centering 
 \begin{subfigure}{1\columnwidth} 
 \includegraphics[width=\linewidth] {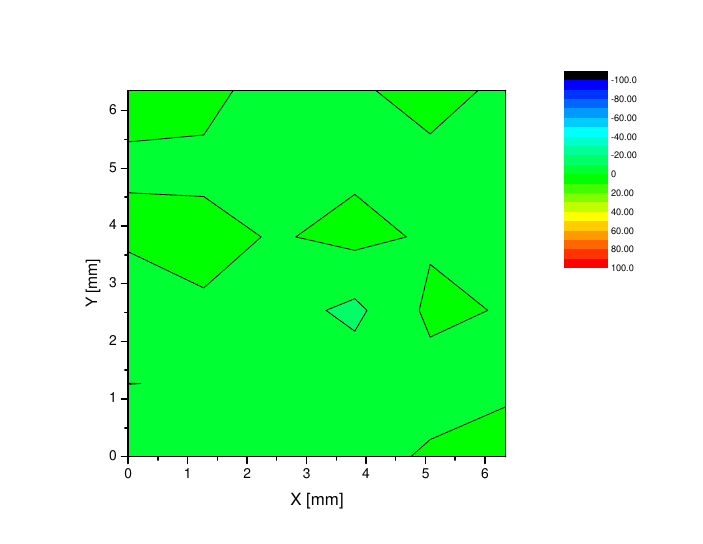} 
\caption{Work function contour plot.}    
 \end{subfigure}     
 \bigskip

 \begin{subfigure}{\columnwidth}     
 \centering     
 \renewcommand\tabularxcolumn[1]{m{#1}}     
 \renewcommand\arraystretch{1.3}     
 \setlength\tabcolsep{2pt}
\begin{tabular}{rrrrrr}
\hline
  2.0 &  3.0 & -4.8 &  -1.5 &  3.7 & -2.2 \\
 -0.9 & -2.0 & -5.7 &  -2.0 & -2.6 & -1.8 \\
  1.3 &  2.4 & -0.8 &   2.7 & -1.3 & -4.6 \\
 -5.3 & -1.1 & -6.2 & -12.4 &  2.1 & -0.7 \\
  0.0 & -0.2 & -3.8 &  -4.1 & -3.7 & -2.0 \\
 -3.2 & -6.0 & -2.6 &  -3.3 &  1.1 &  4.2 \\
\hline
\end{tabular}
\caption{Work function data table.}      
 \end{subfigure}     
 \caption{Piece no.17, the average work function is -1.7\,meV and its standard deviation is 3.3\,meV.}     
 \label{p1}     
 \end{figure}

\begin{figure}[ht] 
 \centering 
 \begin{subfigure}{1\columnwidth} 
 \includegraphics[width=\linewidth] {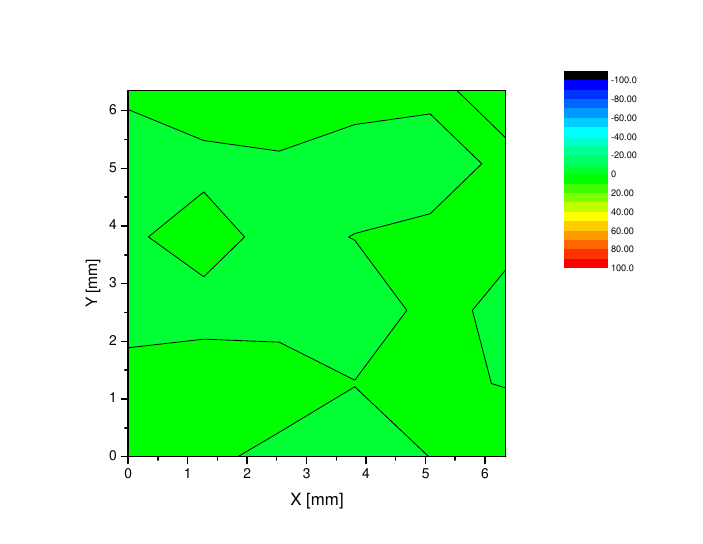} 
\caption{Work function contour plot.}    
 \end{subfigure}     
 \bigskip

 \begin{subfigure}{\columnwidth}     
 \centering     
 \renewcommand\tabularxcolumn[1]{m{#1}}     
 \renewcommand\arraystretch{1.3}     
 \setlength\tabcolsep{2pt}
\begin{tabular}{rrrrrr}
\hline
  1.3 &  3.1 &  1.1 &  3.0 &  1.3 & -2.3 \\
 -3.6 & -1.4 & -0.2 & -3.4 & -2.7 &  1.3 \\
 -0.8 &  2.3 & -1.9 &  0.2 &  1.3 &  1.1 \\
 -3.3 & -1.9 & -3.1 & -3.7 &  1.7 & -1.3 \\
  3.2 &  3.0 &  4.1 &  0.2 &  1.4 & -0.3 \\
 -0.0 &  1.7 & -2.0 & -3.7 &  0.1 &  5.1 \\
\hline
\end{tabular}
\caption{Work function data table.}      
 \end{subfigure}     
 \caption{Piece no.26, the average work function is 0.0\,meV and its standard deviation is 2.4\,meV.}     
 \label{p1}     
 \end{figure}

\begin{figure}[ht] 
 \centering 
 \begin{subfigure}{1\columnwidth} 
 \includegraphics[width=\linewidth] {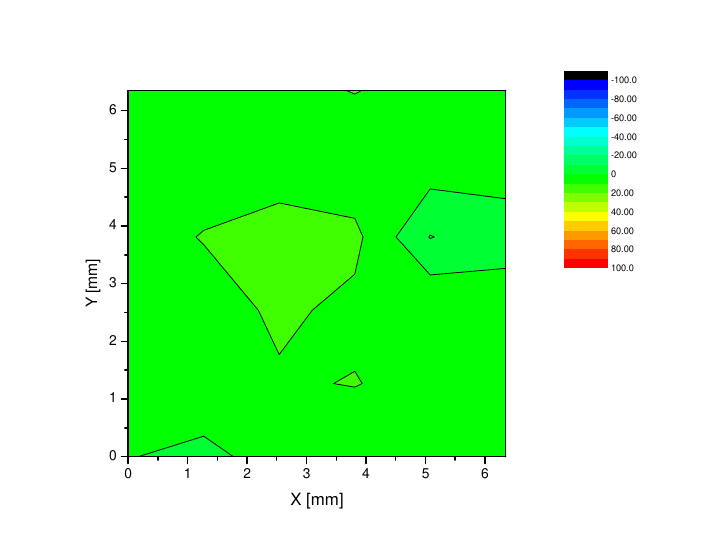} 
\caption{Work function contour plot.}    
 \end{subfigure}     
 \bigskip

 \begin{subfigure}{\columnwidth}     
 \centering     
 \renewcommand\tabularxcolumn[1]{m{#1}}     
 \renewcommand\arraystretch{1.3}     
 \setlength\tabcolsep{2pt}
\begin{tabular}{rrrrrr}
\hline
 2.7 &  5.1 &  6.9 & 10.4 &   6.1 &  1.0 \\
 3.3 &  4.2 &  7.3 &  2.8 &   5.5 &  2.6 \\
 4.9 & 10.6 & 12.4 & 12.5 & -10.4 & -2.8 \\
 0.5 &  5.1 & 11.9 &  7.6 &   9.7 &  3.8 \\
 7.3 &  6.5 &  8.8 & 10.5 &   5.6 &  2.4 \\
 0.4 & -2.5 &  3.9 &  0.8 &   4.1 &  7.7 \\
\hline
\end{tabular}
\caption{Work function data table.}      
 \end{subfigure}     
 \caption{Piece no.27, the average work function is 5.0\,meV and its standard deviation is 4.6\,meV.}     
 \label{p1}     
 \end{figure}

\begin{figure}[ht] 
 \centering 
 \begin{subfigure}{1\columnwidth} 
 \includegraphics[width=\linewidth] {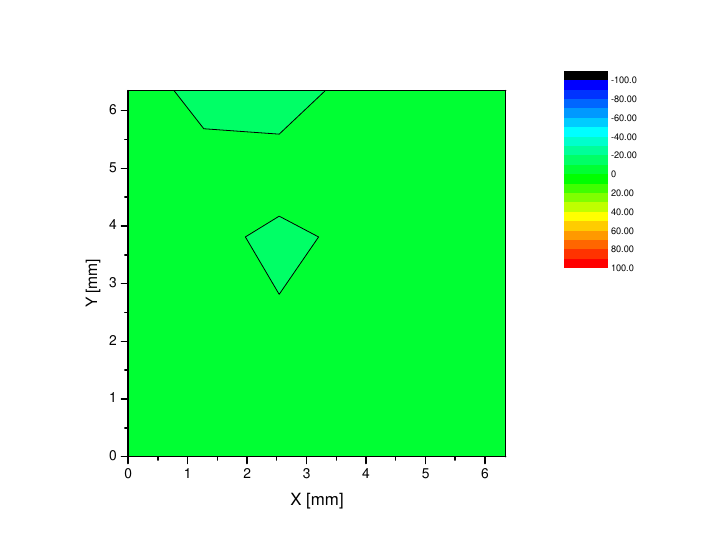} 
\caption{Work function contour plot.}    
 \end{subfigure}     
 \bigskip

 \begin{subfigure}{\columnwidth}     
 \centering     
 \renewcommand\tabularxcolumn[1]{m{#1}}     
 \renewcommand\arraystretch{1.3}     
 \setlength\tabcolsep{2pt}
\begin{tabular}{rrrrrr}
\hline
 -7.0 & -11.9 & -12.3 & -8.5 & -3.0 & -8.1 \\
 -9.9 &  -8.2 &  -8.4 & -7.4 & -6.6 & -6.5 \\
 -8.6 &  -9.2 & -10.6 & -9.4 & -2.7 & -9.0 \\
 -5.3 &  -7.9 &  -9.8 & -9.3 & -5.5 & -6.6 \\
 -8.8 &  -3.3 &  -2.5 & -5.4 & -7.1 & -9.8 \\
 -8.6 &  -9.6 &  -6.0 & -6.1 & -5.1 & -7.2 \\
\hline
\end{tabular}
\caption{Work function data table.}      
 \end{subfigure}     
 \caption{Piece no.28, the average work function is -7.6\,meV and its standard deviation is 2.4\,meV.}     
 \label{p1}     
 \end{figure}

\begin{figure}[ht] 
 \centering 
 \begin{subfigure}{1\columnwidth} 
 \includegraphics[width=\linewidth] {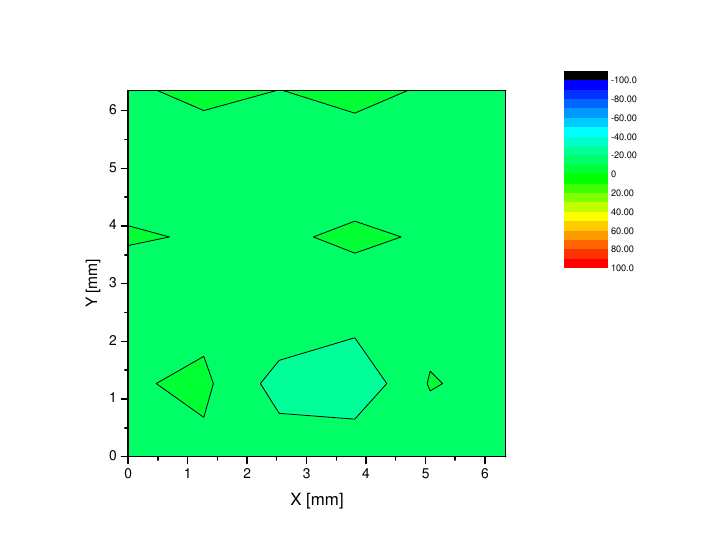} 
\caption{Work function contour plot.}    
 \end{subfigure}     
 \bigskip

 \begin{subfigure}{\columnwidth}     
 \centering     
 \renewcommand\tabularxcolumn[1]{m{#1}}     
 \renewcommand\arraystretch{1.3}     
 \setlength\tabcolsep{2pt}
\begin{tabular}{rrrrrr}
\hline
 -12.1 &  -6.6 & -10.1 &  -7.6 & -11.0 & -14.4 \\
 -14.8 & -18.9 & -11.9 & -15.3 & -13.5 & -15.7 \\
  -9.1 & -10.7 & -11.2 &  -8.5 & -10.9 & -13.5 \\
 -16.6 & -13.5 & -11.3 & -15.2 & -13.8 & -10.6 \\
 -11.2 &  -7.9 & -24.0 & -27.9 &  -9.2 & -13.9 \\
 -14.2 & -12.4 & -14.1 & -11.6 & -16.9 & -11.0 \\
\hline
\end{tabular}
\caption{Work function data table.}      
 \end{subfigure}     
 \caption{Piece no.29, the average work function is -13.1\,meV and its standard deviation is 4.2\,meV.}     
 \label{p1}     
 \end{figure}

\begin{figure}[ht] 
 \centering 
 \begin{subfigure}{1\columnwidth} 
 \includegraphics[width=\linewidth] {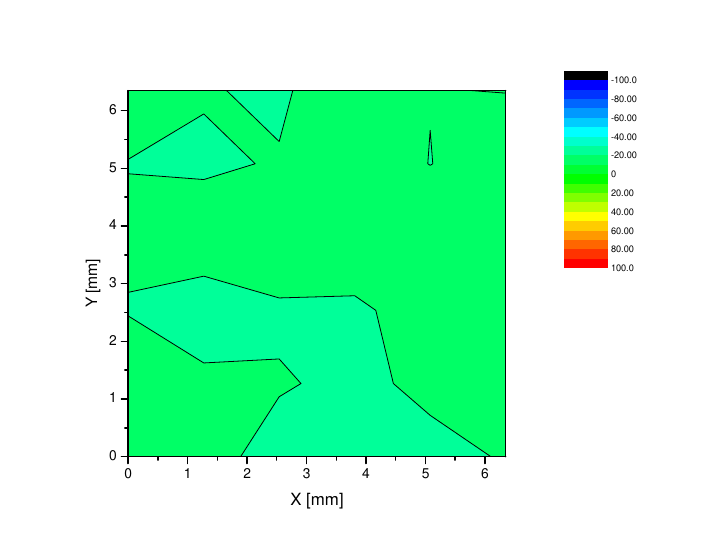} 
\caption{Work function contour plot.}    
 \end{subfigure}     
 \bigskip

 \begin{subfigure}{\columnwidth}     
 \centering     
 \renewcommand\tabularxcolumn[1]{m{#1}}     
 \renewcommand\arraystretch{1.3}     
 \setlength\tabcolsep{2pt}
\begin{tabular}{rrrrrr}
\hline
 -17.0 & -19.4 & -21.4 & -13.6 & -19.9 & -20.1 \\
 -20.2 & -21.3 & -19.4 & -17.3 & -20.1 & -17.5 \\
 -18.8 & -15.3 & -12.1 & -16.4 & -16.0 & -19.4 \\
 -20.4 & -24.1 & -21.6 & -20.9 & -17.7 & -18.5 \\
 -15.3 & -18.4 & -19.2 & -22.0 & -18.1 & -19.0 \\
 -17.0 & -16.5 & -23.7 & -22.7 & -22.5 & -19.4 \\
\hline
\end{tabular}
\caption{Work function data table.}      
 \end{subfigure}     
 \caption{Piece no.30, the average work function is -18.9\,meV and its standard deviation is 2.7\,meV.}     
 \label{p1}     
 \end{figure}

\begin{figure}[ht] 
 \centering 
 \begin{subfigure}{1\columnwidth} 
 \includegraphics[width=\linewidth] {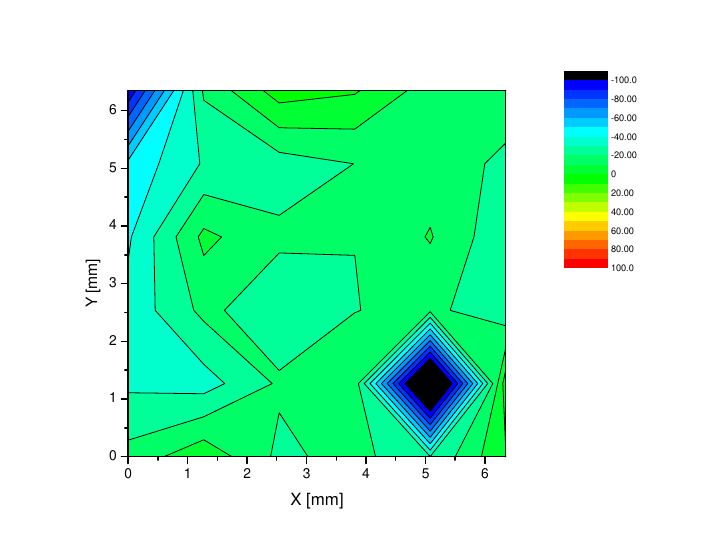} 
\caption{Work function contour plot.}    
 \end{subfigure}     
 \bigskip

 \begin{subfigure}{\columnwidth}     
 \centering     
 \renewcommand\tabularxcolumn[1]{m{#1}}     
 \renewcommand\arraystretch{1.3}     
 \setlength\tabcolsep{2pt}
\begin{tabular}{rrrrrr}
\hline
 -98.0 & -18.6 &   5.0 &   1.1 &  -14.9 & -15.0 \\
 -47.6 & -29.0 & -24.6 & -20.0 &  -14.8 & -22.0 \\
 -41.4 &  -7.5 & -18.1 & -20.0 &   -9.3 & -27.8 \\
 -37.0 & -17.4 & -26.9 & -20.2 &  -17.5 & -27.0 \\
 -31.9 & -34.4 & -18.5 & -13.5 & -145.0 &   6.1 \\
 -16.5 &  -2.7 & -22.2 & -16.4 &  -29.5 &  -0.7 \\
\hline
\end{tabular}
\caption{Work function data table.}      
 \end{subfigure}     
 \caption{Piece no.31, the average work function is -24.8\,meV and its standard deviation is 26.8\,meV.}     
 \label{p1}     
 \end{figure}

\begin{figure}[ht] 
 \centering 
 \begin{subfigure}{1\columnwidth} 
 \includegraphics[width=\linewidth] {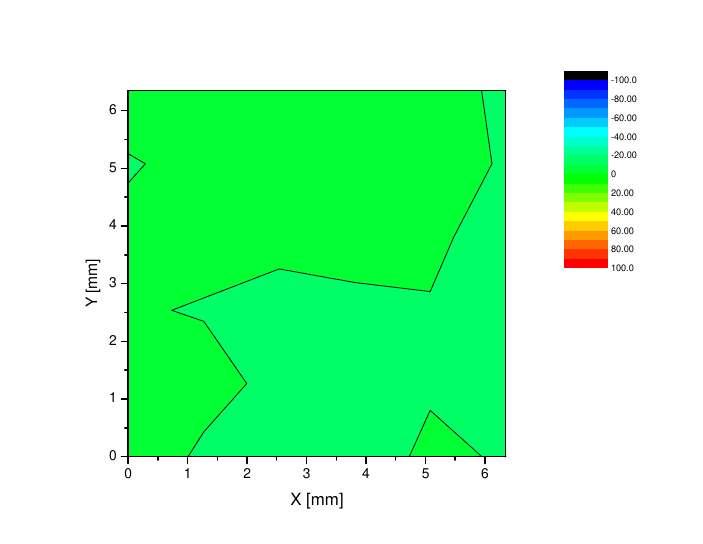} 
\caption{Work function contour plot.}    
 \end{subfigure}     
 \bigskip

 \begin{subfigure}{\columnwidth}     
 \centering     
 \renewcommand\tabularxcolumn[1]{m{#1}}     
 \renewcommand\arraystretch{1.3}     
 \setlength\tabcolsep{2pt}
\begin{tabular}{rrrrrr}
\hline
  -4.4 &  -7.0 &  -6.9 &  -7.8 &  -7.9 & -11.0 \\
 -10.9 &  -7.0 &  -9.3 &  -8.8 &  -6.0 & -10.9 \\
  -7.5 &  -7.1 &  -8.4 &  -6.4 &  -8.3 & -13.8 \\
  -9.2 & -10.6 & -12.1 & -12.2 & -10.6 & -10.9 \\
  -6.7 &  -6.7 & -12.5 & -12.1 & -11.0 & -14.1 \\
  -3.6 & -11.7 & -11.3 & -14.4 &  -8.3 & -10.8 \\
\hline
\end{tabular}
\caption{Work function data table.}      
 \end{subfigure}     
 \caption{Piece no.32, the average work function is -9.4\,meV and its standard deviation is 2.7\,meV.}     
 \label{p1}     
 \end{figure}

\begin{figure}[ht] 
 \centering 
 \begin{subfigure}{1\columnwidth} 
 \includegraphics[width=\linewidth] {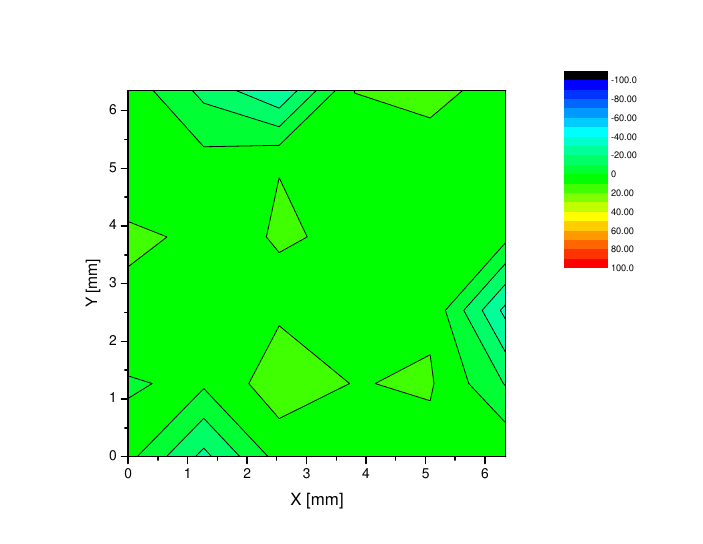} 
\caption{Work function contour plot.}    
 \end{subfigure}     
 \bigskip

 \begin{subfigure}{\columnwidth}     
 \centering     
 \renewcommand\tabularxcolumn[1]{m{#1}}     
 \renewcommand\arraystretch{1.3}     
 \setlength\tabcolsep{2pt}
\begin{tabular}{rrrrrr}
\hline
  6.4 & -12.9 & -29.5 & 10.2 & 13.9 &   4.8 \\
  2.2 &   3.9 &   9.9 &  4.1 &  3.5 &   5.9 \\
 12.1 &   8.0 &  10.4 &  9.3 &  1.3 &   2.6 \\
  6.9 &   9.3 &   8.5 &  3.6 &  8.3 & -32.9 \\
 -0.8 &   1.7 &  15.6 &  9.6 & 11.1 & -10.7 \\
  3.1 & -22.9 &   3.9 &  3.2 &  6.4 &   9.1 \\
\hline
\end{tabular}
\caption{Work function data table.}      
 \end{subfigure}     
 \caption{Piece no.33, the average work function is 2.8\,meV and its standard deviation is 11.0\,meV.}     
 \label{p1}     
 \end{figure}

\begin{figure}[ht] 
 \centering 
 \begin{subfigure}{1\columnwidth} 
 \includegraphics[width=\linewidth] {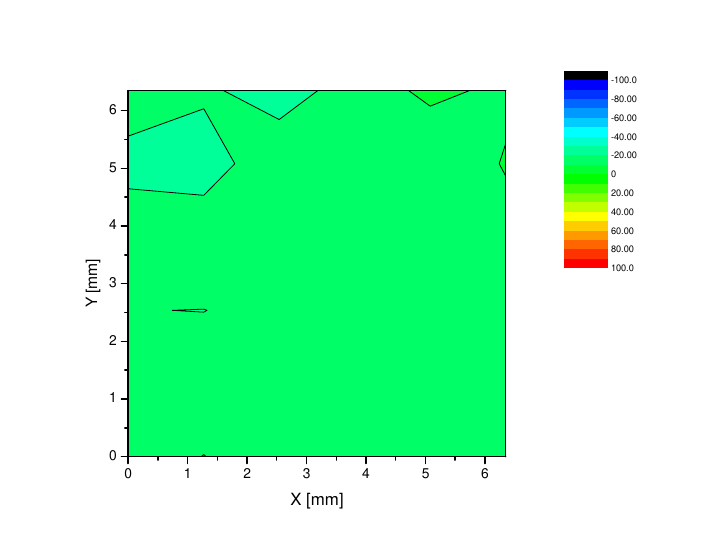} 
\caption{Work function contour plot.}    
 \end{subfigure}     
 \bigskip

 \begin{subfigure}{\columnwidth}     
 \centering     
 \renewcommand\tabularxcolumn[1]{m{#1}}     
 \renewcommand\arraystretch{1.3}     
 \setlength\tabcolsep{2pt}
\begin{tabular}{rrrrrr}
\hline
 -15.3 & -18.0 & -25.6 & -14.6 &  -8.1 & -11.7 \\
 -22.8 & -26.0 & -11.4 & -11.9 & -16.9 &  -9.3 \\
 -14.5 & -12.0 & -13.6 & -13.5 & -13.5 & -13.3 \\
 -19.8 & -20.1 & -17.3 & -12.3 & -19.8 & -13.0 \\
 -19.0 & -15.4 & -15.5 & -12.1 & -14.2 & -10.6 \\
 -15.1 &  -9.8 & -14.9 & -13.0 & -12.0 & -13.2 \\
\hline
\end{tabular}
\caption{Work function data table.}      
 \end{subfigure}     
 \caption{Piece no.18, the average work function is -15.0\,meV and its standard deviation is 4.1\,meV.}     
 \label{p1}     
 \end{figure}

\begin{figure}[ht] 
 \centering 
 \begin{subfigure}{1\columnwidth} 
 \includegraphics[width=\linewidth] {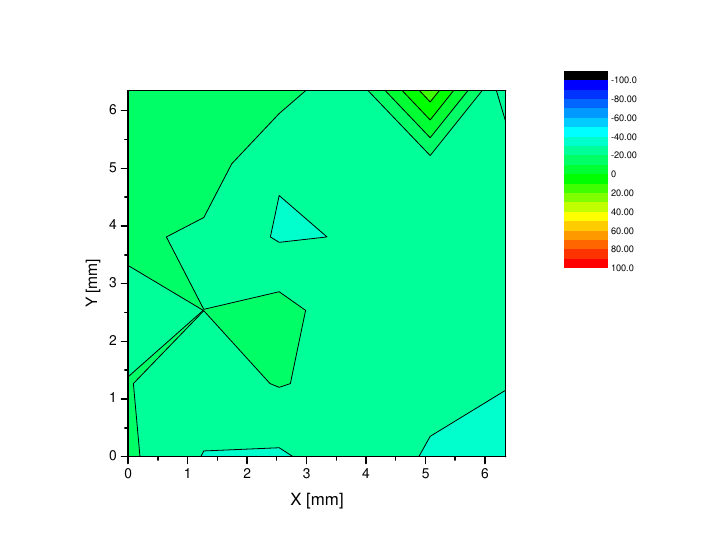} 
\caption{Work function contour plot.}    
 \end{subfigure}     
 \bigskip

 \begin{subfigure}{\columnwidth}     
 \centering     
 \renewcommand\tabularxcolumn[1]{m{#1}}     
 \renewcommand\arraystretch{1.3}     
 \setlength\tabcolsep{2pt}
\begin{tabular}{rrrrrr}
\hline
 -17.5 & -18.3 & -15.8 & -27.8 &  16.4 & -36.6 \\
 -17.1 & -14.6 & -29.2 & -21.7 & -24.6 & -21.3 \\
 -18.0 & -22.0 & -31.1 & -29.4 & -27.8 & -26.5 \\
 -23.2 & -20.0 & -16.3 & -27.0 & -27.7 & -23.5 \\
 -19.7 & -24.5 & -19.4 & -23.6 & -27.0 & -29.4 \\
 -18.1 & -30.5 & -31.5 & -23.5 & -31.2 & -36.2 \\
\hline
\end{tabular}
\caption{Work function data table.}      
 \end{subfigure}     
 \caption{Piece no.20, the average work function is -23.2\,meV and its standard deviation is 8.7\,meV.}     
 \label{p1}     
 \end{figure}

\begin{figure}[ht] 
 \centering 
 \begin{subfigure}{1\columnwidth} 
 \includegraphics[width=\linewidth] {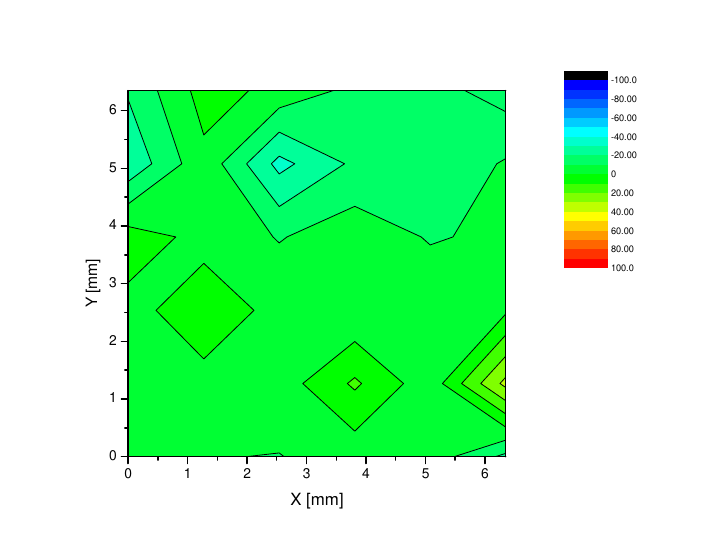} 
\caption{Work function contour plot.}    
 \end{subfigure}     
 \bigskip

 \begin{subfigure}{\columnwidth}     
 \centering     
 \renewcommand\tabularxcolumn[1]{m{#1}}     
 \renewcommand\arraystretch{1.3}     
 \setlength\tabcolsep{2pt}
\begin{tabular}{rrrrrr}
\hline
 -18.9 &  4.0 &  -2.8 & -12.9 & -16.2 & -24.6 \\
 -27.9 & -2.7 & -33.2 & -18.0 & -18.4 &  -8.9 \\
   4.6 & -2.8 & -10.7 &  -4.3 & -10.8 &  -8.2 \\
  -2.9 &  4.8 &  -2.5 &  -8.7 &  -3.5 &  -1.8 \\
  -9.1 & -2.5 &  -5.3 &  11.6 &  -6.4 &  33.0 \\
  -5.7 & -9.7 & -10.3 &  -6.3 &  -4.3 & -22.6 \\
\hline
\end{tabular}
\caption{Work function data table.}      
 \end{subfigure}     
 \caption{Piece no.22, the average work function is -7.3\,meV and its standard deviation is 11.5\,meV.}     
 \label{p1}     
 \end{figure}

\begin{figure}[ht] 
 \centering 
 \begin{subfigure}{1\columnwidth} 
 \includegraphics[width=\linewidth] {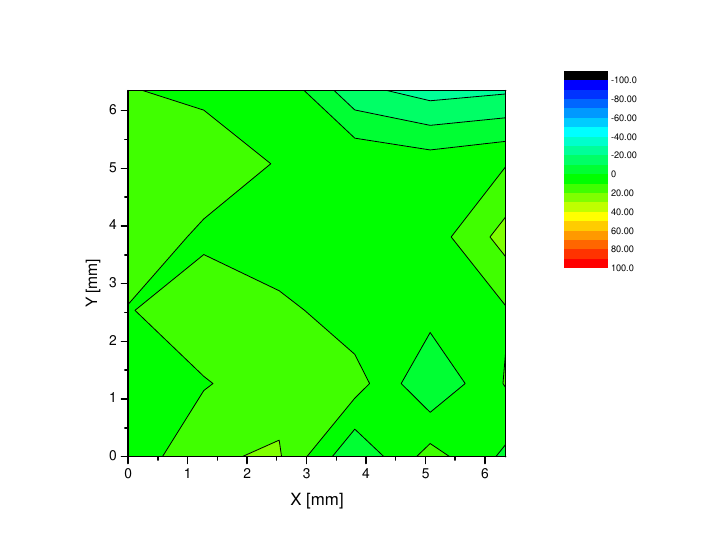} 
\caption{Work function contour plot.}    
 \end{subfigure}     
 \bigskip

 \begin{subfigure}{\columnwidth}     
 \centering     
 \renewcommand\tabularxcolumn[1]{m{#1}}     
 \renewcommand\arraystretch{1.3}     
 \setlength\tabcolsep{2pt}
\begin{tabular}{rrrrrr}
\hline
 10.7 &  6.9 &  8.3 & -16.9 & -24.2 & -21.5 \\
 13.4 & 18.4 &  8.9 &   9.1 &   5.6 &   9.4 \\
 19.8 &  7.2 &  0.0 &   8.3 &   4.6 &  24.0 \\
  9.1 & 18.7 & 13.6 &   2.8 &   4.0 &   9.1 \\
  7.6 &  9.0 & 16.8 &  14.7 &  -9.3 &  10.7 \\
  2.5 & 19.0 & 20.9 &  -9.0 &  14.2 &  -2.2 \\
\hline
\end{tabular}
\caption{Work function data table.}      
 \end{subfigure}     
 \caption{Piece no.23, the average work function is 6.8\,meV and its standard deviation is 11.1\,meV.}     
 \label{p1}     
 \end{figure}

\begin{figure}[ht] 
 \centering 
 \begin{subfigure}{1\columnwidth} 
 \includegraphics[width=\linewidth] {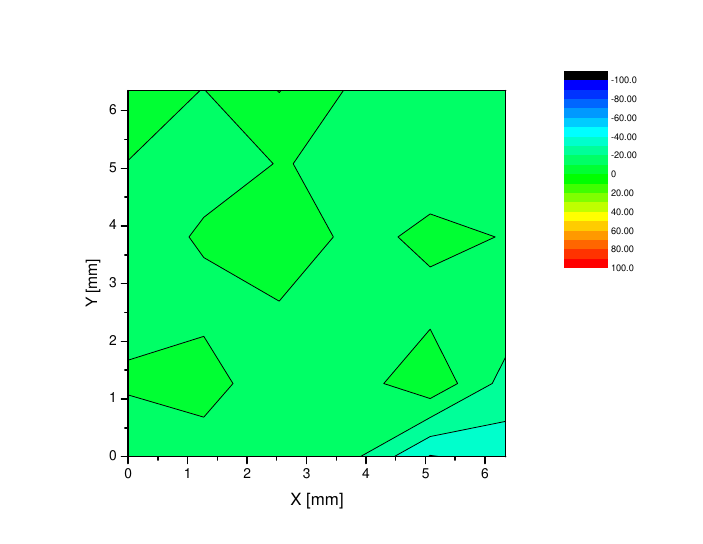} 
\caption{Work function contour plot.}    
 \end{subfigure}     
 \bigskip

 \begin{subfigure}{\columnwidth}     
 \centering     
 \renewcommand\tabularxcolumn[1]{m{#1}}     
 \renewcommand\arraystretch{1.3}     
 \setlength\tabcolsep{2pt}
\begin{tabular}{rrrrrr}
\hline
  -5.5 & -10.2 &   0.3 & -11.8 & -11.3 & -14.1 \\
 -10.2 & -12.1 &  -9.8 & -10.8 & -14.1 & -14.3 \\
 -13.2 &  -9.2 &  -3.6 & -12.5 &  -8.1 & -10.3 \\
 -12.5 & -12.0 & -10.9 & -13.4 & -12.7 & -13.2 \\
  -8.8 &  -6.3 & -15.8 & -14.9 &  -2.1 & -23.9 \\
 -16.4 & -14.3 & -15.8 & -18.4 & -40.7 & -35.7 \\
\hline
\end{tabular}
\caption{Work function data table.}      
 \end{subfigure}     
 \caption{Piece no.24, the average work function is -13.0\,meV and its standard deviation is 7.6\,meV.}     
 \label{p1}     
 \end{figure}

\begin{figure}[ht] 
 \centering 
 \begin{subfigure}{1\columnwidth} 
 \includegraphics[width=\linewidth] {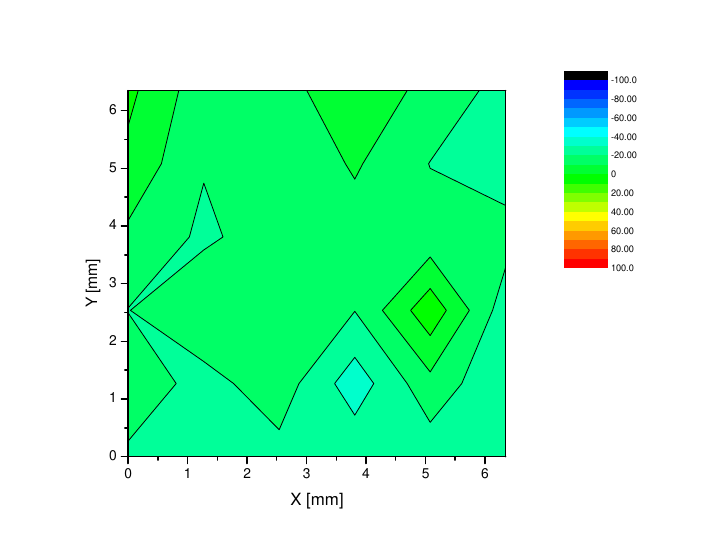} 
\caption{Work function contour plot.}    
 \end{subfigure}     
 \bigskip

 \begin{subfigure}{\columnwidth}     
 \centering     
 \renewcommand\tabularxcolumn[1]{m{#1}}     
 \renewcommand\arraystretch{1.3}     
 \setlength\tabcolsep{2pt}
\begin{tabular}{rrrrrr}
\hline
   2.4 & -16.2 & -14.6 &  -2.1 & -13.5 & -23.6 \\
  -2.7 & -19.4 & -18.8 &  -8.8 & -20.3 & -25.3 \\
 -12.2 & -21.9 & -14.7 & -14.7 & -16.4 & -16.1 \\
 -20.3 & -11.5 & -18.7 & -19.8 &   6.9 & -25.8 \\
 -13.7 & -23.7 & -14.5 & -35.7 & -13.2 & -29.6 \\
 -21.8 & -27.1 & -23.3 & -22.6 & -26.1 & -30.0 \\
\hline
\end{tabular}
\caption{Work function data table.}      
 \end{subfigure}     
 \caption{Piece no.25, the average work function is -17.4\,meV and its standard deviation is 8.8\,meV.}     
 \label{p1}     
 \end{figure}

\begin{figure}[ht] 
 \centering 
 \begin{subfigure}{1\columnwidth} 
 \includegraphics[width=\linewidth] {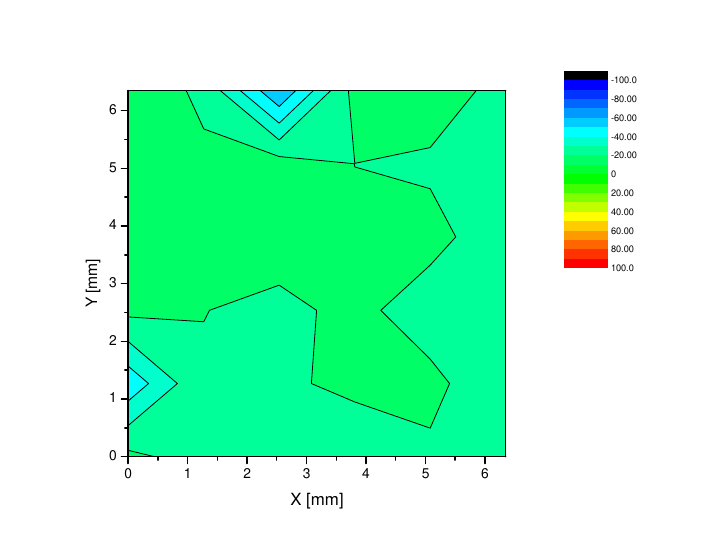} 
\caption{Work function contour plot.}    
 \end{subfigure}     
 \bigskip

 \begin{subfigure}{\columnwidth}     
 \centering     
 \renewcommand\tabularxcolumn[1]{m{#1}}     
 \renewcommand\arraystretch{1.3}     
 \setlength\tabcolsep{2pt}
\begin{tabular}{rrrrrr}
\hline
 -14.3 & -21.7 & -59.5 & -16.3 & -15.3 & -23.0 \\
 -15.7 & -18.4 & -15.7 & -20.0 & -21.3 & -21.9 \\
 -17.1 & -15.0 & -15.9 & -19.5 & -17.4 & -25.0 \\
 -17.3 & -19.8 & -22.1 & -17.8 & -24.1 & -24.4 \\
 -47.0 & -20.9 & -21.3 & -18.2 & -17.9 & -26.0 \\
 -17.4 & -24.7 & -26.2 & -25.3 & -21.3 & -22.4 \\
\hline
\end{tabular}
\caption{Work function data table.}      
 \end{subfigure}     
 \caption{Piece no.34, the average work function is -21.9\,meV and its standard deviation is 8.5\,meV.}     
 \label{p1}     
 \end{figure}

\begin{figure}[ht] 
 \centering 
 \begin{subfigure}{1\columnwidth} 
 \includegraphics[width=\linewidth] {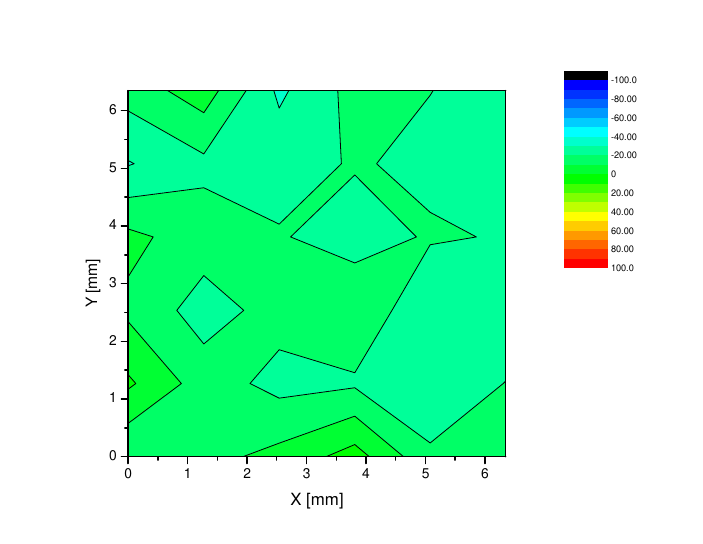} 
\caption{Work function contour plot.}    
 \end{subfigure}     
 \bigskip

 \begin{subfigure}{\columnwidth}     
 \centering     
 \renewcommand\tabularxcolumn[1]{m{#1}}     
 \renewcommand\arraystretch{1.3}     
 \setlength\tabcolsep{2pt}
\begin{tabular}{rrrrrr}
\hline
 -15.9 &  -4.7 & -32.0 & -16.6 & -19.9 & -23.7 \\
 -30.7 & -22.4 & -23.9 & -19.2 & -22.1 & -20.6 \\
  -7.5 & -15.2 & -19.2 & -24.7 & -19.0 & -20.7 \\
 -12.2 & -24.4 & -16.2 & -11.5 & -28.8 & -29.8 \\
   1.6 & -14.9 & -23.3 & -21.5 & -29.7 & -19.7 \\
 -19.8 & -13.3 &  -7.0 &   4.2 & -17.8 & -14.6 \\
\hline
\end{tabular}
\caption{Work function data table.}      
 \end{subfigure}     
 \caption{Piece no.35, the average work function is -18.2\,meV and its standard deviation is 8.2\,meV.}     
 \label{p1}     
 \end{figure}


\include{appb}

\cleardoublepage
\pdfbookmark[0]{Bibliography}{Bibliography}


\begin{singlespace}
\bibliography{main}

\begin{thebibliography}{10}

\bibitem{10.1093/ptep/ptaa104}
P.A. Zyla \textit{et al.} (Particle Data Group), Prog. Theor. Exp. Phys. \textbf{2020}, 083C01 (2020).

\bibitem{geant4}
S.~Agostinelli, J.~Allison, K.~Amako, J.~Apostolakis, H.~Araujo, P.~Arce, M.~Asai, D.~Axen, S.~Banerjee, G.~Barrand, F.~Behner, L.~Bellagamba, J.~Boudreau, L.~Broglia, A.~Brunengo, H.~Burkhardt, S.~Chauvie, J.~Chuma, R.~Chytracek, G.~Cooperman, G.~Cosmo, P.~Degtyarenko, A.~Dell'Acqua, G.~Depaola, D.~Dietrich, R.~Enami, A.~Feliciello, C.~Ferguson, H.~Fesefeldt, G.~Folger, F.~Foppiano, A.~Forti, S.~Garelli, S.~Giani, R.~Giannitrapani, D.~Gibin, J.~{Gómez Cadenas}, I.~González, G.~{Gracia Abril}, G.~Greeniaus, W.~Greiner, V.~Grichine, A.~Grossheim, S.~Guatelli, P.~Gumplinger, R.~Hamatsu, K.~Hashimoto, H.~Hasui, A.~Heikkinen, A.~Howard, V.~Ivanchenko, A.~Johnson, F.~Jones, J.~Kallenbach, N.~Kanaya, M.~Kawabata, Y.~Kawabata, M.~Kawaguti, S.~Kelner, P.~Kent, A.~Kimura, T.~Kodama, R.~Kokoulin, M.~Kossov, H.~Kurashige, E.~Lamanna, T.~Lampén, V.~Lara, V.~Lefebure, F.~Lei, M.~Liendl, W.~Lockman, F.~Longo, S.~Magni, M.~Maire, E.~Medernach, K.~Minamimoto, P.~{Mora de Freitas}, Y.~Morita, K.~Murakami, M.~Nagamatu,
  R.~Nartallo, P.~Nieminen, T.~Nishimura, K.~Ohtsubo, M.~Okamura, S.~O'Neale, Y.~Oohata, K.~Paech, J.~Perl, A.~Pfeiffer, M.~Pia, F.~Ranjard, A.~Rybin, S.~Sadilov, E.~{Di Salvo}, G.~Santin, T.~Sasaki, N.~Savvas, Y.~Sawada, S.~Scherer, S.~Sei, V.~Sirotenko, D.~Smith, N.~Starkov, H.~Stoecker, J.~Sulkimo, M.~Takahata, S.~Tanaka, E.~Tcherniaev, E.~{Safai Tehrani}, M.~Tropeano, P.~Truscott, H.~Uno, L.~Urban, P.~Urban, M.~Verderi, A.~Walkden, W.~Wander, H.~Weber, J.~Wellisch, T.~Wenaus, D.~Williams, D.~Wright, T.~Yamada, H.~Yoshida, and D.~Zschiesche.
\newblock Geant4—a simulation toolkit.
\newblock {\em Nuclear Instruments and Methods in Physics Research Section A: Accelerators, Spectrometers, Detectors and Associated Equipment}, 506(3):250--303, 2003.

\bibitem{Alarcon2010FundingPF}
R.~Alarcon, L.~Alonzi, S.~Bae{\ss{}}ler, S.~Balascuta, J.~Bowman, M.~Bychkov, J.~Byrne, J.~Calarco, T.~V. Cianciolo, C.~Crawford, E.~Frlez, M.~Gericke, F.~Gl{\"u}ck, G.~Greene, R.~Grzywacz, V.~Gudkov, F.~Hersman, A.~Klein, M.~Lehman, J.~Martin, S.~McGovern, S.~Page, A.~Palladino, S.~Penttil{\"a}, D.~Počanić, K.~Rykaczewski, W.~S. Wilburn, and A.~Young.
\newblock {Precise Measurement of $\lambda =G_A/G_V$ and Search for Non-(V$-$A) Weak Interaction Terms in Neutron Decay}.
\newblock 2010.

\bibitem{electronenergydetermination}
S.~Bae{\ss{}}ler.
\newblock {Electron energy determination. Research report, University of Virginia, unpublished}.
\newblock 2014.

\bibitem{parametric}
S.~Bae{\ss{}}ler.
\newblock {Method B and parametric studies. Research report, University of Virginia, unpublished}.
\newblock 2014.

\bibitem{1930ZPhy...66..289B}
W.~{Bothe} and H.~{Becker}.
\newblock {K{\"u}nstliche Erregung von Kern-{\ensuremath{\gamma}}-Strahlen}.
\newblock {\em Zeitschrift f\"ur Physik}, 66(5-6):289--306, May 1930.

\bibitem{BROUSSARD201783}
L.~Broussard, B.~Zeck, E.~Adamek, S.~Bae{\ss}ler, N.~Birge, M.~Blatnik, J.~Bowman, A.~Brandt, M.~Brown, J.~Burkhart, et~al.
\newblock Detection system for neutron $\beta$ decay correlations in the {UCNB} and {N}ab experiments.
\newblock {\em Nuclear Instruments and Methods in Physics Research Section A: Accelerators, Spectrometers, Detectors and Associated Equipment}, 849:83--93, 2017.

\bibitem{electrodeab}
A.~Bryant.
\newblock {{Nab Spectrometer Electrostatic Simulation Summary. Research report, University of Virginia, unpublished}}, 2021.

\bibitem{Chadwick:262756}
J.~Chadwick.
\newblock {{Intensitätsverteilung im magnetischen Spectrum der $\beta$-Strahlen von Radium B + C}}.
\newblock {\em Verhandl. Dtsch. Phys. Ges.}, 16:383, 1914.

\bibitem{Chadwick:1932ma}
J.~Chadwick.
\newblock {Possible Existence of a Neutron}.
\newblock {\em Nature}, 129:312, 1932.

\bibitem{Chadwick1934}
J.~{Chadwick} and M.~{Goldhaber}.
\newblock {A Nuclear Photo-effect: Disintegration of the Diplon by $\gamma$-Rays}.
\newblock {\em Nature}, 134(3381):237--238, Sept. 1934.

\bibitem{1935RSPSA.151..479C}
J.~{Chadwick} and M.~{Goldhaber}.
\newblock {The Nuclear Photoelectric Effect}.
\newblock {\em Proceedings of the Royal Society of London Series A}, 151(873):479--493, Sept. 1935.

\bibitem{mgc}
M.~Chemicals.
\newblock 843ar - super shield silver-coated copper conductive paint, 2021.

\bibitem{DAVIS199413}
R.~Davis.
\newblock A review of the homestake solar neutrino experiment.
\newblock {\em Progress in Particle and Nuclear Physics}, 32:13--32, 1994.

\bibitem{2014}
D.~Dubbers, L.~Raffelt, B.~Märkisch, F.~Friedl, and H.~Abele.
\newblock The point spread function of electrons in a magnetic field, and the decay of the free neutron.
\newblock {\em Nuclear Instruments and Methods in Physics Research Section A: Accelerators, Spectrometers, Detectors and Associated Equipment}, 763:112–119, Nov 2014.

\bibitem{1934ZPhy...88..161F}
E.~{Fermi}.
\newblock {{Versuch einer Theorie der {\ensuremath{\beta}}-Strahlen. I}}.
\newblock {\em Zeitschrift f\"ur Physik}, 88(3-4):161--177, Mar. 1934.

\bibitem{PhysRev.109.193}
R.~P. Feynman and M.~Gell-Mann.
\newblock {Theory of the Fermi Interaction}.
\newblock {\em Phys. Rev.}, 109:193--198, Jan 1958.

\bibitem{1937ZPhy..104..553F}
M.~{Fierz}.
\newblock {{Zur Fermischen Theorie des {\ensuremath{\beta}}-Zerfalls}}.
\newblock {\em Zeitschrift f\"ur Physik}, 104(7-8):553--565, July 1937.

\bibitem{Fry:2018kvq}
J.~Fry et~al.
\newblock The {N}ab {E}xperiment: {A} {P}recision {M}easurement of {U}npolarized {N}eutron {B}eta {D}ecay.
\newblock EPJ Web Conf. 219, 04002 (2019) doi:10.1051/epjconf/201921904002 [arXiv:1811.10047 [nucl-ex]].

\bibitem{PhysRevLett.81.1562}
Y.~Fukuda, T.~Hayakawa, E.~Ichihara, K.~Inoue, K.~Ishihara, H.~Ishino, Y.~Itow, T.~Kajita, J.~Kameda, S.~Kasuga, K.~Kobayashi, Y.~Kobayashi, Y.~Koshio, M.~Miura, M.~Nakahata, S.~Nakayama, A.~Okada, K.~Okumura, N.~Sakurai, M.~Shiozawa, Y.~Suzuki, Y.~Takeuchi, Y.~Totsuka, S.~Yamada, M.~Earl, A.~Habig, E.~Kearns, M.~D. Messier, K.~Scholberg, J.~L. Stone, L.~R. Sulak, C.~W. Walter, M.~Goldhaber, T.~Barszczxak, D.~Casper, W.~Gajewski, P.~G. Halverson, J.~Hsu, W.~R. Kropp, L.~R. Price, F.~Reines, M.~Smy, H.~W. Sobel, M.~R. Vagins, K.~S. Ganezer, W.~E. Keig, R.~W. Ellsworth, S.~Tasaka, J.~W. Flanagan, A.~Kibayashi, J.~G. Learned, S.~Matsuno, V.~J. Stenger, D.~Takemori, T.~Ishii, J.~Kanzaki, T.~Kobayashi, S.~Mine, K.~Nakamura, K.~Nishikawa, Y.~Oyama, A.~Sakai, M.~Sakuda, O.~Sasaki, S.~Echigo, M.~Kohama, A.~T. Suzuki, T.~J. Haines, E.~Blaufuss, B.~K. Kim, R.~Sanford, R.~Svoboda, M.~L. Chen, Z.~Conner, J.~A. Goodman, G.~W. Sullivan, J.~Hill, C.~K. Jung, K.~Martens, C.~Mauger, C.~McGrew, E.~Sharkey, B.~Viren,
  C.~Yanagisawa, W.~Doki, K.~Miyano, H.~Okazawa, C.~Saji, M.~Takahata, Y.~Nagashima, M.~Takita, T.~Yamaguchi, M.~Yoshida, S.~B. Kim, M.~Etoh, K.~Fujita, A.~Hasegawa, T.~Hasegawa, S.~Hatakeyama, T.~Iwamoto, M.~Koga, T.~Maruyama, H.~Ogawa, J.~Shirai, A.~Suzuki, F.~Tsushima, M.~Koshiba, M.~Nemoto, K.~Nishijima, T.~Futagami, Y.~Hayato, Y.~Kanaya, K.~Kaneyuki, Y.~Watanabe, D.~Kielczewska, R.~A. Doyle, J.~S. George, A.~L. Stachyra, L.~L. Wai, R.~J. Wilkes, and K.~K. Young.
\newblock {Evidence for Oscillation of Atmospheric Neutrinos}.
\newblock {\em Phys. Rev. Lett.}, 81:1562--1567, Aug 1998.

\bibitem{Glck2011AXISYMMETRICEF}
F.~Gl{\"u}ck.
\newblock {Axisymmetric Electric Field Calculation with Zonal Harmonic Expansion}.
\newblock {\em Progress in Electromagnetics Research B}, 32:319--350, 2011.

\bibitem{Glck2011AXISYMMETRICMF}
F.~Gl{\"u}ck.
\newblock {Axisymmetric Magnetic Field Calculation with Zonal Harmonic Expansion}.
\newblock {\em Progress in Electromagnetics Research B}, 32:351--388, 2011.

\bibitem{PhysRev.109.1015}
M.~Goldhaber, L.~Grodzins, and A.~W. Sunyar.
\newblock {Helicity of Neutrinos}.
\newblock {\em Phys. Rev.}, 109:1015--1017, Feb 1958.

\bibitem{1979STMP...85....1H}
J.~{H{\"o}lzl} and F.~K. {Schulte}.
\newblock {\em {{Work function of metals}}}, volume~85, page~1.
\newblock 1979.

\bibitem{JACKSON1957206}
J.~Jackson, S.~Treiman, and H.~Wyld.
\newblock Coulomb corrections in allowed beta transitions.
\newblock {\em Nuclear Physics}, 4:206--212, 1957.

\bibitem{Jackson:1998nia}
J.~D. Jackson.
\newblock {\em {{Classical Electrodynamics}}}.
\newblock Wiley, 1998.

\bibitem{PhysRev.106.517}
J.~D. Jackson, S.~B. Treiman, and H.~W. Wyld.
\newblock {Possible Tests of Time Reversal Invariance in Beta Decay}.
\newblock {\em Phys. Rev.}, 106:517--521, May 1957.

\bibitem{SKPmanual}
{KP Technology Ltd.}
\newblock {SKP Kelvin Probe}.
\newblock Manual Version SKP KP 4.5.

\bibitem{1956PhRv..104..254L}
T.~D. {Lee} and C.~N. {Yang}.
\newblock {{Question of Parity Conservation in Weak Interactions}}.
\newblock {\em Physical Review}, 104(1):254--258, Oct. 1956.

\bibitem{kp:sean}
S.~McGovern.
\newblock {{On the Determination of Surface Potential Variations with a Kelvin Probe. Research report, University of Virginia, unpublished}}, 2010.

\bibitem{NabSim}
D.~McLaughlin.
\newblock {{Monte Carlo Simulation of the Nab Experiment. Research report, University of Virginia, unpublished}}, 2014.

\bibitem{completekp}
H.~N. McMurray and G.~Williams.
\newblock {Probe diameter and probe-specimen distance dependence in the lateral resolution of a scanning Kelvin probe}.
\newblock {\em Journal of Applied Physics}, 91:1673--1679, 2002.

\bibitem{PhysRevD.83.073006}
G.~Mention, M.~Fechner, T.~Lasserre, T.~A. Mueller, D.~Lhuillier, M.~Cribier, and A.~Letourneau.
\newblock Reactor antineutrino anomaly.
\newblock {\em Phys. Rev. D}, 83:073006, Apr 2011.

\bibitem{michaelson1977work}
H.~B. Michaelson.
\newblock {The work function of the elements and its periodicity}.
\newblock {\em Journal of Applied Physics}, 48(11):4729--4733, 1977.

\bibitem{pich2012standard}
A.~{Pich}.
\newblock {The Standard Model of Electroweak Interactions}.
\newblock {\em arXiv e-prints}, page arXiv:1201.0537, Jan. 2012.

\bibitem{riviere1969work}
J.~C. Riviere and M.~Green.
\newblock {Work function: measurements and results, solid state surface science}.
\newblock {\em Ed. M Green}, 1, 1969.

\bibitem{10.2307/93888}
E.~Rutherford.
\newblock {Bakerian Lecture. Nuclear Constitution of Atoms}.
\newblock {\em Proceedings of the Royal Society of London. Series A, Containing Papers of a Mathematical and Physical Character}, 97(686):374--400, 1920.

\bibitem{maescott}
E.~M. Scott.
\newblock {\em {{Effects of the Nab Spectrometer on the Measurement of the Electron-Antineutrino Correlation Parameter a}}}.
\newblock PhD thesis, University of Tennessee, 2020.

\bibitem{2007}
M.~Vilches, S.~García-Pareja, R.~Guerrero, M.~Anguiano, and A.~Lallena.
\newblock {Monte Carlo simulation of the electron transport through thin slabs: A comparative study of penelope, geant3, geant4, egsnrc and mcnpx}.
\newblock {\em Nuclear Instruments and Methods in Physics Research Section B: Beam Interactions with Materials and Atoms}, 254(2):219–230, Jan 2007.

\bibitem{WILKINSON1982474}
D.~Wilkinson.
\newblock {Analysis of neutron $\beta$-decay}.
\newblock {\em Nuclear Physics A}, 377(2):474--504, 1982.

\bibitem{1957PhRv..105.1413W}
C.~S. {Wu}, E.~{Ambler}, R.~W. {Hayward}, D.~D. {Hoppes}, and R.~P. {Hudson}.
\newblock {{Experimental Test of Parity Conservation in Beta Decay}}.
\newblock {\em Physical Review}, 105(4):1413--1415, Feb. 1957.

\end{thebibliography}
\bibliographystyle{abbrv}
\end{singlespace}


\cleardoublepage

\end{document}